\documentclass[12pt]{article} 
\usepackage{amsmath} 
\usepackage{amssymb,amsfonts,euscript} 
\usepackage{hyperref} 
\renewcommand{\baselinestretch}{1.2} 
\newcommand\nn{\nonumber}

\newcommand{\Ord}{{\cal{O}}} 
\newcommand{\C}{{\cal C}} 
 
\newcommand{\D}{{\cal D}}

\newcommand{\cL}{\cal{L}} 
\newcommand{\B}{{\cal B}} 
 
\newcommand{\tg}{\tilde g} 
\newcommand{\Zint}{\mathbb{Z}} 
 
\newcommand{\bJ}{\bar{J}} 
\newcommand{\bz}{\bar{z}} 
\newcommand{\bw}{\bar{w}}

\def\a{\alpha} 
\def\be{\beta} 
 
\def\m{\mu} 
\def\n{\nu} 
\def\r{\rho} 

\def\de{\delta}

\def\tr{\mbox{Tr}}

\def\hg{\hat{g}}

\def\part{\partial}

\def\thickone{{\rm 1\mskip-4.5mu l}}

\newcommand{\sa}{\mathop{\vtop{\ialign{##\crcr 
$\hfil\displaystyle{\longrightarrow}\hfil$\crcr\noalign{\kern-1pt\nointerlineskip} 
\hspace{.12in}$^\sigma$\hskip6pt\crcr\noalign{\kern3pt}}}}} 
\newcommand{\slra}{\mathop{\vtop{\ialign{##\crcr 
$\hfil\displaystyle{\longleftrightarrow}\hfil$\crcr\noalign{\kern-1pt\nointerlineskip} 
\hspace{.12in}$^\sigma$\hskip6pt\crcr\noalign{\kern3pt}}}}} 
\newcommand{\sat}{\mathop{\vtop{\ialign{##\crcr 
$\hfil\displaystyle{\longrightarrow}\hfil$\crcr\noalign{\kern-1pt\nointerlineskip} 
\hspace{.12in}$^\sigma$\hskip6pt\crcr\noalign{\kern3pt}}}}} 
\newcommand{\pa}{\mathop{\vtop{\ialign{##\crcr 
$\hfil\displaystyle{\oplus}\hfil$\crcr\noalign{\kern+1pt\nointerlineskip} 
\hspace{.08in}$^{\alpha=0}$\hskip6pt\crcr\noalign{\kern3pt}}}}} 
\newcommand{\pan}{\mathop{\vtop{ialgin{##\crcr 
$\hfil\displaystyle{\oplus}\hfil$\crcr\noaligan{\kern+2pt\nointerlinkeskip} 
\hspace{.03in} $^{\alpha}$\hskip6pt\crcr\noalign{\kern3pit}}}}} 
\newcommand{\ka}{\mathop{\vtop{\ialign{##\crcr 
$\hfil\displaystyle{\longleftrightarrow}\hfil$\crcr\noalign{\kern-1pt\nointerlineskip} 
\hspace{.12in}$^K$\hskip6pt\crcr\noalign{\ker3pt}}}}} 
\newcommand{\bp}{\mathop{\vtop{ialign{##\crcr 
$\hfil\displaystyle{}\hfil$\crcr\noalign{\kern-13pt\nointerlineskip} 
\big{(}\hskip0pt\crcr\noalign{\kern3pt}}}}} 
\newcommand{\cbp}{\mathop{\vtop{ialign{##\crcr 
$\hfil\displaystyle{}\hfil$\crcr\noalign{\kern-13pt\nointerlineskip} 
\big{)}\hskip0pt\crcr\noalign{\kern3pt}}}}}

\newcommand{\s}{\sigma}

\newcommand{\ws}{\omega (h_\s)}

\newcommand{\hc}{$\hat{J}_{\gst}$} 
 
\newcommand{\tp}{{2\pi i}}

\newcommand{\sgb}{{\mbox{\scriptsize{\gb}}}}

\newcommand{\gfraks}{{\mbox{\scriptsize{\mbox{${\mathfrak g}$}}}}} 
\def\gb            {\mbox{$\hat{\mathfrak g}$}} 
 
\def\sm#1      {\mbox{\scriptsize $#1$}}

\def\d             {\mbox{$\mathbb D$}}

\def\srac#1#2{\smal{\frac{#1}{#2}}} 
 
\def\foot#1{\mbox{\footnotesize $#1$}} 
 
\def\tyny#1{\mbox{\tiny $#1$}} 
\def\smal#1{\mbox{\small $#1$}} 
\def\big#1{\mbox{\large $#1$}} 
\def\Big#1{\mbox{\Large $#1$}} 
\def\BIG#1{\mbox{\Huge $#1$}}

\def\hjb{\hat{\bar{J}}}

\def\gfrakh{\hat{\mathfrak g}}

\def\dual{\underset{\s}{\longrightarrow}}

\def\sg{\smal{\EuScript{G}}} 
\def\sj{{\cal J}}

\def\hc{^\dagger} 
 
\def\one{{\mathchoice {\rm 1\mskip-4mu l} {\rm 1\mskip-4mu} {\rm 1\mskip-4.5mu l} 
{\rm 1\mskip-5mu l}}} 
\def\d{\delta}

\def\nrm{{n(r)\m}}

\def\mnrn{{-n(r),\n}} 
\def\nsn{{n(s)\n}} 
\def\ntd{{n(t)\d}}

\def\mnnrnsrs{{m\!+\!n\!+\!\srac{n(r)+n(s)}{\r(\s)}}} 
\def\mnnrnsrsf{{m+n+\frac{n(r)+n(s)}{\r(\s)}}} 
\def\mnrrs{{m+\srac{n(r)}{\r(\s)}}}

\def\nrrs{{\srac{n(r)}{\r(\s)}}}
\def\nsrs{{\srac{n(s)}{\r(\s)}}}
\def\scf{{\cal F}} 
\def\sG{{\cal G}}

\def\gfrak{\mbox{$\mathfrak g$}} 
 
\def\hj{\hat{J}} 
 
\def\nsn{{n(s)\n}} 
\def\schi{{\foot{\chi}}} 
\def\schisig{{\foot{\chi(\s)}}} 

\def\ntd{{n(t)\delta}} 
\def\hc{^\dagger} 
\def\st{{\cal T}}
\def\0b{\ } 
\def\pl{\partial} 
 
\def\Nrm{{N(r)\m}}

\def\Nsn{{N(s)\n}} 
\def\Ntd{{N(t)\d}}

\def\sm{{\cal M}}

\def\sx{\smal{\EuScript{X}}}

\newcommand{\+}{\hspace{-.03in}+\hspace{-.02in}}

\def\hx{{\hat{x}}}

\def\bpl{{\bar{\pl}}}

\def\su{{ \mathfrak{su} }} 
\def\so{{ \mathfrak{so} }} 
 
\def\bigspc{{ \quad \quad \quad \quad}}
\def\gscfwt{{ \hat{\Delta}_0 (\s)}}

\def\hU{{\hat{U}}}
\def\bT{{\bar{T}}}
\def\nrrsf{{ \frac{n(r)}{\r(\s)}}}

\def\lD{{ \overleftarrow{D} }}

\def\lpl{{ \overleftarrow{\pl}}}
\def\lplw{{ \overleftarrow{\pl_w}}}

\def\sDg{{ \D_{\sgb (\s)} (\st (T,\s)) }}
\def\lsD{{ \overleftarrow{\D} }}
\def\tws{{ \hat{\omega} (\hat{h}_\s)}}
\def\tWsT{{ \hat{W} (\hat{h}_\s;T)}}
\def\tU{{ \hat{U}}}
\def\tE{{ \hat{E}}}
\def\tnrm{{ \hat{n}(r)\hat{\m}}}
\def\tNrm{{ \hat{N}(r)\hat{\m}}}
\def\tr{{ \hat{\r}(\s)}}
\def\tR{{ \hat{R}(\s)}}
\def\Id{{ \dot{I}}}
\def\Jd{{ \dot{J}}}
\def\Kd{{ \dot{K}}}
\def\tg{{ \tilde{g}}}
\def\nrmu{{ n(r)\m u}}
\def\nsnv{{ n(s)\n v}}
\def\ntdw{{ n(t)\de w}}

\def\sU{{U_{\!P}}}

\def\eps{{ \epsilon}}

\makeatletter 
\renewcommand{\@makefnmark}{\mbox{$^{\ddagger\@thefnmark}$}} 
\renewcommand{\subsection}{\@startsection 
{subsection}{2}{0pt
}{-\baselineskip}{0.5\baselineskip} 
{\normalfont\normalsize\bf}} 
\renewcommand{\section}{\@startsection 
{section}{2}{0pt
}{-\baselineskip}{0.5\baselineskip} 
{\bf\large}} 

\makeatother 
\numberwithin{equation}{section} 
\numberwithin{table}{section}

\newcounter{myfigctr}

\newcommand{\publititle}[8] 
{ 
  \vspace*{-3cm} 
  \begin{flushright} #1 \\ {\tt #2} \end{flushright} 
  \vfill 
  \begin{center}{\Large
    \bfseries #3}\end{center} 
  \vskip 8mm 
  \begin{center}{\large #4}\end{center} 
  \begin{center}{\normalsize #5}\end{center} 
  \vskip 8mm 
  \nopagebreak 
  \noindent #6 
  \vfill 
  \begin{flushleft} #7 
  \end{flushleft} 
  \hrule width 6.7cm \vskip.1mm 
  {\small #8} 
  \thispagestyle{empty} 
  \clearpage 
}

\begin{document} 
 
\publititle{ ${}$ \\ UCB-PTH-03/09 \\ LBNL-52799 \\ hep-th/0306014}{}{Twisted Open Strings from Closed Strings: \\ The WZW Orientation Orbifolds
}{M.B.Halpern$^{a}$ and C. Helfgott$^{b}$} 
{{\em Department of Physics, University of California and \\
Theoretical Physics Group,  Lawrence Berkeley National 
Laboratory \\ 
University of California, Berkeley, California 94720, USA}
\\[2mm]} {Including {\it world-sheet orientation-reversing automorphisms} $\hat{h}_\s \!\in \!H_-$ in the orbifold program, we construct the operator
algebras and twisted KZ systems of the general WZW {\it orientation orbifold} $A_g(H_-)/H_-$. We find that the orientation-orbifold sectors corresponding 
to each $\hat{h}_\s \!\in \!H_-$ are {\it twisted open} WZW strings, whose properties are quite distinct from conventional open-string orientifold sectors. 
As simple illustrations, we also discuss the classical (high-level) limit of our construction and free-boson examples on abelian $g$. 
} {$^a${\tt halpern@physics.berkeley.edu} \\ $^b${\tt helfgott@socrates.berkeley.edu} 
} 
 
\clearpage 
 
\renewcommand{\baselinestretch}{.4}\rm 
{\footnotesize 
\tableofcontents 
} 
\renewcommand{\baselinestretch}{1.2}\rm 
 
\section{Introduction}

In the last few years there has been a quiet revolution in the local theory of {\it current-algebraic orbifolds}.
Building on the discovery of {\it orbifold affine algebras} \cite{Chr,TVME} in the cyclic
permutation orbifolds, Refs.~[3-5] gave the twisted currents and stress tensor in each twisted sector of any 
current-algebraic orbifold $A(H)/H$ - where $A(H)$ is any current-algebraic conformal field theory [6-12] with a discrete 
symmetry group $H$. The construction treats all current-algebraic orbifolds at the same time, using the method of {\it eigenfields} 
and the {\it principle of local isomorphisms} to map OPEs in the symmetric theory to OPEs in the orbifold. The orbifold results 
are expressed in terms of a set of twisted tensors or {\it duality transformations}, which are discrete Fourier transforms 
constructed from the eigendata of the $H$-{\it eigenvalue problem}.

More recently, the special case of the WZW orbifolds 
\begin{gather}
\frac{A_g(H)}{H} ,\quad H\subset Aut(g) \label{Eq1.1}
\end{gather}
was worked out in further detail [13-16], introducing the {\it extended $H$-eigenvalue problem} and the {\it linkage relation} to 
include the {\it twisted affine primary fields}, the twisted vertex operator equations and the {\it twisted KZ equations} of the WZW orbifolds. 
Ref.~\cite{so2n} includes a review of the general left- and right-mover twisted KZ systems. For detailed 
information on particular classes of WZW orbifolds, we direct the reader to the following references:

$\bullet$ the WZW permutation orbifolds [13-15]

$\bullet$ the inner-automorphic WZW orbifolds \cite{Big, Perm}

$\bullet$ the (outer-automorphic) charge conjugation orbifold on $\su (n\geq 3)$ \cite{Big'}

$\bullet$ the outer-automorphic WZW orbifolds on $\so (2n)$, including the triality orbifolds \linebreak
\indent $\quad$ on $\so(8)$ \cite{so2n}.

\noindent Ref.~\cite{Big'} also solved the twisted vertex operator equations and the twisted KZ systems in an abelian 
limit\footnote{An abelian twisted KZ equation for the inversion orbifold $x \rightarrow -x$ was given earlier in Ref.~\cite{Froh}.}
to obtain the {\it twisted vertex operators} for each sector of a large class of orbifolds on abelian $g$. Moreover, Ref.~\cite{Perm} 
found the {\it general orbifold Virasoro algebra} (twisted Virasoro operators \cite{Chr,DV2}) of the WZW permutation 
orbifolds and used the general twisted KZ system to study {\it reducibility} of the general twisted affine primary field. Recent progress at 
the level of characters has been also reported in Refs.~\cite{KacTod,Chr,Ban,Birke}.

In addition to the operator formulation, there have been a number of discussions of {\it orbifold geometry} at the action level. In particular, 
Ref.~\cite{Big} also gave the {\it general WZW orbifold action}, special cases of which are further discussed in 
Refs.~\cite{Big',so2n}. The general WZW orbifold action provides the classical description of each sector of every WZW orbifold 
$A_g (H)/H$ in terms of appropriate {\it group orbifold elements} with diagonal monodromy, which are the classical (high-level) limit of the twisted affine 
primary fields. Moreover, Ref.~\cite{Fab} gauged the general WZW orbifold action by general twisted gauge groups to obtain the {\it 
general coset orbifold action}, which describes each sector of the general coset orbifold $A_{g/h} (H)/H$. Finally, the geometric description was 
extended in Ref.~\cite{Geom} to include a large class of {\it sigma-model orbifolds} and their corresponding {\it twisted Einstein equations}.

In the present paper, we extend the orbifold program beyond the space-time orbifolds above to construct a new class of orbifolds that we will call the 
{\it WZW orientation orbifolds}:
\begin{gather}
\frac{A_g (H_-)}{H_-} ,\quad H_- \subset Aut (g\oplus g) \,. \label{Eq1.2}
\end{gather}
Here $H_-$ is any automorphism group which contains {\it world-sheet orientation-reversing automorphisms}. Although we begin as usual with closed-string
WZW models, the orientation-orbifold sectors which arise by twisting the orientation-reversing automorphisms are in fact {\it twisted open WZW strings}, 
for which we work out the relevant twisted (fractionally-moded) operator algebras and {\it twisted open-string KZ systems}.

The fundamental reason that these sectors are open strings is that each possesses only a {\it single untwisted Virasoro algebra}, and the reader may prefer
to begin with the simple explanation of this mechanism given in Subsec.~$3.1$. Similarly, the reader will find that the open-string nature of these sectors
is particularly transparent in Subsec.~$6.2$, where we have worked out the general free-boson example on abelian $g$ (see e.~g. Eqs.~\eqref{hx-BCs}, \eqref{hx-Mode}). 

At a more technical level, we find that our open-string orientation-orbifold sectors exhibit all relevant open-string structure, including what is 
known from the open-string picture of untwisted open WZW strings \cite{Giusto}. Such features include world-sheet identifications, simultaneous left- and
right-mover actions of the twisted currents and the twisted stress tensors on the twisted affine primary fields, and the interaction of charges with
image charges in the twisted KZ systems.

The twisted open-string sectors of the orientation orbifolds are however {\it not} open-string sectors of conventional orientifolds [25-27] 
-- in part because conventional orientifolds do not involve fractional moding. Other differences are noted in Subsecs.~$3.1$, $3.4$ and $6.3$.

\section{General Orientation-Reversing Automorphisms}

\subsection{The Primary Fields of Affine $(g \oplus g)$}

For any Lie algebra $g=\oplus_I \gfrak^I$ with simple compact $\gfrak^I$ we have the OPEs of affine $(g\oplus g)$ \cite{Kac,Moody,BH,ICFT}:
\vspace{-0.07in}
\begin{subequations} 
\label{Eq2.1}
\begin{gather}
J_a (z) J_b (w) = \frac{G_{ab}}{(z-w)^2} + \frac{if_{ab}{}^c J_c (w)}{z-w} + \Ord (z-w)^0 \\
\bJ_a (\bz) \bJ_b (\bw) = \frac{G_{ab}}{(\bz-\bw)^2} + \frac{if_{ab}{}^c \bJ_c (\bw)}{\bz-\bw} + \Ord (\bz-\bw)^0 \\
J_a(z) \bJ_b(\bw) = \Ord (z-\bw)^0 ,\quad a,b,c=1,\ldots ,\text{dim }g
\end{gather}
\begin{gather}
J_a (z) g(T,\bw,w) = \frac{g(T,\bw,w) T_a}{z-w} +\Ord (z-w)^0 \\
\bJ_a (\bz) g(T,\bw,w) = -\frac{T_a g(T,\bw,w)}{\bz-\bw} +\Ord (z-\bw)^0 \\
G_{ab} =\oplus_I k_I \eta_{a(I)b(I)}^I ,\quad f_{ab}{}^c = \oplus_I f_{a(I)b(I)}^I{}^{c(I)} ,\quad T_a =\oplus_I T_a^I ,\quad [T_a ,T_b] =if_{ab}{}^c T_c \,.
\end{gather}
\end{subequations}
These and all following OPEs hold for $|z| \!>\! |w|$. Here $g(T,\bw,w)$ is the affine primary field in representation $T$ of affine $(g \oplus g)$. Our 
normalization convention for the representation matrices $T$ is given in Refs.~[13-15,23]. The index structure of the affine primary 
fields is
\begin{subequations}
\label{Eq2.2}
\begin{gather}
g(T,\bz,z)_{\a(I)I}{}^{\be(J)J} =\de_I{}^J g_I (T^I ,\bz,z)_{\a(I)}{}^{\be(I)} \\
(T_a)_{\a(I)I}{}^{\be(J)J} =\de_I{}^J (T_a^I )_{\a(I)}{}^{\be(I)} ,\quad \a (I) ,\be(I) =1,\ldots ,\text{dim }T^I 
\end{gather}
\end{subequations}
where $I$ labels the simple components $\gfrak^I$ of $g$.

To study orientation-reversing automorphisms of affine $(g\oplus g)$, we will need in fact an {\it extended}, partly redundant set of OPEs:
\begin{subequations}
\label{Eq2.3}
\begin{gather}
J_a (z) J_b (w) = \frac{G_{ab}}{(z-w)^2} + \frac{if_{ab}{}^c J_c (w)}{z-w} + \Ord (z-w)^0 \\
\bJ_a (z) \bJ_b (w) = \frac{G_{ab}}{(z-w)^2} + \frac{if_{ab}{}^c \bJ_c (w)}{z-w} + \Ord (z-w)^0 \\
J_a(z) \bJ_b(w) = \Ord (z-w)^0 
\end{gather}
\begin{gather}
J_a (z) g(T,\bw,w) = \frac{g(T,\bw,w) T_a}{z-w} +\Ord (z-w)^0 \\
\bJ_a (z) g(T,\bw,w) = -\frac{T_a g(T,\bw,w)}{z-\bw} +\Ord (z-w)^0 \\
J_a (z) g(\bT,w,\bw)^t = -\frac{T_a g(\bT,w,\bw)^t}{z-\bw} + \Ord (z-w)^0 \\
\bJ_a (z) g(\bT,w,\bw)^t = \frac{g(\bT,w,\bw)^t T_a}{z-w} +\Ord (z-w)^0 
\end{gather}
\begin{gather}
\bT_a \equiv -(T_a)^t ,\quad (T_a^t )_{\a (I)I}{}^{\be (J)J} \!\equiv \!(T_a)_{\be (J)J}{}^{\a(I)I} \\
(g(T,\bz,z)^t )_{\a (I)I}{}^{\be (J)J} = g(T,\bz,z)_{\be (J)J}{}^{\a(I)I} \,.
\end{gather}
\end{subequations}
Here we have everywhere replaced $\bz$ by $z$, superscript $t$ is matrix transpose, and the OPEs of the transposed operators $g(\bT)^t$
follow from those of the original operators. 

We consider now the action of a general world-sheet orientation-reversing automorphism $\hat{h}_\s$:
\vspace{-0.05in}
\begin{gather}
\label{Eq2.4}
\hat{h}_\s \equiv \tau_1 \times h_\s ,\quad \hat{h}_\s \in H_- \subset Aut (g\oplus g) ,\quad h_\s \in H \subset Aut (g)
\end{gather}\vspace{-0.15in}
\begin{subequations}
\label{Eq2.5}
\begin{gather}
J_a (z)' = \ws_a{}^b \bJ_b (z) ,\quad \bJ_a (z)' =\ws_a{}^b J_b (z) \label{Eq 2.5a} \\
g(T,\bz,z)' = W(h_\s ;T) g(\bT,z,\bz)^t W\hc (h_\s;T) \label{Eq 2.5b} \\
g(\bT,z,\bz)^{t\,\prime}  =W(h_\s;T) g(T,\bz,z) W\hc (h_\s;T) \label{Eq 2.5c}
\end{gather}
\begin{gather}
\ws_a{}^c \ws_b{}^d G_{cd} = G_{ab} ,\quad \ws_a{}^d \ws_b{}^e f_{de}{}^c = f_{ab}{}^d \ws_d{}^c \label{Eq 2.5d} \\
W\hc (h_\s;T) T_a W(h_\s;T) = \ws_a{}^b T_b ,\quad W\hc (h_\s ;\bT) \bT_a W(h_\s ;\bT) = \ws_a{}^b \bT_b \label{Eq 2.5e} \\
W(h_\s ;\bT) =W^\ast (h_\s ;T) \,. \label{Eq 2.5f}
\end{gather}
\end{subequations}
Here $\tau_1$ is world-sheet parity, $h_\s$ is any Lie automorphism and the automorphism group $H_-$ is any subgroup of $Aut (g\oplus g)$ which contains 
orientation-reversing automorphisms. The matrices $\ws$ and $W(h_\s;T)$ are the actions of $h_\s \!\in \!Aut(g)$ in the adjoint representation and in matrix 
rep $T$ respectively. These unitary matrices are familiar in the orbifold program \cite{Dual,More,Big}, where the relations in Eq.~\eqref{Eq 2.5e} are known 
as the {\it linkage relations} \cite{Big} for reps $T$ and $\bT$. Consistency of Eqs.~\eqref{Eq 2.5b} and \eqref{Eq 2.5c} follows from Eq.~\eqref{Eq 2.5f}, 
which we have chosen as the solution to the linkage relation for rep $\bT$. The reader may find it interesting to check explicitly that the general 
transformation in \eqref{Eq2.5} is indeed an automorphism of affine $(g\oplus g)$. 

We emphasize that, thanks to the $\bz \rightarrow z$ device for $\bJ$, the mode form of Eq.~\eqref{Eq 2.5a} is:
\begin{subequations}
\label{Eq2.6}
\begin{gather}
J_a (z) =\sum_m J_a(m) z^{-m-1} ,\quad \bJ_a (z) =\sum_m \bJ_a (m) z^{-m-1} \\
J_a (m)' =\ws_a{}^b \bJ_b (m) ,\quad \bJ_a (m)' =\ws_a{}^b J_b (m).
\end{gather}
\end{subequations}
As we will explain in Subsec.~$6.1$, the classical (high-level) limit of Eq.~\eqref{Eq 2.5b} can also be written as
\vspace{-0.05in}
\begin{gather}
g(T,\bz,z)' =W(h_\s;T) g^{-1} (T,z,\bz) W\hc (h_\s;T) \label{Eq2.7}
\end{gather}
where $g(T,\bz,z)$ is now the WZW group element in rep $T$.

The simple case with $h_\s =1$
\begin{subequations}
\label{Eq2.8}
\begin{gather} 
\ws = W(h_\s;T) =W(h_\s;\bT)= \thickone \\
J(z) \, \leftrightarrow \,\bJ(z) \\
g(T,\bz,z) \,\leftrightarrow \,g(\bT,z,\bz)^t \label{Eq 2.8c}
\end{gather}
\end{subequations}
will be called the {\it basic} orientation-reversing automorphism of affine $(g\oplus g)$ (see also Subsec.~$5.2$). The transformation in \eqref{Eq 2.8c} is 
the behavior of the affine primary fields under {\it world-sheet parity} $z\leftrightarrow \bz$.

In what follows we discuss only the orientation-reversing automorphisms above, which as we shall see, are associated to the {\it open-string sectors} 
of the WZW orientation orbifolds. We mention however that the full automorphism group $H_- \!\subset \!Aut (g\oplus g)$ contains a subgroup $H^0$ of 
orientation-preserving automorphisms
\begin{gather}
H^0 \subset H_- ,\quad H^0 \subset Aut(g) ,\quad |H_- | =2|H^0 | \label{Eq2.9}
\end{gather}
whose associated closed-string sectors have already been constructed in the orbifold program. Moreover, the general principles of the orbifold 
program tell us that all the twisted tensors of all orbifold sectors are class functions, so that one need only keep one representative
from each conjugacy class of $H_-$.

\subsection{Two-Component Fields}

The currents and affine primary fields of the orbifold program are defined to be irreducible under the action of the automorphism group, which may
require {\it reducible} representations of the affine Lie algebra. This has been studied extensively for the case of complex representations in ordinary
outer-automorphic orbifolds \cite{Big',so2n}.

To accommodate the action \eqref{Eq2.5} of the orientation-reversing automorphism, we therefore introduce the following {\it two-component notation}
\begin{subequations}
\label{Eq2.10}
\begin{gather}
J_{a\Id} (z) : \quad J_{a0} (z) \equiv J_a (z) ,\quad J_{a1} (z) \equiv \bJ_a (z) \\
\tg (T,\bz,z) \equiv \left( \begin{array}{cc} g(T,\bz,z) &0\\0& g(\bT,z,\bz)^t \end{array} \right) \label{Eq 2.10b} 
\end{gather}
\end{subequations}
where $J_{a\Id}(z),\, \Id =0,1$ are the two-component currents and the reducible quantity $\tg(T,\bz,z)$ will be called the {\it matrix affine primary field} 
in rep $T$ of $g$. In this notation, the OPE system \eqref{Eq2.3} takes the simple form:
\begin{subequations}
\label{Eq2.11}
\begin{gather}
J_{a\Id} (z) J_{b\Jd}(w) = \left( \frac{G_{a\Id;b\Jd}}{(z-w)^2} +\frac{if_{a\Id;b\Jd}{}^{c\Kd} J_{c\Kd}(w)}{z-w} \right) +\Ord (z-w)^0 \nn \\
\bigspc \quad = \de_{\Id \Jd} \left( \frac{G_{ab}}{(z-w)^2} + \frac{if_{ab}{}^c J_{c\Id} (w)}{z-w} \right) + \Ord (z-w)^0 \\
J_{a\Id} (z) \tg(T,\bw,w) =\frac{\tg(T,\bw,w) T_a \r_\Id}{z-w} - \frac{(\thickone -\r_\Id) T_a \tg(T,\bw,w)}{z-\bw}  \nn \\
   \quad \quad \quad + \Ord (z-w)^0 + \Ord (z-\bw)^0 
\end{gather}
\begin{gather}
G_{a\Id;b\Jd} =\de_{\Id \Jd} G_{ab} ,\quad f_{a\Id;b\Jd}{}^{c\Kd} =(\r_\Id)_\Jd{}^\Kd f_{ab}{}^c ,\quad \Id,\Jd,\Kd =0,1 \\
\r_0 \equiv \left( \begin{array}{cc} 1&0\\0&0 \end{array} \right) ,\quad \r_1 \equiv \left( \begin{array}{cc} 0&0\\0&1 \end{array} \right) ,\quad
   \one =\left( \begin{array}{cc} 1&0\\0&1 \end{array} \right) \,.
\end{gather}
\end{subequations}
We also note the {\it constraints} on the matrix affine primary field
\begin{subequations}
\label{Eq2.12}
\begin{gather}
\tg (\bT,\bz,z)^t =\tau_1 \tg (T,z,\bz) \tau_1 ,\quad [\tau_3 ,\tg(T,\bz,z)]= [\r_\Id ,\tg(T,\bz,z)]=0 \label{Eq 2.12a} \\
\vec{\tau} =\text{ Pauli matrices } ,\quad \left( \begin{array}{cc} A&B\\C&D \end{array} \right)^t = \left( \begin{array}{cc} 
   A^t &C^t \\B^t &D^t \end{array} \right) 
\end{gather}
\end{subequations}
which follow directly from the definition in Eq.~\eqref{Eq 2.10b}. The first constraint in \eqref{Eq 2.12a} is the behavior of 
the matrix affine primary field under world-sheet parity $z \leftrightarrow \bz$.

In this two-component notation, the action of the orientation-reversing automorphism $\hat{h}_\s =\tau_1 \!\times \!h_\s$ also takes a simple form:
\begin{subequations}
\label{Eq2.13}
\begin{gather}
J_{a\Id} (z)' = (\tau_1 )_{\Id}{}^\Jd \ws_a{}^b J_{b\Jd} (z) \\
\tg(T,\bz,z)' = \tau_1 W(h_\s;T) \tg(T,\bz,z) W\hc (h_\s;T) \tau_1 \label{Eq 2.13b} \\
\tau_1 W(h_\s;T) \!=\! W(h_\s;T) \tau_1 \!\equiv \! \tau_1 \otimes W(h_\s;T) \,.
\end{gather}
\end{subequations}
The automorphic responses of the two-component fields are now in the form required for the orbifold program, with ``total" actions $\hat{w} =\tau_1 
\!\otimes \!w$ and $\hat{W}=\tau_1 \!\otimes \!W$ (see App.~B).

In our development, we will also need the well-known (see e.~g.~Ref.~\cite{H+O}) vertex operator equations
\vspace{-0.10in}
\begin{subequations}
\label{Eq2.14}
\begin{gather}
\pl g(T,\bz,z) =2L_g^{ab} :\!J_a (z) g(T,\bz,z) T_b \!: \\
\bpl g(T,\bz,z) =-2L_g^{ab} :\!T_a \bJ_b (\bz) g(T,\bz,z)\!: \\
L_g^{ab} =\oplus_I \frac{\eta^{a(I)b(I)}_I}{2k_I +Q_I} ,\quad L_g^{cd} \ws_c{}^a \ws_d{}^b =L_g^{ab} \label{Eq 2.14c}
\end{gather}
\end{subequations}
which are the standard operator equations of motion of the corresponding WZW model on $g$. Here the symbol $: \cdot :$ is operator-product normal 
ordering and $L_g^{ab}$ is the inverse inertia tensor of the affine-Sugawara construction [6,7,31-33,12] on $g$. The vertex operator equations may 
be written in the two-component notation as\footnote{Our convention is that repeated indices are summed, but we will also show explicitly the sums 
over the two-component indices $\Id ,\Jd$.}
\begin{subequations}
\label{Eq2.15}
\begin{gather}
\pl \tg(T,\bz,z) =2L_g^{ab} \sum_{\Id =0}^1 :\!J_{a\Id}(z) \tg(T,\bz,z) T_b \r_\Id \!: \label{Eq 2.15a} \\
\bpl \tg(T,\bz,z) =-2L_g^{ab} \sum_{\Id =0}^1 :\!T_a (\one -\r_\Id) J_{b\Id} (\bz) \tg(T,\bz,z)\!: \label{Eq 2.15b}
\end{gather}
\end{subequations}
where $: \cdot :$ is still operator-product normal ordering. We also note that Eqs.~\eqref{Eq 2.15a} and \eqref{Eq 2.15b} are redundant 
in the following sense: Using the identity \eqref{A1} and the world-sheet parity in \eqref{Eq 2.12a}, we find that Eq.~\eqref{Eq 2.15b} follows 
from \eqref{Eq 2.15a} and vice-versa. It is also not difficult to check with the linkage relations \eqref{Eq 2.5e} that each vertex operator 
equation in \eqref{Eq2.15} is individually form-invariant under the general orientation-reversing automorphism \eqref{Eq2.13}.

\subsection{Eigenfields}

The next step in the orbifold program is the construction of the eigenfield basis \cite{Dual,More,Big}, which diagonalizes the action of each automorphism. 
(The eigenfields are the prerequisite for an application of the principle of local isomorphisms in Sec.~3.)

To begin, we will need the solutions (for particular $h_\s \!\in \!Aut(g)$) of the $H$-eigenvalue problem of Ref.~\cite{More} and the extended 
$H$-eigenvalue problem of Ref.~\cite{Big}
\begin{subequations}
\label{Eq2.16}
\begin{gather}
\ws U\hc (\s) = U\hc (\s) E(\s),\quad E (\s)_\nrm{}^\nsn = \de_\nrm{}^\nsn e^{-\tp \nrrsf} \label{Eq 2.16a} \\
W(h_\s;T) U\hc (T,\s) = U\hc (T,\s) E(T,\s) ,\quad E(T,\s)_\Nrm{}^{\!\!\Nsn} \equiv \de_\Nrm{}^\Nsn e^{-\tp \frac{N(r)}{R(\s)}} \label{Eq 2.16b} \\
\de_\nrm{}^\nsn \equiv \de_\m{}^\n \de_{n(r)-n(s),0\, \text{mod }\r(\s)} ,\quad \de_\Nrm{}^\Nsn \equiv \de_\m{}^\n \de_{N(r)-N(s),0\, 
   \text{mod }R(\s)} \\
U\hc(\s) =\{ U\hc (\s)_a{}^\nrm \} ,\quad U\hc (T,\s) =\{ U\hc (T,\s)_{\a(I)I}{}^\Nrm \} \\
\bar{n}(r) \in \{0,\ldots ,\r(\s)\!-\! 1\} ,\quad \bar{N}(r) \in \{0,\ldots ,R(\s)\!-\!1\} 
\end{gather}
\end{subequations}
which are defined in the orbifold program for each $h_\s \!\in \!H \!\subset \!Aut (g)$. Here $\r(\s)$ and $R(\s)$ are the orders of $h_\s$ acting in 
the adjoint rep and rep $T$ respectively. The (unitary) eigenvector matrices $U\hc (\s), U\hc (T,\s)$ are periodic 
\begin{gather}
n(r) \rightarrow n(r)\! \pm \!\r(\s) ,\quad N(r) \rightarrow N(r) \!\pm \!R(\s) \label{Eq2.17}
\end{gather}
in their respective spectral indices $n(r)$ and $N(r)$, and the same is true below for any object with these indices. The barred indices $\bar{n}(r),
\bar{N}(r)$ are the pullbacks of the spectral indices to their fundamental ranges
\begin{gather}
\srac{\bar{n}(r)}{\r(\s)} =\srac{n(r)}{\r(\s)} -\Big{\lfloor} \srac{n(r)}{\r(\s)} \Big{\rfloor} ,\quad 
\srac{\bar{N}(r)}{R(\s)} =\srac{N(r)}{R(\s)} -\Big{\lfloor} \srac{N(r)}{R(\s)} \Big{\rfloor} \label{Eq2.18}
\end{gather}
where $\lfloor x \rfloor$ is the floor of $x$.

For the basic automorphism types, these eigenvalue problems have been well-studied in the orbifold program:

$\bullet$ permutations [3,5,13-15]

$\bullet$ inner automorphisms of simple $g$ \cite{More,Big,Perm}

$\bullet$ outer automorphisms of simple $g$ \cite{Big',so2n}.

\noindent The eigenvalue problems for compositions of these basic types have not yet been extensively studied (see however Refs.~\cite{Coset,Fab}).

It is not difficult to verify that the solution of the extended $H$-eigenvalue problem \eqref{Eq 2.16b} for rep $\bar{T}$ can be taken as
\begin{subequations}
\label{Eq2.19}
\begin{gather}
W(h_\s;\bT) U\hc (\bT,\s) = U\hc (\bT,\s) E(\bT,\s) \\
U(\bT,\s) = U^\ast(T,\s) ,\quad E(\bT,\s) =E^\ast(T,\s) \label{Eq 2.19b}
\end{gather}
\end{subequations}
where we have used the form of $W(h_\s;\bT)$ chosen in Eq.~\eqref{Eq 2.5f}.

We will also need the standard {\it duality transformations} \cite{Dual,More,Big}
\begin{subequations}
\label{Eq2.20}
\begin{gather}
\sG_{\nrm;\nsn} (\s) \equiv \schisig_\nrm \schisig_\nsn U (\s)_\nrm{}^a U(\s)_\nsn{}^b G_{ab} \nn \\
\bigspc = \de_{n(r)+n(s), 0\,\text{mod }\r(\s)} \sG_{\nrm;\mnrn} (\s) \label{Eq2.20a} \\
\scf_{\nrm;\nsn}{}^{\!\!\ntd} (\s) \!\equiv \!\schisig_\nrm \schisig_\nsn \schisig^{-1}_\ntd U(\s)_\nrm{}^a U(\s)_\nsn{}^b f_{ab}{}^c
   U\hc (\s)_c{}^{\!\ntd} \quad \nn \\
\bigspc =\de_{n(r)+n(s)-n(t) ,0\,\text{mod } \r(\s)} \scf_{\nrm;\nsn}{}^{n(r)+n(s),\de} (\s) \label{Eq 2.20b} \\
{\cL}_{\sgb (\s)}^{\nrm;\nsn}(\s) \equiv \schisig^{-1}_\nrm \schisig^{-1}_\nsn U\hc(\s)_a{}^\nrm U\hc (\s)_b{}^\nsn L_g^{ab} \nn \\
   = \de_{n(r)+n(s) ,0\,\text{mod }\r(\s)} {\cL}_{\sgb(\s)}^{\nrm;\mnrn}(\s) \label{Eq 2.20c}
\end{gather}
\begin{gather}
\st_\nrm (T,\s) \equiv \schisig_\nrm U(\s)_\nrm{}^a U(T,\s) T_a U\hc (T,\s) \label{Eq 2.20d} \\
e^{\tp \nrrsf} \st_\nrm (T,\s) = E(T,\s) \st_\nrm (T,\s) E^\ast (T,\s) \label{Eq 2.20e} \\
[\st_\nrm ,\st_\nsn] =i\scf_{\nrm;\nsn}{}^{n(r)+n(s),\de} (\s) \st_{n(r)+n(s),\de} \label{Eq 2.20f}
\end{gather}
\end{subequations}
or {\it twisted tensors} of current-algebraic space-time orbifold theory. Here $\sG(\s), \scf(\s),{\cL}(\s)$ and $\st(T,\s)$ are called respectively the 
(symmetric) twisted metric, the (anti-symmetric) twisted structure constants, the (symmetric) twisted inverse inertia tensor and the twisted representation 
matrices of sector $\hat{h}_\s$. The algebra \eqref{Eq 2.20f} of the twisted representation matrices is called the orbifold Lie algebra $\hg(\s)$ of sector 
$\hat{h}_\s$, and the normalization of the twisted representation matrices is given in Ref.~[13-15,23]. The duality transformations in 
\eqref{Eq2.20} have been explicitly evaluated in the following cases

\noindent $\bullet$ the WZW permutation orbifolds [3,5,13-15]

\noindent $\bullet$ the inner-automorphic WZW orbifolds \cite{More,Big,Perm} on simple $g$

\noindent $\bullet$ the charge conjugation orbifolds on $\su(n)$ \cite{Big'}

\noindent $\bullet$ the outer-automorphic WZW orbifolds on $\so (2n)$ (and $\so (8)$) \cite{so2n}

\noindent as well as for certain classes of coset orbifolds \cite{Coset,Fab} which involve the composition of permutations with inner automorphisms 
of simple $\gfrak$. 

We will also introduce the twisted representation matrices corresponding to rep $\bT$ of $g$
\begin{subequations}
\label{Eq2.21}
\begin{align}
\st_\nrm (\bT,\s) &\equiv \st_\nrm (T,\s) |_{T \rightarrow \bT} = \schisig_\nrm U(\s)_\nrm{}^a U(\bT,\s) \bT_a U\hc (\bT,\s) 
   \label{bst-Defn} \\
&= -\st_\nrm (T,\s)^t \label{Eq 2.21b}
\end{align}
\end{subequations}
which also satisfy the same orbifold Lie algebra \eqref{Eq 2.20f}. Eq.~\eqref{Eq 2.19b} was used to obtain the final form in \eqref{Eq 2.21b}. In what follows, 
the twisted tensors above will be called the {\it ordinary} twisted tensors of space-time orbifold theory.

We are now prepared to define the {\it eigenfields} \cite{Dual,More,Big}
\begin{subequations}
\label{Eq2.22}
\begin{gather}
\sj_\nrmu (z,\s) \equiv \schisig_\nrm U(\s)_\nrm{}^a (\sqrt{2} U_u{}^\Id) J_{a\Id} (z) \label{Eq 2.22a} \\
\sg (\st(T,\s),\bz,z,\s) \equiv UU(T,\s) \tg(T,\bz,z) U\hc (T,\s) U\hc \nn \\
\bigspc \bigspc =\{ \sg (\st(T,\s),\bz,z,\s)_{\Nrm u}{}^{\Nsn v} \} \\
U = U\hc \equiv \srac{1}{\sqrt{2}} \left( \begin{array}{cc} 1 & \,1 \\ 1 & -1 \end{array} \right) ,\quad \tau_1 U\hc =U\hc \tau_3 ,\quad 
   \bar{u},\bar{v},\bar{w} \in \{ 0,1\} 
\end{gather}
\end{subequations}
corresponding to each $\hat{h}_\s$ for each two-component field. Here $\schisig$ is an arbitrary normalization and $U$ is the analogue in the 2x2 space of 
the unitary matrix $U(\s)$ in ordinary space-time orbifold theory. Note also that the index $u$ is periodic $u \rightarrow u\pm 2$, and $\bar{u},\bar{v},
\bar{w}$ are the pullbacks of $u,v,w$ to the fundamental range. In terms of the eigenfields, the constraints \eqref{Eq2.12} become
\begin{subequations}
\label{Eq2.23}
\begin{gather}
\tau_3 \sg (\st (T,\s),\bz,z,\s) \tau_3 =\sg (\st(\bT,\s),z,\bz,\s)^t  \label{Eq 2.23a} \\
[\tau_1 ,\sg(\st(T,\s),\bz,z,\s) ]=0
\end{gather}
\end{subequations}
where we have used Eqs.~\eqref{Eq 2.19b} and \eqref{A2}. The constraint in Eq.~\eqref{Eq 2.23a} is the behavior of the eigengroup element under world-sheet 
parity.

The eigenfields of orbifold theory are defined to diagonalize the automorphic responses, and in this case we find the action of the orientation-reversing 
automorphism $\hat{h}_\s$:
\begin{subequations}
\label{Eq2.24}
\begin{gather}
\sj_\nrmu (z,\s)' = e^{-\tp (\nrrs +\srac{u}{2})} \sj_\nrmu (z,\s) \label{Eq 2.24a}  \\
\sg (\st(T,\s),z,\s)' = \tau_3 E(T,\s) \sg(\st(T,\s),z,\s) E^\ast (T,\s) \tau_3 \,.  \label{Eq 2.24b}
\end{gather}
\end{subequations}
These responses are easily verified from Eqs.~\eqref{Eq2.13}, \eqref{Eq2.16} and the definitions in Eq.~\eqref{Eq2.22}.

Following Refs.~\cite{Dual,More,Big}, the OPEs of the eigenfields may now be computed in terms of the twisted tensors. For example, the $\sj \sg$ OPE reads
\begin{gather}
\sj_\nrmu (z,\s) \sg(\st(T,\s),\bz,z,\s) = \frac{\sg (\st(T,\s),\bz,z,\s) (\st_\nrm (T,\s) \tau_u )}{z-w}  \bigspc \nn \\
\bigspc -\frac{(\st_\nrm (T,\s) (-1)^u \tau_u )\sg(\st(T,\s),\bz,z,\s)}{z-\bw} +\Ord (z-w)^0 +\Ord (\bz- \bw)^0 \label{Eq2.25}
\end{gather}
where $\tau_0 =\thickone_2$ and the ordinary twisted representation matrices $\st_\nrm (T,\s)$ are defined in \eqref{Eq 2.20d}. We note, however, that these 
ordinary $\st$'s are only factors in the {\it total} twisted representation matrices 
\vspace{-0.10in}
\begin{subequations}
\label{Eq2.26}
\begin{gather}
\st_\nrmu (T,\s) \equiv \schisig_\nrm U(\s)_\nrm{}^a (\sqrt{2} U_u{}^\Id )U U(T,\s) (T_a \r_\Id) U\hc(T,\s)U\hc \nn \\ 
  =\st_\nrm(T,\s)\tau_u =\st_\nrm (T,\s) \otimes \tau_u ,\quad u=0,1 \label{Eq 2.26a} \\
\tilde{\st}_\nrmu (T,\s) \equiv \schisig_\nrm U(\s)_\nrm{}^a (\sqrt{2} U_u{}^\Id )U U(T,\s) (T_a (\one-\r_\Id)) U\hc(T,\s)U\hc \nn \\ 
  =\st_\nrm(T,\s) (-1)^u \tau_u \\
[\st_\nrmu(T,\s) ,\st_\nsnv (T,\s)] =i\scf_{\nrm;\nsn}{}^{n(r)+n(s),\de}(\s) \st_{n(r)+n(s),\de,u+v} (T,\s) \label{Eq 2.26c} \\
[\tilde{\st}_\nrmu (T,\s) ,\tilde{\st}_\nsnv (T,\s)] =i\scf_{\nrm;\nsn}{}^{n(r)+n(s),\de}(\s) \tilde{\st}_{n(r)+n(s),\de,u+v} (T,\s) 
\end{gather}
\end{subequations}
which appear in Eq.~\eqref{Eq2.25}. The total twisted representation matrices are those which would be obtained by treating the total automorphism in 
the standard language of current-algebraic orbifold theory (see also App.~B). Note that both sets of total twisted representation matrices $\st$ and 
$\tilde{\st}$ satisfy the same algebra ($2.26$c,d), which is the total {\it orbifold Lie algebra} $\hg_O (\s)$ of sector $\hat{h}_\s$.

\subsection{Stress Tensors and the Stress-Tensor Eigenfields}

We finally consider the stress tensors in the untwisted WZW theory, which are the standard affine-Sugawara constructions [6,7,31-33,12] on $g$
\begin{subequations}
\label{Eq2.27}
\begin{gather}
T_g(z) =L_g^{ab} :\!J_a (z) J_b (z) \!: ,\quad \bT_g(\bz) = L_g^{ab} :\!\bJ_a (\bz) \bJ_b (\bz)\!: \\
\bar{c}_g =c_g =2G_{ab} L_g^{ab}= \sum_I \frac{2k_I \text{dim }\gfrak^I}{2k_I +Q_I} 
\end{gather}
\end{subequations}
where $g= \oplus_I \gfrak^I$ and $L_g^{ab}$ is the inverse inertia tensor in Eq.~\eqref{Eq 2.14c}. 

The $T_g,\bT_g$ OPEs with themselves and with the currents are well-known:
\begin{subequations}
\label{Eq2.28}
\begin{gather}
T_g(z) T_g(w) = \frac{c_g/2}{(z-w)^4} +\left( \frac{2}{(z-w)^2} +\frac{\pl_w}{z-w} \right) T_g(w) +\Ord (z-w)^0 \\
\bT_g (z) \bT_g(w) = \frac{c_g/2}{(z-w)^4} +\left( \frac{2}{(z-w)^2} +\frac{\pl_w}{z-w} \right) \bT_g(w) +\Ord (z-w)^0 
\end{gather}
\begin{gather}
T_g(z) J_a (w) =\left( \frac{1}{(z-w)^2} +\frac{\pl_w}{z-w} \right) J_a (w) +\Ord (z-w)^0 \\
\bT_g(z) \bJ_a (w) =\left( \frac{1}{(z-w)^2} +\frac{\pl_w}{z-w} \right) \bJ_a (w) +\Ord (z-w)^0 \\
T_g(z) \bT_g(w) =T_g(z) \bJ_a (w) =\bT_g(z) J_a (w) = \Ord (z-w)^0 \,.
\end{gather}
\end{subequations}
Following the discussion of Subsec.~$2.1$, we will also need the extended set of OPEs with the affine primary fields
\begin{subequations} 
\label{Eq2.29}
\begin{gather}
T_g(z) g(T,\bw,w) =g(T,\bw,w) \lD (z,w) +\Ord(z-w)^0 \\
\bT_g(z) g(T,\bw,w) =D(z,\bw) g(T,\bw,w) +\Ord(z-\bw)^0 \\
T_g(z) g(\bT,w,\bw)^t =D(z,\bw) g(\bT,w,\bw)^t +\Ord(z-\bw)^0 \label{Eq 2.29c} \\
\bT_g(z) g(\bT,w,\bw)^t =g(\bT,w,\bw)^t \lD (z,w) +\Ord (z-w)^0 \label{Eq 2.29d}
\end{gather}
\begin{gather}
D (z,w) \equiv \frac{ D_g (T)}{(z-w)^2} + \frac{1}{z-w}\pl_w ,\quad \lD (z,w) \equiv \frac{ D_g (T)}{(z-w)^2} + \lplw \frac{1}{z-w} \\
D_g (\bT) = D_g (T) = L^{ab}_g T_a T_b ,\quad [D_g (T), g(T,\bz,z)] =0 
\end{gather}
\end{subequations}
where $D_g (T)$ is the conformal weight matrix of rep $T$ under the affine-Sugawara construction. Here we have again replaced $\bz \rightarrow z$ everywhere,
and the OPEs with transposed operators in Eqs.~\eqref{Eq 2.29c}, \eqref{Eq 2.29d} are redundant.

The general orientation-reversing automorphism $\hat{h}_\s$ acts on the affine-Sugawara stress tensors as
\vspace{-0.10in}
\begin{subequations}
\label{Eq2.30}
\begin{gather}
T_g(z)' =L_g^{ab} :\!J_a (z)' J_b(z)' \!: = L_g^{ab} :\!\bJ_a (z) \bJ_b(z) \!: =\bT_g(z) \\
\bT_g(z)' =L_g^{ab} :\!\bJ_a (z)' \bJ_b(z)' \!: = L_g^{ab} :\!J_a (z) J_b(z) \!: =T_g(z)
\end{gather}
\end{subequations}
where we have used the invariance \eqref{Eq 2.14c} of $L_g^{ab}$. Thanks again to our $\bz \rightarrow z$ device for $\bT_g$, the mode form of this 
action is:
\begin{subequations}
\label{Eq2.31}
\begin{gather}
T_g(z) =\sum_{m\in \Zint} L_g(m) z^{-m-2} ,\quad \bT_g (z)=\sum_{m\in \Zint} \bar{L}_g (m) z^{-m-2} \\
L_g(m)' =\bar{L}_g(m) ,\quad \bar{L}_g(m)' =L_g(m) \,.
\end{gather}
\end{subequations}
It is not difficult to check that the action in Eqs.~\eqref{Eq2.5} and \eqref{Eq2.30} indeed constitutes an automorphism of the stress-tensor 
OPE system \eqref{Eq2.28}, \eqref{Eq2.29}.

As above, we now introduce a two-component notation for the stress tensors:
\begin{subequations}
\label{Eq2.32}
\begin{gather}
T_0 (z) \equiv T_g(z) ,\quad T_1 (z) \equiv \bT_g (z) ,\quad T_\Id (z) = L_g^{ab} :\! J_{a\Id} (z) J_{b\Id} (z) \!: \\
T_\Id (z)' = (\tau_1 )_\Id {}^\Jd T_\Jd (z) ,\quad \Id=0,1 \,. \label{Eq 2.32b}
\end{gather}
\end{subequations}
Then the entire system \eqref{Eq2.28}, \eqref{Eq2.29} of stress tensor OPEs becomes
\begin{subequations}
\label{Eq2.33}
\begin{gather}
T_\Id (z) T_\Jd (w) = \de_{\Id \Jd} \left( \frac{c_g/2}{(z-w)^4} + \left( \frac{2}{(z-w)^2} + \frac{\pl_w}{z-w} \right) T_\Jd (w) \right) +
   \Ord (z-w)^0 \\
T_\Id (z) J_{a\Jd} (w) =\de_{\Id \Jd} \left( \frac{1}{(z-w)^2} + \frac{\pl_w}{z-w} \right) J_{a\Jd} (w) + \Ord (z-w)^0 \\
T_\Id (z) \tg (T,\bw,w) = \tg (T,\bw,w) \lD (z,w) \r_\Id + D(z,\bw) (\one- \r_\Id) \tg (T,\bw,w) \nn \\
+ \Ord (z-w)^0 +\Ord (z-\bw)^0
\end{gather}
\end{subequations}
where the matrix affine primary fields $\tg$ are defined in Eq.~\eqref{Eq 2.10b}. In the two-component notation, it is not difficult to check that
\eqref{Eq2.13}, \eqref{Eq 2.32b} is an automorphism of the full WZW system.

The associated stress-tensor eigenfields and their automorphic responses are
\begin{subequations}
\label{Eq2.34}
\begin{gather}
\Theta_u (z,\s) \equiv T_0 (z)+ (-1)^u T_1 (z) ,\quad \quad \bar{u}=0,1  \label{Eq 2.34a} \\
\Theta_u (z,\s)' = (-1)^u \Theta_u (z,\s)  \label{Eq 2.34b} \\
H= \oint \frac{dz}{\tp} z \Theta_0 (z,\s) =\bar{L}_g (0) +L_g (0) ,\quad H' =H \label{Eq 2.34c}
\end{gather}
\end{subequations}
and we note that (as it should be in orbifold theory) the Hamiltonian in \eqref{Eq 2.34c} is invariant under the orientation-reversing automorphisms. 
Finally, the stress-tensor eigenfields can be rewritten in terms of the eigencurrents
\begin{gather}
\Theta_u (z,\s) = \srac{1}{2} {\cL}_{\sgb(\s)}^{\nrm;\mnrn}(\s) \sum_{v=0}^1 :\!\sj_{\nrm v} (z,\s) \sj_{\mnrn ,u-v} (z,\s) \!: \label{Eq2.35} 
\end{gather}
where the twisted inverse inertia tensor ${\cL}$ is defined in Eq.~\eqref{Eq 2.20c}.

\section{The Operator Algebra of Open-String Orbifold Sector $\hat{h}_\s$}

\subsection{Automorphisms, Twists and Open Strings}

The method of eigenfields and local isomorphisms \cite{Dual,More,Big} maps each untwisted CFT $A(H)$ with a discrete symmetry $H$ to the twisted sectors of 
the orbifold $A(H)/H$, where each twisted sector corresponds to a conjugacy class of $H$. In the present case, we shall see that the sector $\hat{h}_\s 
=\tau_1 \!\times \!h_\s$ of the WZW orientation orbifold $A_g(H_-)/H_-$ is a twisted open string. The mode form of the twisted open-string algebra of 
these sectors is given in Subsec.~$3.4$.

Before working out the detailed dynamics of these twisted sectors, we want to emphasize that the open-string character of these sectors is easily 
understood as follows.

Given the left- and right-mover Virasoro generators $L(m)$ and $\bar{L}(m)$ of {\it any closed-string CFT} with central charges $\bar{c}=c$, we know 
that any orientation-reversing automorphism of the underlying fields will induce the automorphism of $Vir \oplus Vir$:
\begin{gather}
L (m)' =\bar{L}(m) ,\quad \bar{L}(m)' =L(m) \,. \label{Eq3.1}
\end{gather}
Our orbifolding procedure follows the well-known prescription in mathematics for twisting an algebra: The automorphic response of the eigenstates 
\begin{gather}
L_{\pm}(m) \equiv L(m) \pm \bar{L}(m) ,\quad L_{\pm}(m)' =\pm L_{\pm}(m) \label{Eq3.2}
\end{gather}
tells us immediately that we may consistently assign integral and half-integral moding respectively to twisted versions of the eigenstates
\begin{gather}
\{ L_+ (m) \} \,\rightarrow \,\{ \hat{L}_0 (m) \} ,\quad \{ L_- (m) \} \,\rightarrow \,\{ \hat{L}_1 (m \!+\!\srac{1}{2}) \}  \label{Eq3.3}
\end{gather}
which in this case satisfy a simple {\it orbifold Virasoro algebra} (see Eq.~\eqref{Eq 3.21a} and Refs.~\cite{Chr,DV2,Perm}) in twisted sector $\hat{h}_\s$
of the orientation orbifold. Since the orbifold Virasoro algebra contains only a {\it single} untwisted Virasoro subalgebra generated by $\{ \hat{L}_0(m)\}$, 
the corresponding twisted sector of the orientation orbifold is an {\it open string}. 

Moreover, central charges do not change in the orbifolding process, so the central charge of the single untwisted Virasoro is the same as that of the 
eigenstate $\{L_+ (m)\}$:
\begin{gather}
\hat{c}=2c.  \label{Eq3.4}
\end{gather}
It follows that our construction is not an open-string sector of a conventional orientifold [25-27], which in any case would not involve 
fractional moding. 

In the case of the WZW models, the eigenstates \eqref{Eq3.2} are the mode form of the stress-tensor eigenfields \eqref{Eq 2.34a}. More generally,
the method of eigenfields and local isomorphisms \cite{Dual,More,Big} is a local version (with OPEs and monodromies) of the arguments above, 
extended to include all the underlying fields and normal ordering of all relevant composite operators at once.

\subsection{The Twisted Currents and Twisted Affine Primary Fields}

The principle of local isomorphisms maps the eigenfields to the twisted fields, for example 
\begin{subequations}
\label{Eq3.5}
\begin{gather}
\sj_\nrmu (z,\s) \dual \hj_\nrmu (z,\s) ,\quad \Theta_u (z,\s) \dual \hat{\Theta}_u (z,\s) \\
\sg(\st,\bz,z,\s) \dual \hg(\st,\bz,z,\s) 
\end{gather}
\end{subequations}
in each orientation-orbifold sector $\hat{h}_\s$. In this map, the principle tells us first that the singular terms of OPEs of the twisted fields are 
the same as those of the eigenfields in the untwisted closed WZW theory. Moreover, the principle tells us that the automorphic responses \eqref{Eq 2.24a},
\eqref{Eq 2.34b} of the current and stress-tensor eigenfields $\sj$ and $\Theta$ become the monodromies of the corresponding twisted fields. 

This gives for the twisted currents $\hj$ and the twisted affine primary fields $\hg$
\begin{subequations}
\label{Eq3.6}
\begin{gather}
\hj_\nrmu (ze^{\tp},\s) = e^{-\tp (\nrrs +\srac{u}{2})} \hj_\nrmu (z,\s) \label{Eq 3.6a}
\end{gather}
\begin{align}
&\hj_\nrmu (z) \hj_\nsnv (w) = \frac{2\de_{u+v ,0\, \text{mod }2} \de_{n(r)+n(s),0\,\text{mod }\r(\s)} \sG_{\nrm;\mnrn}(\s)}{(z-w)^2} \bigspc \nn \\
&\bigspc + \frac{ i\scf_{\nrm;\nsn}{}^{\!\!\!n(r)+n(s),\de}(\s) \hj_{n(r)+n(s),\de,u+v} (w)}{z-w} + :\!\hj_\nrmu (z)
   \hj_\nsnv (w)\!: \label{Eq 3.6b} 
\end{align}
\begin{align}
&\hj_\nrmu (z) \hg(\st(T),\bw,w) \!=\frac{\hg(\st(T),w) \st_\nrmu(T,\s)}{z-w} -\frac{\tilde{\st}_\nrmu(T,\s) \hg(\st(T),\bw,w)}{z-\bw} \quad \nn \\
& \bigspc \bigspc \bigspc +:\!\hj_\nrmu (z) \hg(\st(T),\bw,w)\!: \label{Eq 3.6c} \\
& \bigspc \bigspc \hg(\st(T),\bz,z) =\{ \hg(\st(T,\s),\bz,z,\s)_{\Nrm u}{}^{\Nsn v} \} \label{Eq 3.6d}
\end{align}
\end{subequations}
Here $\st$ and $\bar{\st}$ in \eqref{Eq 3.6c} are the total twisted representation matrices in Eq.~\eqref{Eq2.26}, and the symbol $:\cdot :$ denotes 
operator-product normal ordering in the twisted sector. Moreover, the ordinary twisted tensors $\sG(\s)$ and $\scf(\s)$ are defined in 
Eq.~\eqref{Eq2.20}, and we have suppressed some $\s$-dependence for brevity. As in the case of $\st$ and $\tilde{\st}$, the numerical factors in 
Eq.~\eqref{Eq 3.6b} are nothing but the {\it total} twisted metric and {\it total} twisted structure constants
\begin{subequations}
\label{Eq3.7}
\begin{align}
\sG_{\nrmu;\nsnv} (\s) &\equiv \schisig_\nrm \schisig_\nsn U(\s)_\nrm{}^a U(\s)_\nsn{}^b (\sqrt{2} U_u{}^\Id ) (\sqrt{2} U_v{}^\Jd) 
   G_{a\Id;b\Jd} \\
&= (2\de_{u+v,0\,\text{mod }2}) \sG_{\nrm;\nsn}(\s) \nn \\
&= 2\de_{u+v,0\,\text{mod }2} \de_{n(r)+n(s),0\,\text{mod }\r(\s)} \sG_{\nrm;\mnrn}(\s)  
\end{align}
\begin{align}
\scf_{\nrmu;\nsnv}{}^{\!\!\ntdw}(\s) &\equiv \schisig_\nrm \schisig_\nsn \schisig_\ntd^{-1} U(\s)_\nrm{}^a U(\s)_\nsn{}^b \bigspc \nn \\
&\quad \quad \quad \times (\sqrt{2} U_u{}^\Id) (\sqrt{2} U_v{}^\Jd) f_{a\Id ;b\Jd}{}^{c\Kd} (\srac{1}{\sqrt{2}} (U\hc)_\Kd{}^w ) U\hc(\s)_c{}^\ntd \\
&=(\de_{u+v+w,0\, \text{mod }2}) \scf_{\nrm;\nsn}{}^\ntd (\s) \nn \\
&= \de_{u+v+w,0\, \text{mod }2} \de_{n(r)+n(s)-n(t),0\, \text{mod }\r(\s)} \scf_{\nrm;\nsn}{}^{n(r)+n(s),\de} (\s)
\end{align}
\end{subequations}
which would appear in the standard language of the orbifold program (see also App.~B). As shown in Eq.~\eqref{Eq 3.6d}, the index structure of the twisted 
affine primary fields is the same as that of the eigenprimary fields. We also emphasize that, as is appropriate for an open WZW string (see Ref.~\cite{Giusto}), 
the singularities of the OPE \eqref{Eq 3.6c} show the current at $z$ interacting both with a charge at $w$ and its {\it image charge} at $\bw$. 

The principle of local isomorphisms also gives the constraints 
\begin{subequations}
\label{Eq3.8}
\begin{gather}
[\tau_1 ,\hg(\st(T,\s),\bz,z,\s)]=0  \label{Eq 3.8a} \\
\hg(\st(\bT,\s),z,\bz,\s)^t =\tau_3 \hg(\st(T,\s),\bz,z,\s) \tau_3 \label{Eq 3.8b}
\end{gather}
\end{subequations}
as the image of the constraints \eqref{Eq2.23} of the eigenprimary fields. The constraint in \eqref{Eq 3.8b} is the behavior of the twisted affine 
primary field under {\it world-sheet parity} $z \leftrightarrow \bz$, which can be understood as a {\it world-sheet identification}: The twisted affine 
primary fields for $\st(T)$ and $\st(\bT)$ are independent only when we restrict attention to the upper half-plane. The implications of this world-sheet 
identification are more transparent in the classical (high-level) limit of Subsec.~$6.1$ and the free-boson examples of Subsec.~$6.2$.

As noted above, the twisted fields and duality transformations are periodic in all the spectral indices
\begin{subequations}
\label{Eq3.9}
\begin{gather}
\hj_\nrmu (z,\s) = \hj_{n(r) \pm \r(\s),\m u} (z,\s) = \hj_{\nrm ,u\pm 2} (z,\s) =\hj_{\nrm ,-u} (z,\s) \label{Eq 3.9a} \\
\sG_{\nrm;\nsn} (\s) = \sG_{n(r)\pm \r(\s),\m ;\nsn} (\s) = \sG_{\nrm;n(s) \pm \r(\s),\n} (\s) \\
\hg(\st,\bz,z,\s)_{\Nrm u}{}^{\Nsn v} =\hg(\st,\bz,z,\s)_{N(r) \pm R(\s),\m u}{}^{\!\!\Nsn v} =\hg(\st,\bz,z,\s)_{\Nrm u}{}^{\Nsn ,v\pm 2} \\
\st_\nrmu (T,\s) =\st_{n(r)\pm \r(\s),\m u} (T,\s) = \st_{\nrm ,u\pm 2} (T,\s) \\
\bar{n}(r) \in \{0 ,\ldots ,\r(\s)-1 \} ,\quad \bar{N}(s) \in \{0 ,\ldots ,R(\s)-1\} ,\quad \bar{u} \in \{0,1 \}
\end{gather}
\end{subequations}
and we remind the reader that overbar denotes the pullback of the spectral indices to their fundamental ranges.

Finally, applying the method of eigenfields and local isomorphisms to Eqs.~($2.15$a,b), we obtain the {\it twisted vertex operator 
equations}\footnote{Following convention in the orbifold program, we have suppressed explicit sums $\sum_{r,\m,\n}$ on the right-hand sides of 
Eqs.~($3.10$a,b), but we will show explicit sums over the indices $u,v$ which label the 2x2 subspace.} for 
the twisted affine primary fields: 
\begin{subequations}
\label{Eq3.10}
\begin{align}
&\pl \hg(\st,\bz,z,\s) = {\cL}_{\sgb (\s)}^{\nrm;\mnrn}(\s) :\!\sum_{u=0}^1 \hj_\nrmu (z,\s) \hg(\st,\bz,z,\s) \st_{\mnrn ,-u}\!: \label{Eq 3.10a}\\
&\bar{\pl} \hg(\st,\bz,z,\s) =-{\cL}_{\sgb(\s)}^{\nrm;\mnrn}(\s) :\!\sum_{u=0}^1 \tilde{\st}_{\mnrn ,-u} 
   \hj_\nrmu (\bz,\s) \hg(\st,\bz,z,\s)\!: \,. \label{Eq 3.10b} 
\end{align}
\end{subequations}
Here $\st =\st(T,\s)$, $:\cdot :$ is operator-product normal ordering in the orbifold (see Eq.~\eqref{Eq 3.6c}), and the twisted inverse inertia 
tensor ${\cL}(\s)$ is defined in \eqref{Eq 2.20c}. In parallel to the redundancy of the untwisted vertex operator equations \eqref{Eq2.15}, we have 
checked (see App.~C) that -- given the world-sheet parity \eqref{Eq 3.8b} -- the $\bpl$ equation \eqref{Eq 3.10b} follows from the 
$\pl$ equation \eqref{Eq 3.10a} and vice-versa.

\subsection{The Twisted Stress Tensors}

We turn finally to the stress tensors, for which the principle of local isomorphisms gives the twisted stress tensors $\hat{\Theta}_u (z,\s)$ 
of open-string sector $\hat{h}_\s$ and their monodromies
\begin{subequations}
\label{Eq3.11}
\begin{gather}
\Theta_u(z,\s) \dual \hat{\Theta}_u(z,\s) ,\quad \hat{\Theta}_u (ze^\tp ,\s) = (-1)^u \hat{\Theta}_u (z,\s) \label{Eq 3.11a} \\
\hat{\Theta}_{u \pm 2} (z,\s) = \hat{\Theta}_u (z,\s)  \label{Eq 3.11b}
\end{gather}
\end{subequations}
from the automorphic response \eqref{Eq 2.34b} of the stress-tensor eigenfields.

The OPEs of the twisted stress tensors 
\begin{subequations}
\label{Eq3.12}
\begin{gather}
\hat{\Theta}_u (z,\s) \hat{\Theta}_v (w,\s) \!=\! \frac{c_g\de_{u+v,0\,\text{mod }2}}{(z-w)^4} \!+\!\left( \frac{2}{(z-w)^2} 
   +\frac{\pl_w}{z-w} \right) \hat{\Theta}_{u+v} (w,\s) \nn \\
   +\Ord (z-w)^0 \label{Eq 3.12a}
\end{gather}
\begin{gather}
\hat{\Theta}_u (z,\s) \hj_{\nrm v} (w,\s) =\left( \frac{1}{(z-w)^2} +\frac{\pl_w}{z-w} \right) \hj_{\nrm ,u+v} (w,\s) +\Ord (z-w)^0 \\
\!\hat{\Theta}_u (z,\s) \hg(\st,\bw,w,\s) \!=\!\hg(\st,\bw,w,\s) \lsD(z,w) \tau_u \bigspc \bigspc \bigspc \quad \quad \nn \\
\quad \quad \quad \bigspc -\!(-1)^u \tau_u \D(z,\bw) \hg(\st,\bw,w,\s) +\Ord(z-w)^0 +\Ord (z-\bw)^0 
\end{gather}
\end{subequations}
\begin{subequations}
\label{Eq3.13}
\begin{align}
&\D(z,w) \equiv \frac{\D_{\sgb(\s)}(\st)}{(z-w)^2} +\frac{1}{z-w}\pl_w ,\quad \lsD (z,w) \equiv \frac{\D_{\sgb(\s)}(\st)}{(z-w)^2} 
   +\lplw \frac{1}{z-w} \\
&\sDg \!=\! U(T,\s) D_g (T) U\hc(T,\s) \!=\! {\cL}_{\sgb(\s)}^{\nrm;\nsn}(\s) \st_\nrm (T,\s) \st_\nsn (T,\s) \label{Eq 3.13b}
\end{align}
\begin{gather} 
[\D_{\sgb(\s)}(\st) ,\hg(\st,\bz,z,\s)] = [\D_{\sgb(\s)}(\st) ,\tau_u ]=0
\end{gather}
\end{subequations}
also follow by local isomorphisms from the OPEs of the stress-tensor eigenfields. The quantity $\D_{\sgb(\s)}(\st)$ is called the twisted conformal 
weight matrix \cite{Big} in the orbifold program.

For the special case of permutation-invariant systems
\begin{gather}
g =\oplus_I \gfrak^I ,\quad \gfrak^I \simeq \text{simple } \gfrak ,\quad k_I =k ,\quad T^I \simeq T \label{Eq3.14}
\end{gather}
(with $T$ any rep of $\gfrak$ and any $h_\s \in Aut(g)$) we find the simple result
\begin{equation}
\sDg = D_g (T) = \Delta_{\gfraks} (T) \one ,\quad \Delta_{\gfraks} (T) \one =L_g^{ab} T_a T_b = \oplus_I L_{\gfraks}^{a(I)b(I)} T_a^I T_b^I   \label{Eq3.15}
\end{equation}
where $\Delta_{\gfraks}(T)$ is the conformal weight of rep $T$ of $\gfrak$ under the affine-Sugawara construction [6,7,31-33,12] on $\gfrak$.

The unique physical stress tensor $\hat{T}_\s (z)$ of open-string sector $\hat{h}_\s$ is the component with trivial monodromy
\vspace{-0.1in}
\begin{gather}
\hat{T}_\s (z) \equiv \hat{\Theta}_0 (z,\s) ,\quad \hat{T}_\s (ze^\tp )=\hat{T}_\s (z) \label{Eq3.16}
\end{gather}
and we see from \eqref{Eq 3.12a} that the central charge of $\hat{T}_\s (z)$ is $\hat{c}=2c_g$. The physical stress tensor $\hat{T}_\s$ is the image in the 
orbifold of the eigenfield $(T_g+\bT_g)$, while the image $\hat{\Theta}_1$ of the other eigenfield $(T_g-\bT_g)$ has now picked up a non-trivial monodromy.

Finally, the principle of local isomorphisms applied to the eigenfield form $\Theta_u (\sj)$ in \eqref{Eq2.35} gives the explicit form of the twisted 
stress tensors as the twisted affine-Sugawara constructions:
\begin{gather}
\hat{\Theta}_u (z,\s) = \srac{1}{2} {\cL}_{\sgb (\s)}^{\nrm;\nsn} (\s) \sum_{v=0}^1 :\! \hj_{\nrm v} (z,\s) \hj_{\nsn ,u-v} (z,\s)\!: ,\quad \bar{u}=0,1\,.
  \label{Eq3.17}
\end{gather}
Here the twisted inverse inertia tensor ${\cL} (\s)$ was defined in \eqref{Eq 2.20c} and $: \cdot :$ is operator-product normal ordering in the twisted 
sector. In the standard notation of the orbifold program (see also App.~\ref{OrbNotApp}), the physical stress tensor \eqref{Eq3.16} would be written
\begin{subequations}
\label{Eq3.18}
\begin{gather}
\hat{T}_\s (z) =\hat{\Theta}_0 (z,\s) ={\cL}_{\sgb(\s)}^{\nrmu ;\mnrn ,-u}(\s) :\!\hj_\nrmu (z,\s) \hj_{\mnrn ,-u}(z,\s) \!: \label{Eq 3.18a} 
\end{gather}
\begin{align}
\!{\cL}_{\sgb(\s)}^{\nrmu ;\nsnv}(\s) \!&\equiv \!L_g^{a\Id ;b\Jd} U\hc (\s)_a{}^\nrm \schisig_\nrm^{-1} U\hc (\s)_b{}^\nsn \schisig_\nsn^{-1} 
   (\srac{1}{\sqrt{2}} U\hc {}_\Id{}^u ) (\srac{1}{\sqrt{2}} U\hc {}_\Jd{}^v ) \quad \quad  \nn \\
&=(\srac{1}{2} \de_{u+v ,0\,\text{mod }2}) {\cL}_{\sgb(\s)}^{\nrm;\nsn}(\s) \nn \\
&=\srac{1}{2} \de_{u+v ,0\,\text{mod }2} \de_{n(r)+n(s),0\,\text{mod }\r(\s)} {\cL}_{\sgb(\s)}^{\nrm ;\mnrn}(\s) \\
& \bigspc \bigspc L_g^{a\Id ;b\Jd} \equiv \de^{\Id \Jd} L_g^{ab} 
\end{align}
\end{subequations}
where ${\cL}$ in \eqref{Eq 3.18a} is the {\it total} twisted inverse inertia tensor, and we have suppressed an explicit sum on $u$.

\subsection{Mode Form of the Twisted Open-String Operator Algebra}

The monodromies \eqref{Eq 3.6a}, \eqref{Eq 3.11a} of the twisted fields determine the mode expansions
\begin{subequations}
\label{Eq3.19}
\begin{gather}
\hat{\Theta}_u (z,\s) = \sum_{m \in \Zint} \hat{L}_u (m \!+\!\srac{u}{2}) z^{-(m +\srac{u}{2}) -2} \\
\hat{T}_\s (z) =\hat{\Theta}_0 (z,\s) =\sum_{m \in \Zint} L_\s (m) z^{-m-2} ,\quad L_\s (m) = \hat{L}_0 (m) \\
\hj_\nrmu (z,\s) \equiv \sum_{m \in \Zint} \hj_\nrmu (m\!+\!\nrrs \!+\! \srac{u}{2}) z^{-(\mnrrs +\srac{u}{2})} 
\end{gather} \vspace{-0.27in}
\begin{align}
&\hat{L}_u (m\!+\!\srac{u}{2})= \bigspc \bigspc \nn \\
&\,\,\,\,\quad \srac{1}{2}{\cL}_{\sgb}^{\nrm;\mnrn}(\s) \sum_{v=0}^1 \sum_{p\in \Zint} :\!\hj_{\nrm v} (p\!+\! \nrrs \!+\!\srac{v}{2}) \hj_{\mnrn ,u-v} 
     (m\!-\!p \!-\!\nrrs \!+\!\srac{u-v}{2}) \!: 
\end{align}
\end{subequations}
and the mode periodicities
\begin{subequations}
\label{Eq3.20}
\begin{gather}
\hat{L}_{u\pm 2} (m\! +\!\srac{u\pm 2}{2}) = \hat{L}_u (m\!\pm \!1 \!+\!\srac{u}{2}) 
\end{gather}
\begin{align}
&\!\!\! \hj_{\nrm, u\pm 2} (m\!+\!\nrrs \!+\!\srac{u\pm 2}{2}) \!=\!\hj_{n(r)\pm \r(\s),\m u} (m\!+\!\srac{n(r) \pm \r(\s)}{\r(\s)} 
   \!+\!\srac{u}{2}) \!=\!\hj_\nrmu (m\!\pm \!1 \!+\!\nrrs \!+\!\srac{u}{2}) 
\end{align}
\end{subequations}
follow from the periodicity \eqref{Eq 3.9a}, \eqref{Eq 3.11b} of the local fields.

These mode expansions and the twisted OPEs \eqref{Eq3.6}, \eqref{Eq3.12} then give the mode form of the twisted operator algebra in 
open-string sector $\hat{h}_\s$:
\begin{subequations}
\label{Eq3.21}
\begin{gather}
[\hat{L}_u (m\!+\!\srac{u}{2}) ,\hat{L}_v (n\!+\!\srac{v}{2}) ] =(m\!-\!n \!+\!\srac{u-v}{2}) \hat{L}_{u+v} 
   (m\!+\!n \!+\! \srac{u+v}{2}) \bigspc \quad \quad \nn \\
\bigspc \quad \quad +\de_{m+n+\srac{u+v}{2},0} \frac{2c_g}{12} (m\!+\!\srac{u}{2}) ((m\!+\!\srac{u}{2} )^2 -1) \label{Eq 3.21a} \\
[\hat{L}_u (m\!+\!\srac{u}{2}) ,\hj_{\nrm v} (n\!+\!\nrrs \!+\!\srac{v}{2})] =-(n\!+\!\nrrs \!+\!\srac{v}{2}) \hj_{\nrm ,u+v}
   (m\!+\!n\!+\! \nrrs \!+\!\srac{u+v}{2}) 
\end{gather}
\begin{align}
& [\hat{L}_u (m\!+\!\srac{u}{2}) ,\hg(\st,\bz,z,\s)] =\hg(\st,\bz,z,\s) \tau_u \left( \lpl \!z+ \!(m\!+\!\srac{u}{2}\!+\!1) 
   \D_{\sgb(\s)}(\st) \right) z^{m+\srac{u}{2}}  \nn \\
&\bigspc +(-1)^u \tau_u \bz^{m+\srac{u}{2}} \left( \bz \bpl +(m\!+\!\srac{u}{2} \!+\!1) \D_{\sgb(\s)}(\st) \right) \hg(\st,\bz,z,\s) \label{Eq 3.21c}
\end{align}
\end{subequations}\vspace{-0.15in}
\begin{subequations}
\label{Eq3.22}
\begin{align}
&[ \hj_\nrmu (m\! +\!\nrrs \!+\!\srac{u}{2}) ,\hj_\nsnv (n\!+\!\nsrs \!+\!\srac{v}{2})] \bigspc \bigspc \bigspc \nn \\
&\bigspc = i\scf_{\nrm;\nsn}{}^{\!\!\!n(r)+n(s),\de}(\s) \hj_{n(r)+n(s),\de ,u+v} (\mnnrnsrs \!+\! \srac{u+v}{2}) \nn \\
&\bigspc +(\mnrrs +\srac{u}{2}) \de_{\mnnrnsrsf + \frac{u+v}{2},0} (2\de_{u+v ,0\, \text{mod }2}) \sG_{\nrm;\mnrn}(\s) \label{Eq 3.22a} \\
&[\hj_\nrmu (m\!+\!\nrrs \!+\!\srac{u}{2}) ,\hg(\st,\bz,z,\s) ] \nn \\
&\quad \quad =\hg (\st,\bz,z,\s) \st_\nrmu z^{\mnrrs +\srac{u}{2}} \!-\!\tilde{\st}_\nrmu \hg(\st,\bz,z,\s) \bz^{\mnrrs +\srac{u}{2}} \,. \label{Eq 3.22b} 
\end{align}
\end{subequations}
Here the duality transformations $\sG,\scf ,\st ,\tilde{\st}$ and $\D$ are defined in Eqs.~\eqref{Eq2.20}, \eqref{Eq2.26}, and \eqref{Eq 3.13b}.

We comment first on the {\it orbifold Virasoro algebra} \eqref{Eq 3.21a}, which was originally seen in $\Zint_\lambda$ cyclic orbifolds \cite{Chr,DV2}. 
Generalizations of this algebra are also found in the theory of general WZW permutation orbifolds \cite{Perm}.

In the orbifold Virasoro algebra, we see that twisted sector $\hat{h}_\s$ has only the {\it single} untwisted Virasoro subalgebra (the so-called 
integral Virasoro subalgebra) generated 
by $\{ L_\s (m) =\hat{L}_0 (m) \}$ 
\begin{gather}
[L_\s (m) ,L_\s(n)] =(m-n) L_\s (m+n) +\de_{m+n,0} \frac{2c_g}{12} m(m^2 -1)  \label{Eq3.23}
\end{gather}
which tells us that the twisted sector is an open string. Moreover, the central charge of the single untwisted Virasoro subalgebra is $\hat{c} \!=\! 2c_g$
-- which tells us (along with the fractionally-moded fields) that our construction is not a conventional orientifold [25-27]. We also remark on commutation 
relations such as \eqref{Eq 3.22b} and 
\begin{gather}
[L_\s (m),\hg(\st,\bz,z,\s)] =\hg(\st,\bz,z,\s) \Big{(} \lpl z +(m+1) \D_{\sgb(\s)}(\st) \Big{)} z^m \bigspc \bigspc \nn \\
\bigspc \bigspc + \bz^m \Big{(} \bz \,\bpl +(m+1) \D_{\sgb(\s)}(\st) \Big{)} \hg(\st,\bz,z,\s)  \label{Eq3.24}
\end{gather}
which show the twisted currents and the single untwisted Virasoro acting simultaneously as left- and right-movers. Such actions are well-known in untwisted
open-string WZW theory \cite{Giusto}.

The twisted current algebra $\gfrakh_O (\s)$ in \eqref{Eq 3.22a} provides a large set of examples of so-called `doubly-twisted' current 
algebras \cite{TVME,Coset}, which result more generally from the composition of automorphisms of different types. We also mention the identity
\begin{equation}
(\de_{u+v,0\, \text{mod }2} - \de_{n(r)+n(s),0\, \text{mod }\r(\s)}) \de_{\mnnrnsrsf +\frac{u+v}{2},0} =0  \label{Eq3.25}
\end{equation}
which may be used to obtain other forms of the central term in the twisted current algebra. Note also that (as is well-known in the orbifold program 
\cite{Big,Fab}) the twisted current algebra $\gfrakh_O (\s)$ shares the same twisted structure constants $\scf(\s)$ with the total orbifold 
Lie algebra $\hg_O(\s)$ in \eqref{Eq 2.26c}. 

We finally remark on the integral affine subalgebra (integer-moded subalgebra) of the twisted current algebra \eqref{Eq 3.22a}, whose zero modes $\hj(0)$
generate the residual symmetry algebra in open-string sector $\hat{h}_\s$ of the orientation orbifold. When the order $\r(\s)$ of $h_\s$ is odd, the 
integral affine subalgebra is generated by $\{ \hj_{0,\m,0} (m)\}$ -- but when $\r(\s)$ is even, this subalgebra has two sets of generators:
\begin{equation}
\label{Eq3.26}
\{ \hj_{0,\m,0} (m) \} \,\,\, \text{and} \,\,\, \{ \hj_{\r(\s)/2 ,\m,1} (m+ \srac{\r(\s)/2}{\r(\s)} + \srac{1}{2}) 
   =\hj_{\r(\s)/2,\m,1} (m+1) \} ,\quad \forall m,\m \,. 
\end{equation}
In this case, the integral affine subalgebra takes the form
\begin{subequations}
\label{Eq3.27}
\begin{gather}
[ \hj_{0,\m,0} (m) ,\hj_{0,\n,0} (n)] =i\scf_{0,\m;0,\n}{}^{0,\de}(\s) \hj_{0,\de,0}(m\!+\!n) +2m\de_{m+n,0}\sG_{0,\m;0,\n}(\s) \label{Eq 3.27a} \\
[ \hj_{0,\m,0} (m) ,\hj_{\r(\s)/2,\n,1} (n\+1)] = i\scf_{0,\m;\r(\s)/2,\n}{}^{\r(\s)/2,\de}(\s) \hj_{\r(\s)/2,\de,1} (m+n+1) \\
[\hj_{\r(\s)/2,\m,1} (m\+1) ,\hj_{\r(\s)/2,\n,1} (n\+1)] =i\scf_{\r(\s)/2,\m;\r(\s)/2 ,\n}{}^{0,\de}(\s) \hj_{0,\de,0}(m\+n\+2) \bigspc \quad \nn \\
   \bigspc \bigspc \bigspc +2(m\+1) \de_{m+n+2,0} \sG_{\r(\s)/2,\m ;\r(\s)/2,\n}(\s) 
\end{gather}
\end{subequations}
and the residual symmetry algebra of sector $\hat{h}_\s$ is
\begin{subequations}
\label{Eq 3.28}
\begin{gather}
[ \hj_{0,\m,0} (0) ,\hj_{0,\n,0} (0)] =i\scf_{0,\m;0,\n}{}^{0,\de}(\s) \hj_{0,\de,0}(0) \label{Eq 3.28a} \\
[ \hj_{0,\m,0} (0) ,\hj_{\r(\s)/2,\n,1} (0)] = i\scf_{0,\m;\r(\s)/2,\n}{}^{\r(\s)/2,\de}(\s) \hj_{\r(\s)/2,\de,1} (0) \label{Eq 3.28b} \\
[\hj_{\r(\s)/2,\m,1} (0) ,\hj_{\r(\s)/2,\n,1} (0)] =i\scf_{\r(\s)/2,\m;\r(\s)/2 ,\n}{}^{0,\de}(\s) \hj_{0,\de,0}(0) \label{Eq 3.28c}
\end{gather}
\end{subequations}
where only \eqref{Eq 3.27a}, \eqref{Eq 3.28a} are relevant when $\r(\s)$ is odd. The global Ward identities of Subsec.~$4.3$ are a consequence of these 
residual symmetry algebras.

\subsection{Mode Ordering and the Scalar Twist-Field State}

In each open-string sector $\hat{h}_\s$ of each WZW orientation orbifold, the scalar twist-field state $|0\rangle_\s$ satisfies 
\begin{gather}
\hj_\nrmu (m\!+\!\nrrs \!+\!\srac{u}{2} \geq 0)|0\rangle_\s = {}_\s \langle 0| \hj_\nrmu (m\!+\!\nrrs \!+\!\srac{u}{2} \leq 0) =0 
  \label{Eq3.29}
\end{gather}
and therefore the standard mode normal ordering \cite{Dual,More,Big} of the twisted currents
\begin{subequations}
\label{Eq3.30}
\begin{align}
&:\!\hj_\nrmu (m\!+\!\nrrs \!+\!\srac{u}{2}) \hj_\nsnv (n\!+\!\nsrs \!+\!\srac{v}{2}) \!:_M = \nn \\
& \bigspc \theta (m\!+\!\nrrs \!+\!\srac{u}{2} \geq 0) \hj_\nsnv (n\!+\!\nsrs \!+\!\srac{v}{2}) \hj_\nrmu 
   (m\!+\!\nrrs \!+\!\srac{u}{2}) \nn \\
&\bigspc \quad +\theta (m\!+\!\nrrs \!+\!\srac{u}{2} <0) \hj_\nrmu (m\!+\!\nrrs \!+\!\srac{u}{2}) 
   \hj_\nsnv (n\!+\!\nsrs \!+\!\srac{v}{2}) \label{Eq 3.30a}
\end{align}
\begin{gather}
{}_\s \langle 0| :\!\hj_\nrmu (m\!+\!\nrrs \!+\!\srac{u}{2}) \hj_\nsnv (n\!+\!\nsrs \!+\!\srac{v}{2}) \!:_M |0\rangle_\s =0 \label{Eq 3.30b}
\end{gather}
\end{subequations}
will be useful for computations below.

As a first application, we may use the ordering \eqref{Eq 3.30a} and the twisted current algebra \eqref{Eq 3.22a} to compute the 
exact $\hj \hj$ operator product in open-string sector $\hat{h}_\s$:
\begin{subequations}
\label{Eq3.31}
\begin{align}
\!&\!\!\hj_\nrmu (z,\s) \hj_\nsnv (w,\s) \!=\!\left( \frac{w}{z} \right)^{\!\bar{y}(r,u)} \!\Big{(} \frac{(2\de_{u+v,0\,\text{mod }2})
   \sG_{\nrm;\nsn}(\s)}{(z-w)^2} h (z,w;\bar{y}(r,u)) \quad \nn \\
&\bigspc \quad \quad +\frac{i\scf_{\nrm;\nsn}{}^{n(r)+n(s),\de}(\s) \hj_{n(r)+n(s),\de,u+v} (w,\s)}{z-w} f (z,w;\bar{y}(r,u)) \Big{)} \nn \\
&\bigspc \quad \quad \, +:\!\hj_\nrmu (z,\s) \hj_\nsnv (w,\s)\!:_M \label{Eq 3.31a}
\end{align}
\begin{gather}
h (z,w;\bar{y}) \equiv 1 + \frac{ \bar{y}(z-w)}{w} +\theta (\bar{y} \geq 1) \frac{(\bar{y}-1)(z-w)^2}{w^2} \\
f (z,w;\bar{y}) \equiv 1 + \theta (\bar{y} \geq 1) \frac{z-w}{w} \\ 
\bar{y}(r,u) \equiv \frac{\bar{n}(r)}{\r(\s)} +\frac{\bar{u}}{2} ,\quad 0\leq \bar{y}(r,u) < \frac{3}{2} \,. \label{Eq 3.31d}
\end{gather}
\end{subequations} 
The summation identity \eqref{Inf-Sum} is useful in obtaining this result.

The normal-ordering identity \eqref{Eq 3.30b} and the operator product \eqref{Eq 3.31a} then give the exact twisted current-current correlator in the 
scalar twist-field state
\begin{align}
&\langle \hj_\nrmu(z,\s) \hj_\nsnv(w,\s)\rangle_\s \equiv {}_\s \langle 0|\hj_\nrmu(z,\s) \hj_\nsnv(w,\s) |0\rangle_\s \bigspc \bigspc \nn \\
&\bigspc \quad =\left( \frac{w}{z} \right)^{\!\bar{y}(r,u)} \!2\de_{u+v,0\,\text{mod }2} \de_{n(r)+n(s),0\,\text{mod }\r(\s)} \sG_{\nrm;\mnrn}(\s) \nn \\
&\bigspc \bigspc \quad \times \left( \frac{1}{(z-w)^2} \!+\!\frac{\bar{y}(r,u)}{z-w} \!+\!\theta(\bar{y}(r,u) \geq 1) \frac{\bar{y}(r,u)-1}{w^2} \right)  \label{Eq3.32}
\end{align}
where $\bar{y}$ is defined in \eqref{Eq 3.31d}.

The operator-product expansion of Eq.~\eqref{Eq 3.31a} reproduces the singular terms in the twisted current-current OPE \eqref{Eq 3.6b}, and we also 
find the following local relation between operator-product normal ordering and mode normal ordering:
\begin{align}
&:\!\hj_\nrmu (z,\s) \hj_\nsnv (z,\s)\!: = :\!\hj_\nrmu (z,\s) \hj_\nsnv (z,\s)\!:_M \bigspc \bigspc \bigspc \nn \\
&\quad \quad -i\scf_{\nrm;\nsn}{}^{\!\!\!n(r)+n(s),\de}(\s) \hj_{n(r)+n(s),\de,u+v}(z,\s) \frac{1}{z} (\bar{y}(r,u)-\theta (\bar{y}(r,u) \geq 1)) \nn \\
&\quad \quad +(2\de_{u+v ,0\,\text{mod }2}) \sG_{\nrm;\nsn}(\s) \frac{1-\bar{y}(r,u)}{z^2} (\frac{\bar{y}(r,u)}{2} -\theta (\bar{y}(r,u) \geq 1)) \,.  \label{Eq3.33}
\end{align}
With this relation and Eq.~\eqref{Eq3.17}, we find the mode-ordered form of the twisted stress tensors
\begin{align}
\hat{\Theta}_u(z,\s) =& \srac{1}{2} {\cL}_{\sgb(\s)}^{\nrm;\mnrn}(\s) \sum_{v=0}^1 \Big{\{} :\!\hj_{\nrm,v}(z,\s) 
   \hj_{\mnrn ,u-v}(z,\s)\!:_M  \nn \\
&\quad  -i\scf_{\nrm;\mnrn}{}^{0,\de}(\s) \hj_{0,\de,u}(z,\s) \frac{1}{z}(\bar{y}(r,v) -\theta(\bar{y}(r,v) \geq 1)) \nn \\
&\quad +2\de_{u,0\,\text{mod }2} \sG_{\nrm;\mnrn}(\s) \frac{1-\bar{y}(r,v)}{z^2} (\frac{\bar{y}(r,v)}{2} -\theta(\bar{y}(r,v) \geq 1)) \Big{\}} \label{Eq3.34}
\end{align}
and hence the mode-ordered form of the generators of the orbifold Virasoro algebra:
\begin{align}
&\hat{L}_u (m\!+\!\srac{u}{2})= \bigspc \bigspc \bigspc \bigspc \bigspc \nn \\
& \quad \quad \srac{1}{2} {\cL}_{\sgb(\s)}^{\nrm;\mnrn}(\s) \sum_{v=0}^1 \!\Big{\{} \sum_{p\in \Zint} 
   \!:\!\hj_{\nrm v} (p\!+\!\nrrs \!+\!\srac{v}{2}) \hj_{\mnrn ,u-v} (m\!-\!p\!-\!\nrrs \!+\!\srac{u-v}{2}) \!:_M  \nn \\
&\quad \quad \quad -i\scf_{\nrm;\mnrn}{}^{0,\de}(\s) \hj_{0,\de,u}(m\!+\!\srac{u}{2}) (\bar{y}(r,v) -\theta(\bar{y}(r,v) \geq 1)) \nn \\
&\quad \quad \quad +2\de_{m+\srac{u}{2},0} \sG_{\nrm;\mnrn}(\s) (1-\bar{y}(r,v))(\frac{\bar{y}(r,v)}{2} -\theta(\bar{y}(r,v) \geq 1)) \Big{\}} \,. \label{Eq3.35}
\end{align}
We may then compute the conformal weight $\gscfwt$ of the scalar twist-field state in open-string sector $\hat{h}_\s$ of the orientation orbifold
\begin{subequations}
\label{Eq3.36}
\begin{align}
&\left( \hat{L}_u (m\!+\!\srac{u}{2}\geq 0) -\!\de_{m+\srac{u}{2},0} \gscfwt \right) |0\rangle_\s ={}_\s\langle 0| \left( \hat{L}_u (m\!+\!\srac{u}{2} \leq 0) -\!\de_{m+\srac{u}{2},0} \gscfwt \right) =0 \\
&\gscfwt \equiv \sum_{r,\m,\n,u} \left[ {\cL}_{\sgb(\s)}^{\nrm;\mnrn}(\s) \sG_{\nrm;\mnrn}(\s) (1\!-\!\srac{\bar{n}(r)}{\r(\s)} 
   \!-\!\srac{\bar{u}}{2}) \right. \bigspc \nn \\
&\bigspc \bigspc \bigspc \quad \quad \left. \times \left( \srac{1}{2}(\srac{\bar{n}(r)}{\r(\s)} \!+\! \srac{\bar{u}}{2}) 
   -\theta(\srac{\bar{n}(r)}{\r(\s)}\!+\!\srac{\bar{u}}{2} \geq 1)\right) \right]  \label{Eq 3.36b} \\
& \,\,\, =\!\sum_{r,\m,\n} \!{\cL}_{\sgb(\s)}^{\nrm;\mnrn}(\s) \sG_{\nrm;\mnrn}(\s) \left[ ( \srac{1}{8} \!+\!\srac{\bar{n}(r)}{2\r(\s)}
   \!-\!(\srac{\bar{n}(r)}{\r(\s)} )^2 )+ (\srac{\bar{n}(r)}{\r(\s)} \!-\!\srac{1}{2}) \theta(\srac{\bar{n}(r)}{\r(\s)} \geq \srac{1}{2} ) \right] \label{Eq 3.36c}
\end{align}
\end{subequations}
as a consequence of Eqs.~\eqref{Eq3.35} and the defining relations \eqref{Eq3.29}. 

In the special case of permutation-invariant systems 
\begin{gather}
g =\oplus_I \gfrak^I ,\quad \gfrak^I \simeq \text{simple } \gfrak, \quad k_I =k  \label{Eq3.37}
\end{gather}
we find (for any $h_\s \in Aut(g)$) a simpler form for the twist-field conformal weight:
\begin{subequations}
\label{Eq3.38}
\begin{gather}
{\cL}_{\sgb(\s)}^{\nrm;\nsn}(\s) = \frac{k}{2k+Q_{\gfraks}} \sG^{\nrm;\nsn}(\s) \\
\gscfwt = \frac{c_{\gfraks}}{2} \sum_{r} \frac{\text{dim}[\bar{n}(r)]}{\text{dim} \,\gfrak} \left[ ( \srac{1}{8} \!+\!\srac{\bar{n}(r)}{2\r(\s)} 
   \!-\!(\srac{\bar{n}(r)}{\r(\s)} )^2 )+ (\srac{\bar{n}(r)}{\r(\s)} \!-\!\srac{1}{2}) \theta(\srac{\bar{n}(r)}{\r(\s)} \geq \srac{1}{2} ) \right] \\
\text{dim} [\bar{n}(r)] \equiv \sum_\m \de_\nrm{}^\nrm ,\quad \sum_r \text{dim}[\bar{n}(r)] =\text{dim } g ,\quad c_{\gfraks} =\frac{2k \text{dim}\,\gfrak}
   {2k +Q_{\gfraks}} \,.
\end{gather}
\end{subequations}
Here dim$[\bar{n}(r)]$ is the degeneracy of the energy $E_{n(r)}(\s) \!=\! e^{-\tp \nrrsf}$ in the $H$-eigenvalue problem \eqref{Eq 2.16a}.

\section{The Twisted KZ Systems of the WZW Orientation Orbifolds}

\subsection{The $\hj \hg$ Operator Products}

In parallel with Subsec.~$3.5$, we study here exact operator products and normal orderings in the $\hj \hg$ system.

The basic tool here is the standard \cite{Big} mode normal ordering $:\cdot :_M$ defined for $\hj$ with $\hg$
\vspace{-0.15in}
\begin{subequations}
\label{Eq4.1}
\begin{gather}
:\!\hj_\nrmu (m\!+\!\nrrs \!+\!\srac{u}{2}) \hg(\st,\bz,z,\s) \!:_M \,\equiv \bigspc \bigspc \bigspc \quad \quad \nn \\
   \theta (m\!+\!\nrrs \!+\!\srac{u}{2} \geq 0) \hg(\st,\bz,z,\s) \hj_\nrmu (m\!+\!\nrrs \!+\!\srac{u}{2}) \nn \\
   \bigspc +\theta (m\!+\!\nrrs \!+\!\srac{u}{2} <0) \hj_\nrmu (m\!+\!\nrrs \!+\!\srac{u}{2}) \hg(\st,\bz,z,\s) \\
:\!\hj_\nrmu (z,\s) \hg(\st,\bw,w,\s) \!:_M \,= \hj_\nrmu^{(-)} (z,\s) \hg(\st,\bw,w,\s) \bigspc \quad \quad \nn \\
   \bigspc \bigspc \bigspc \bigspc +\hg(\st,\bw,w,\s) \hj_\nrmu^{(+)} (z,\s) \\
\hj_\nrmu^{(+)} (z,\s) \equiv \sum_{m \in \Zint} \theta (m\!+\!\nrrs \!+\!\srac{u}{2} \geq 0) \hj_\nrmu (m\!+\!\nrrs \!+\!\srac{u}{2}) 
   z^{- (m+\frac{n(r)}{\r(\s)} +\frac{u}{2})-1} \\
\hj_\nrmu^{(-)} (z,\s) \equiv \sum_{m \in \Zint} \theta (m\!+\!\nrrs \!+\!\srac{u}{2} < 0) \hj_\nrmu (m\!+\!\nrrs \!+\!\srac{u}{2}) 
   z^{- (m+\frac{n(r)}{\r(\s)} +\frac{u}{2})-1} 
\end{gather}
\end{subequations}
where $\hj^{(\pm)}$ are called the twisted partial currents. Then we may compute the exact $\hj \hg$ operator product in open-string sector $\hat{h}_\s$
\begin{subequations}
\label{Eq4.2}
\begin{align}
\!\!\!&\!\! \hj_\nrmu (z,\s) \hg(\st,\bw,w,\s) =\left( \frac{w}{z} \right)^{\!\bar{y}(r,u)} \frac{f(z,w;\bar{y}(r,u))}{z-w} \hg(\st,\bw,w,\s) \st_\nrmu \nn \\
&\,\,\,\, -\!\left( \frac{\bw}{z} \right)^{\!\bar{y}(r,u)} \!\!\frac{f(z,\bw;\bar{y}(r,u))}{z-\bw} \tilde{\st}_\nrmu \hg(\st,\bw,w,\s) 
   +\!:\!\hj_\nrmu (z,\s) \hg(\st,\bw,w,\s)\!:_M \label{Eq 4.2a} \\
&\bigspc \bigspc \quad \quad \quad  f (z,w;\bar{y}) = 1+\frac{z-w}{w} \theta (\bar{y} \geq 1)
\end{align}
\end{subequations}
where we have used the commutator \eqref{Eq 3.22b} and the identity \eqref{Inf-Sum}. This is the same function $f$ encountered in Eq.~\eqref{Eq3.31}.
The pullback $\bar{y}(r,u)$ and the twisted representation matrices $\st$ and $\tilde{\st}$ are defined respectively in Eqs.~\eqref{Eq 3.31d} 
and \eqref{Eq2.26}.

The operator-product expansion of Eq.~\eqref{Eq4.2} reproduces the singular terms of the $\hj \hg$ OPE in \eqref{Eq 3.6c}, and one finds moreover 
the following relation between operator product and mode normal ordering:
\begin{align}
&:\! \hj_\nrmu (z,\s) \hg(\st,\bw,w,\s) \!: = :\! \hj_\nrmu (z,\s) \hg(\st,\bw,w,\s) \!:_M \nn \\
&\bigspc +\left\{ \!\left( \frac{w}{z} \right)^{\!\bar{y}(r,u)} \!f(z,w;\bar{y}(r,u))-1 \right\} \frac{1}{z-w} \hg(\st,\bw,w,\s) \st_\nrmu \nn \\
&\bigspc -\left\{ \!\left( \frac{\bw}{z} \right)^{\!\bar{y}(r,u)} \!f(z,\bw;\bar{y}(r,u))-1 \right\} \frac{1}{z-\bw} \tilde{\st}_\nrmu \hg(\st,\bw,w,\s) \,.  \label{Eq4.3}
\end{align}
This exact relation may be analyzed in the limits $z \rightarrow w$ or $\bw$ to obtain the following quasi-local normal-ordering relations for 
$\hj$ with $\hg$:
\begin{subequations}
\label{Eq4.4}
\begin{align}
&:\!\hj_\nrmu(z,\s) \hg(\st,\bz,z,\s)\!: \,=\,:\!\hj_\nrmu(z,\s) \hg(\st,\bz,z,\s)\!:_M \nn \\
& \bigspc -\frac{1}{z} (\bar{y}(r,u)- \theta(\bar{y}(r,u) \geq 1)) \hg(\st,\bz,z,\s) \st_\nrmu \nn \\
& \bigspc -\left\{\!\left( \frac{\bz}{z} \right)^{\!\bar{y}(r,u)} \!f(z,\bz;\bar{y}(r,u)) -1 \right\} \frac{1}{z-\bz} \tilde{\st}_\nrmu \hg(\st,\bz,z,\s) \\
&:\!\hj_\nrmu(\bz,\s) \hg(\st,\bz,z,\s)\!: \,=\, :\!\hj_\nrmu(\bz,\s) \hg(\st,\bz,z,\s)\!:_M \nn \\
& \bigspc +\left\{ \!\left( \frac{z}{\bz} \right)^{\!\bar{y}(r,u)} \!f(\bz,z;\bar{y}(r,u)) -1 \right\} \frac{1}{\bz-z} \hg(\st,\bz,z,\s) \st_\nrmu \nn \\
& \bigspc +\frac{1}{\bz} (\bar{y}(r,u) -\theta(\bar{y}(r,u) \geq 1)) \tilde{\st}_\nrmu \hg(\st,\bz,z,\s) \,.  
\end{align}
\end{subequations}
All these normal-ordered products are finite because $f(z,z;\bar{y})=1$.

\subsection{Mode-Ordered Form of the Twisted Vertex Operator Equations}

Using Eqs.~\eqref{Eq3.10} and \eqref{Eq4.4}, we may now give the mode normal-ordered form of the twisted vertex operator equations: 
\begin{subequations}
\label{Eq4.5}
\begin{align}
&\pl \hg(\st,\bz,z,\s)={\cL}_{\sgb(\s)}^{\nrm;\mnrn} (\s) \sum_{u=0}^1 \Big{\{} :\!\hj_\nrmu (z,\s)\hg(\st,\bz,z,\s) \!:_M  \nn \\
& \bigspc -\left( \frac{\bz}{z} \right)^{\!\bar{y}(r,u)} \frac{f(z,\bz;\bar{y}(r,u))}{z-\bz} \tilde{\st}_\nrmu \hg(\st,\bz,z,\s) \nn \\
&\bigspc -\frac{1}{z} \left( \bar{y}(r,u)-\theta (\bar{y}(r,u) \geq 1) \right) \hg(\st,\bz,z,\s) \st_\nrmu \Big{\}} \st_{\mnrn ,-u} \label{Eq 4.5a}
\end{align}
\begin{align}
&\bpl \hg(\st,\bz,z,\s) \!=\! -{\cL}_{\sgb(\s)}^{\nrm;\mnrn} (\s) \sum_{u=0}^1 \tilde{\st}_{\mnrn ,-u} \Big{\{} \!:\!\hj_\nrmu (\bz,\s) 
   \hg(\st,\bz,z,\s)\!:_M  \quad \quad \nn \\
&\bigspc \quad \quad \quad +\left( \frac{z}{\bz} \right)^{\!\bar{y}(r,u)} \frac{f(\bz,z;\bar{y}(r,u))}{\bz -z} \hg(\st,\bz,z,\s) \st_\nrmu \nn \\
& \bigspc \quad \quad \quad +\frac{1}{\bz} \left( \bar{y}(r,u)-\theta (\bar{y}(r,u) \geq 1) \right) \tilde{\st}_\nrmu \hg(\st,\bz,z,\s) \Big{\}} \,. \label{Eq 4.5b}
\end{align}
\end{subequations}
In stating these results, we simplified the $\tilde{\st} \hg \st$ terms of both equations by the non-trivial substitutions:
\begin{subequations}
\label{Eq4.6}
\begin{gather}
\left\{ \left( \frac{\bz}{z} \right)^{\!\bar{y}(r,u)} \!f(\bz,z;\bar{y}(r,u)) -1 \right\} \,\,\longleftrightarrow \,\, 
    \left( \frac{\bz}{z} \right)^{\!\bar{y}(r,u)} \!f(\bz,z;\bar{y}(r,u)) \\
\left\{ \left( \frac{z}{\bz} \right)^{\!\bar{y}(r,u)} \!f(z,\bz;\bar{y}(r,u)) -1 \right\} \,\,\longleftrightarrow \,\, 
    \left( \frac{z}{\bz} \right)^{\!\bar{y}(r,u)} \!f(z,\bz;\bar{y}(r,u)) \,.
\end{gather}
\end{subequations}
These substitutions are possible because of the identity 
\begin{gather}
[\hg ,\tau_1] =0 \,\Rightarrow \, \sum_{u=0}^1 \tilde{\st}_\nrmu \hg \st_{\mnrn ,-u} =\sum_{u=0}^1 (-1)^u \st_\nrm \hg \st_\mnrn=0  \label{Eq4.7}
\end{gather}
which follows from the $\tau_1$-constraint \eqref{Eq 3.8a} and the definitions \eqref{Eq2.26} of $\st$ and $\tilde{\st}$. We note in particular that the 
twisted vertex operator equations \eqref{Eq4.5} are {\it non-singular} as $z \rightarrow \bz$ (as they must be according to Eq.~\eqref{Eq4.6}) and this 
can also be verified directly by doing the sums on $u$ (see Subsec.~$4.4$).

Following Refs.~\cite{H+O,Big',Perm}, the questions of reducibility and monodromy of the twisted affine primary fields can be studied with these twisted vertex 
operator equations, but these topics are beyond the scope of the present paper. Monodromy is discussed for the classical (high-level) limit in Subsec.~$6.1$.

\subsection{The Twisted KZ System of Open-String Sector $\hat{h}_\s$}

For this computation, we will need the following commutators of the twisted partial currents with the twisted affine primary fields:
\begin{subequations}
\label{Eq4.8}
\begin{gather}
[\hj_\nrmu^{(+)}(z_i,\s) ,\hg(\st^{(j)},\bz_j ,z_j,\s)] =\!\left( \frac{z_j}{z_i} \right)^{\!\!\bar{y}(r,u)} \!\frac{f(z_i,z_j ;\bar{y}(r,u))}{z_i-z_j} 
   \hg(\st^{(j)},\bz_j ,z_j,\s) \st_\nrmu^{(j)} \bigspc  \nn \\
\bigspc \quad \quad -\!\left( \frac{\bz_j}{z_i} \right)^{\!\bar{y}(r,u)} \frac{f(z_i ,\bz_j ;\bar{y}(r,u))}{z_i -\bz_j} \tilde{\st}_\nrmu^{(j)} 
   \hg(\st^{(j)},\bz_j,z_j,\s) ,\quad |z_i| >|z_j| 
\end{gather}
\begin{gather}
[\hj_\nrmu^{(-)} (z_i,\s) ,\hg(\st^{(j)},\bz_j ,z_j,\s)] =-\!\left( \frac{z_j}{z_i} \right)^{\!\bar{y}(r,u)} \frac{f(z_i ,z_j ;\bar{y}(r,u))}{z_i-z_j} 
   \hg(\st^{(j)},\bz_j ,z_j,\s) \st_\nrmu^{(j)} \bigspc \nn \\
\bigspc \quad \quad +\!\left( \frac{\bz_j}{z_i} \right)^{\!\bar{y}(r,u)} \frac{f(z_i ,\bz_j ;\bar{y}(r,u))}{z_i -\bz_j} \tilde{\st}_\nrmu^{(j)} 
   \hg(\st^{(j)},\bz_j,z_j,\s) ,\quad |z_j| >|z_i| 
\end{gather}
\begin{gather}
[\hj_\nrmu^{(+)}(\bz_i,\s) ,\hg(\st^{(j)},\bz_j ,z_j,\s)] =\left( \frac{z_j}{\bz_i} \right)^{\!\bar{y}(r,u)} \frac{f(\bz_i ,z_j ;\bar{y}(r,u))}{\bz_i-z_j}
   \hg(\st^{(j)},\bz_j ,z_j,\s) \st_\nrmu^{(j)} \bigspc  \nn \\
\bigspc \quad \quad -\!\left( \frac{\bz_j}{\bz_i} \right)^{\!\bar{y}(r,u)} \frac{f(\bz_i ,\bz_j ;\bar{y}(r,u))}{\bz_i -\bz_j} \tilde{\st}_\nrmu^{(j)} 
   \hg(\st^{(j)},\bz_j,z_j,\s) ,\quad |z_i| >|z_j| \\
[\hj_\nrmu^{(-)}(\bz_i,\s) ,\hg(\st^{(j)},\bz_j,z_j,\s)] =-\!\left( \frac{z_j}{\bz_i} \right)^{\!\bar{y}(r,u)} \frac{f(\bz_i ,z_j ;\bar{y}(r,u))}{\bz_i-z_j} 
   \hg(\st^{(j)},\bz_j ,z_j,\s) \st_\nrmu^{(j)}  \bigspc \nn \\
\bigspc \quad \quad +\left( \frac{\bz_j}{\bz_i} \right)^{\!\bar{y}(r,u)} \frac{f(\bz_i ,\bz_j ;\bar{y}(r,u))}{\bz_i -\bz_j} \tilde{\st}_\nrmu^{(j)} 
   \hg(\st^{(j)},\bz_j,z_j,\s) ,\quad |z_j| >|z_i| \\
\st^{(j)} \equiv \st(T^{(j)}\!,\s) ,\quad \tilde{\st}^{(j)} \equiv \tilde{\st}(T^{(j)}\!,\s) \,.
\end{gather}
\end{subequations}
These relations are obtained from the mode algebra \eqref{Eq 3.22b} and the summation identity \eqref{Inf-Sum}, and the function $f$ is defined 
in Eq.~\eqref{Eq3.31}.

We next define the twisted open-string correlators in the scalar twist-field state (see Subsec.~$3.5$) of sector $\hat{h}_\s$:
\begin{subequations}
\label{Eq4.9}
\begin{gather}
\hat{A}_\s (\st,\bz,z) \equiv \langle \hg(\st^{(1)},\bz_1 ,z_1,\s) \ldots \hg (\st^{(n)} ,\bz_n ,z_n,\s) \rangle_\s \\
\hat{A}_\s (\st,\bz,z) \equiv \left\{ \hat{A}_\s (\st(T,\s),\bz,z,\s)_{N(r_1)\m_1 u_1 ,\ldots ,N(r_n)\m_n u_n}{}^{\!\!\!N(s_1)\n_1 v_1 ,\ldots
   ,N(s_n)\n_n v_n} \right\} \,. \label{Eq 4.9b}
\end{gather}
\end{subequations}
Using the commutators \eqref{Eq4.8} and the mode-ordered forms \eqref{Eq4.5} of the twisted vertex operator equations, we may now 
differentiate the correlators to obtain the {\it twisted KZ equations} for open-string sector $\hat{h}_\s$ of the orientation orbifold:
\begin{subequations}
\label{Eq4.10}
\begin{align}
& \pl_i \hat{A}_\s (\st,\bz,z) =\! {\cL}_{\sgb(\s)}^{\nrm;\mnrn}(\s) \nn \\
& \quad \quad \times \sum_{u=0}^1 \left[ \sum_{j \neq i} \!\left( \frac{z_j}{z_i} 
  \right)^{\!\bar{y}(r,u)} \!\frac{f(z_i ,z_j ;\bar{y}(r,u))}{z_i-z_j} \hat{A}_\s (\st,\bz,z) \st_\nrmu^{(j)} \right. \nn \\
&\bigspc \bigspc \left. -\sum_j \left( \frac{\bz_j}{z_i} \right)^{\!\bar{y}(r,u)} \frac{f(z_i ,\bz_j ;\bar{y}(r,u))}{z_i -\bz_j} 
  \tilde{\st}_\nrmu^{(j)} \hat{A}_\s(\st,\bz,z) \right. \nn \\
& \bigspc \bigspc \left. -\frac{1}{z_i} (\bar{y}(r,u) -\theta(\bar{y}(r,u) \geq 1)) \hat{A}_\s(\st,\bz,z) \st_\nrmu^{(i)} \right]
   \st_{\mnrn ,-u}^{(i)} \label{Eq 4.10a}
\end{align}
\begin{align} 
&\bpl_i \hat{A}_\s (\st,\bz,z) ={\cL}_{\sgb(\s)}^{\nrm;\mnrn}(\s) \nn \\
& \quad \quad \times \sum_{u=0}^1 \tilde{\st}_{\mnrn ,-u}^{(i)} \left[ \sum_{j \neq i} \left( \frac{\bz_j}{\bz_i} \right)^{\!\bar{y}(r,u)} 
  \frac{f(\bz_i ,\bz_j ,\bar{y}(r,u))}{\bz_i -\bz_j} \tilde{\st}_\nrmu^{(j)} \hat{A}_\s (\st,\bz,z) \right. \nn \\
&\bigspc \bigspc \quad \quad \left. -\sum_j \left( \frac{z_j}{\bz_i} \right)^{\!\bar{y}(r,u)} \frac{f(\bz_i ,z_j ;\bar{y}(r,u))}{\bz_i -z_j} 
   \hat{A}_\s (\st,\bz,z) \st_\nrmu^{(j)} \right. \nn \\
&\bigspc \bigspc \quad \quad \left. -\frac{1}{\bz_i} (\bar{y}(r,u) -\theta(\bar{y}(r,u)\geq 1)) \tilde{\st}_\nrmu^{(i)} \hat{A}_\s(\st,\bz,z) \right] \label{Eq 4.10b}
\end{align}
\begin{gather}
\st_\nrmu^{(i)} = \st_\nrm (T^{(i)}\!,\s) \tau_u^{(i)} ,\quad \tilde{\st}_\nrmu^{(i)} = \st_\nrm (T^{(i)}\!,\s) (-1)^u \tau_u^{(i)} \\
f(z,w;\bar{y}) = 1+ \frac{z-w}{w} \theta (\bar{y} \geq 1) ,\quad \bar{y}(r,u) = \srac{\bar{n}(r)}{\r(\s)} +\srac{\bar{u}}{2} \\
\pl_i \equiv \frac{\pl}{\pl z_i} ,\quad \bpl_i \equiv \frac{\pl}{\pl \bz_i} ,\quad i=1,\ldots ,n \,.
\end{gather}
\end{subequations}
Here the ordinary twisted inverse inertia tensor ${\cL}_{\sgb(\s)}(\s)$ and the ordinary twisted representation matrices $\st(T,\s)$ are defined in 
Eq.~\eqref{Eq2.20}. Note that the $\tilde{\st}\hat{A} \st$ terms in Eq.~\eqref{Eq4.10} are summed over all $j$, where the $j=i$ terms come from 
the $\tilde{\st}\hg\st$ terms of the twisted vertex operator equations \eqref{Eq4.5}. 

We emphasize that similar KZ systems with terms of the form
\begin{gather}
\frac{T_a^{(j)} A T_b^{(i)}}{z_i -\bz_j } ,\quad \frac{T_a^{(i)} A T_b^{(j)}}{\bz_i -z_j} \nn
\end{gather}
are known in untwisted open-string WZW theory \cite{Giusto} and, following this reference, we may interpret the corresponding $\tilde{\st} \hat{A} \st$ 
terms in the twisted KZ equations \eqref{Eq4.10} as the interaction of charges at $z_i$ with {\it image charges} at $\bz_j$.

In addition to the twisted KZ equations, we find the global Ward identities of open-string sector $\hat{h}_\s$
\begin{subequations}
\label{Eq4.11}
\begin{gather}
\langle [\hj_{0,\m,0} (0),\hg(\st^{(1)},\bz_1 ,z_1,\s) \ldots \hg(\st^{(n)},\bz_n ,z_n,\s) ]\rangle_\s =0 \\
\Rightarrow \hat{A}_\s (\st,\bz,z) \left( \sum_{i=1}^n \st_{0,\m,0}^{(i)} \right) -\left( \sum_{i=1}^n \tilde{\st}_{0,\m,0}^{(i)} \right) 
    \hat{A}_\s (\st,\bz,z) =0 ,\quad \forall \,\m \quad \nn \\
\quad \Rightarrow \hat{A}_\s (\st,\bz,z) \left( \sum_{i=1}^n \st_{0,\m}^{(i)} \right) -\left( \sum_{i=1}^n \st_{0,\m}^{(i)} \right) 
    \hat{A}_\s (\st,\bz,z) =0 ,\quad \forall \,\m 
\end{gather}
\end{subequations}
which enforce the residual symmetries associated to the zero modes $\hj_{0,\m,0}(0)$ of the twisted currents. We remind the reader that 
\begin{gather}
\st_{0,\m}^{(i)} =\st_{0,\m}(T^{(i)}\!,\s)  \label{Eq4.12}
\end{gather}
are the special cases with $\bar{n}(r)=0$ of the twisted representation matrices \eqref{Eq 2.20d} of ordinary space-time orbifold theory. Recalling 
the discussion in Subsec.~$(3.4)$, we also note the additional global Ward identities
\begin{subequations}
\label{Eq4.13}
\begin{gather}
\langle [\hj_{\r(\s)/2,\m,1} (0),\hg(\st^{(1)},\bz_1 ,z_1,\s) \ldots \hg(\st^{(n)},\bz_n ,z_n,\s) ]\rangle_\s =0 \\
\Rightarrow \hat{A}_\s (\st,\bz,z) \left( \sum_{i=1}^n \st_{\r(\s)/2,\m,1}^{(i)} \right) -\left( \sum_{i=1}^n \tilde{\st}_{\r(\s)/2,\m,1}^{(i)} 
   \right) \hat{A}_\s (\st,\bz,z) =0 ,\quad \forall \,\m \quad \nn \\
\quad \Rightarrow \hat{A}_\s (\st,\bz,z) \left( \sum_{i=1}^n \st_{\r(\s)/2,\m}^{(i)} \tau_1^{(i)} \right) +\left( \sum_{i=1}^n 
  {\st}_{\r(\s)/2,\m}^{(i)} \tau_1^{(i)} \right) \hat{A}_\s (\st,\bz,z) =0 ,\quad \forall \,\m 
\end{gather}
\end{subequations}
which hold for any open-string sector $\hat{h}_\s$ when the order $\r(\s)$ of $h_\s \in H$ is even.

We turn finally to discuss the consequences for the system of the constraints \eqref{Eq3.8} found above, beginning with the $\tau_1$-constraint:
\begin{gather}
[\tau_1 ,\hg(\st(T,\s),\bz,z,\s) ]=0 \,. \label{Eq4.14}
\end{gather}
The solution to \eqref{Eq4.14} is
\begin{gather}
\hg(\st(T,\s),\bz,z,\s)_{\Nrm u}{}^{\Nsn v} = \hg_{(u-v)} (\st(T,\s),\bz,z,\s)_\Nrm{}^\Nsn \label{Eq4.15}
\end{gather}
which tells us that the twisted correlators $\hat{A}_\s$ in Eq.~\eqref{Eq4.9} are functions of the differences $\{ (u_i -v_i) ,\,i=1\ldots n \}$ only. 
As discussed in Ref.~\cite{Perm}, this constraint is consistent with the twisted KZ equations and the global Ward identities.

The remaining constraint is the world-sheet parity in Eq.~\eqref{Eq 3.8b}
\begin{gather}
\hg(\st(\bT,\s),z,\bz,\s)^t =\tau_3 \hg(\st(T,\s),\bz,z,\s) \tau_3 \label{Eq4.16} 
\end{gather}
which gives correlator relations of the form
\begin{gather}
\hat{A}_\s (\st(T),\bz,z) =\tau_3^{(i)} \hat{A}_\s (\st(T'),\bw,w)^{t_{(i)}} \tau_3^{(i)} \BIG{|}_{\stackrel{T^{(i)}{}' =\bT^{(i)} ,\,\, w_i =\bz_i \quad \quad}
  {\tyny{T^{(j)}{}' =T^{(j)} ,\,\,w_j =z_j ,\,\, j\neq i }}} ,\quad i=1\ldots n \label{Eq4.17}
\end{gather}
where superscript $t_{(i)}$ is transpose in the $i$th subspace. These relations are the reflection in the correlators of the world-sheet identification
discussed in Subsec.~$3.2$. The relations \eqref{Eq4.17} are also consistent with the twisted KZ equations in the following sense: As discussed above for
the twisted vertex operator equations \eqref{Eq3.10}, one finds correspondingly (see App.~C) that the $\pl_i$ and $\bpl_i$ twisted KZ equations are 
also {\it redundant under world-sheet parity}. More precisely, the $\bpl_i$ twisted KZ equations can be ignored in favor of Eq.~\eqref{Eq4.17} and the 
$\pl_i$ twisted KZ equations and vice-versa:
\begin{subequations}
\label{Eq4.18}
\begin{gather}
\text{world-sheet parity } +\pl_i \hat{A}_\s \text{ KZ } \Longrightarrow \bpl_i \hat{A}_\s \text{ KZ} \\
\,\text{world-sheet parity } +\bpl_i \hat{A}_\s \text{ KZ } \Longrightarrow \pl_i \hat{A}_\s \text{ KZ.}
\end{gather}
\end{subequations}
On the other hand, the set of {\it both} $\pl_i$ and $\bpl_i$ twisted KZ equations is {\it not} sufficient to imply world-sheet parity, which must 
therefore be included separately in the system.

\subsection{The One-Sided Form of the Open-String Twisted KZ System}

Following Ref.~\cite{Giusto}, we may define a {\it one-sided} notation for open-string sector $\hat{h}_\s$ as follows:
\vspace{-0.15in}
\begin{subequations}
\label{Eq4.19}
\begin{gather}
\tilde{\hg}(\st,\bz,z,\s)^{\Nsn v ;\Nrm u} \equiv \hg(\st,\bz,z,\s)_{\Nrm u}{}^{\Nsn v} \label{Eq 4.19a} \\
(A \hg(\st,\bz,z,\s) B )_{\Nrm u}{}^{\Nsn v} = A_{\Nrm u}{}^{\Ntd w} \hg_{\Ntd w}{}^{N(t')\de 'w'} B_{N(t')\de 'w'}{}^{\Nsn v} \bigspc \bigspc \nn \\
\bigspc =\!-\tilde{\hg}^{N(t')\de 'w';\Ntd w} B_{N(t')\de 'w'}{}^{\!\!\!\Nsn v} (\bar{A})_{\Ntd w}{}^{\!\!\!\Nrm u} \label{Eq 4.19b} \\
=\!-(\tilde{\hg}(\st,\bz,z,\s) B \otimes \bar{A} )^{\Nsn v;\Nrm u} \quad \, \label{Eq 4.19c} \\
\bar{A}=-A^t ,\quad (A^t )_{\Nrm u}{}^{\Nsn v} \equiv A_{\Nsn v}{}^{\Nrm u} \,.
\end{gather}
\end{subequations}
In our application, we will then need to evaluate the following twisted representation matrices $\bar{\st}(\bT) ,\,\forall \bT$
\begin{subequations}
\label{Eq4.20}
\begin{gather}
\bar{\st}_\nrmu (\bT,\s) \equiv -\tilde{\st}_\nrmu (T,\s)^t = -\st_\nrm (T,\s)^t (-1)^u \tau_u^t = \st_\nrm (\bT,\s) (-1)^u \tau_u \label{Eq 4.20a} \\
[\bar{\st}_\nrmu (\bT,\s) ,\bar{\st}_\nsnv (\bT,\s)] =i\scf_{\nrm;\nsn}{}^{n(r)+n(s),\de}(\s) \bar{\st}_{n(r)+n(s),\de,u+v}(\bT,\s) \label{Eq 4.20b}
\end{gather}
\end{subequations}
where the matrix $-\bar{\st}(\bT,\s)$ is the image on the right of our total twisted representation matrix $\tilde{\st}(T,\s)$ on the left. The 
result in \eqref{Eq 4.20a} follows from the identity in Eq.~\eqref{Eq2.21}, and, according to Eq.~\eqref{Eq 4.20b}, the matrices $\bar{\st}(\bT,\s)$ 
satisfy the same total orbifold Lie algebra $\hg_O(\s)$ in Eq.~\eqref{Eq 2.26c}. We note in particular that (as shown in \eqref{Eq 4.19b}, \eqref{Eq 4.19c}) 
$\bar{\st}(\bT,\s)$ acts on the right indices of $\tilde{\hg}$, while $\st(T,\s)$ acts on the left indices. This $B\!\otimes \!\bar{A}$ bookkeeping 
should be born in mind even though we sometimes neglect the ordering in the tensor product
\vspace{-0.12in}
\begin{gather}
\bar{\st} (\bT,\s) \!\otimes \!\st (T,\s) \simeq \st (T,\s) \!\otimes \!\bar{\st} (\bT,\s)  \label{Eq4.21}
\end{gather}
for notational convenience. 

The one-sided notation allows us to rewrite all our results for the twisted affine primary fields in a one-sided form. As examples, the $\hj \hg$ OPE 
\eqref{Eq 3.6c} and the mode-ordered form \eqref{Eq4.5} of the twisted vertex operator equations now take the one-sided form
\begin{subequations}
\label{Eq4.22}
\begin{gather}
\hj_\nrmu (z,\s) \tilde{\hg}(\st(T),\bw,w,\s) = \tilde{\hg}(\st(T),\bw,w,\s) \left( \frac{\st_\nrmu (T)}{z-w} +\frac{\bar{\st}_\nrmu (\bT)}{z-\bw} \right) \bigspc \nn \\
  \bigspc \bigspc +\Ord ((z-w)^0 ,(z-\bw)^0 ) \\
[\hj_\nrmu (m\!+\!\nrrs \!+\!\srac{u}{2}) ,\tilde{\hg}(\st(T),\bz,z,\s)]= \tilde{\hg}(\st(T),\bz,z,\s) \Big{(} \st_\nrmu (T) z^{m+\nrrsf +\frac{u}{2}} \bigspc \nn \\
   \bigspc \bigspc \bigspc \bigspc \bigspc +\bar{\st}_\nrmu (\bT) \bz^{m+\nrrsf +\frac{u}{2}} \Big{)}
\end{gather}\vspace{-0.12in}
\begin{align}
&\pl \tilde{\hg}(\st(T),\bz,z,\s) ={\cL}_{\sgb(\s)}^{\nrm;\mnrn} (\s) \sum_{u=0}^1 \Big{\{} :\!\hj_\nrmu (z,\s) \tilde{\hg}(\st(T),\bz,z,\s) \!:_M  \nn \\
& \bigspc +\left( \frac{\bz}{z} \right)^{\!\bar{y}(r,u)} \frac{f(z,\bz;\bar{y}(r,u))}{z-\bz} \tilde{\hg}(\st(T),\bz,z,\s) \bar{\st}_\nrmu (\bT) \nn \\
&\bigspc -\frac{1}{z} \left( \bar{y}(r,u)-\theta (\bar{y}(r,u) \geq 1) \right) \tilde{\hg}(\st(T),\bz,z,\s) \st_\nrmu(T) \Big{\}} \st_{\mnrn ,-u}(T)
\end{align}\vspace{-0.12in}
\begin{align}
&\bpl \tilde{\hg}(\st(T),\bz,z,\s) ={\cL}_{\sgb(\s)}^{\nrm;\mnrn}(\s) \sum_{u=0}^1 \Big{\{} :\!\hj_\nrmu (\bz,\s) \tilde{\hg}(\st(T),\bz,z,\s)\!:_M \nn \\
&\bigspc +\left( \frac{z}{\bz} \right)^{\!\bar{y}(r,u)} \frac{f(\bz,z;\bar{y}(r,u))}{\bz -z} \tilde{\hg}(\st(T),\bz,z,\s) \st_\nrmu (T) \nn \\
& \bigspc -\frac{1}{\bz} \left( \bar{y}(r,u)-\theta (\bar{y}(r,u) \geq 1) \right) \tilde{\hg}(\st(T),\bz,z,\s) \bar{\st}_\nrmu (\bT) \Big{\}} \bar{\st}_{\mnrn ,-u}(\bT)  
\end{align}
\end{subequations}
where we have suppressed the $\s$-dependence of $\st, \bar{\st}$, and $\bar{\st}(\bT)$ is defined in Eq.~\eqref{Eq4.20}.

We turn next to the twisted KZ system itself. In the one-sided notation, the correlators of the $\tilde{\hg}$ fields are written as:
\begin{subequations}
\label{Eq4.23}
\begin{gather}
\tilde{\hat{A}}_\s (\st,\bz,z) \equiv \langle \tilde{\hg}(\st^{(1)},\bz_1,z_1,\s) \ldots \tilde{\hg}(\st^{(n)},\bz_n ,z_n,\s) \rangle_\s \\
\hat{A}_\s (\st)_{N(r_1)\m_1 u_1 ,\ldots ,N(r_n)\m_n u_n}^{\quad N(s_1)\n_1 v_1 ,\ldots, N(s_n),\n_n ,v_n} =
   \tilde{\hat{A}}_\s (\st)^{N(s_1)\n_1 v_1 ,\ldots N(s_n)\n_n v_n; N(r_1)\m_1 u_1 ,\ldots ,N(r_n)\m_n u_n} \,.
\end{gather}
\end{subequations}
Then we find that the twisted KZ system \eqref{Eq4.9}-\eqref{Eq4.13} takes the {\it one-sided} or {\it chiral form} in $2n$ variables $z_i ,\bz_i ,i=1\ldots n$:
\begin{subequations}
\label{Eq4.24}
\begin{align}
&\!\!\pl_i \tilde{\hat{A}}_\s (\st,\bz,z) \!=\! \tilde{\hat{A}}_\s (\st,\bz,z) \hat{W}_i (\st,\bz,z,\s) ,\,\,\,
   \bpl_i \tilde{\hat{A}}_\s (\st,\bz,z) \!=\! \tilde{\hat{A}}_\s (\st,\bz,z) \hat{\bar{W}}_{\!i} (\st,\bz,z,\s) 
\end{align}
\begin{align}
\hat{W}_i (\st,\bz,z,\s) \equiv & {\cL}_{\sgb(\s)}^{\nrm;\mnrn}(\s) \sum_{u=0}^1 \BIG{[} \BIG{(} \sum_{j\neq i} \left( \frac{z_j}{z_i} 
  \right)^{\!\!\bar{y}(r,u)} \!\frac{f (z_i,z_j ;\bar{y}(r,u))}{z_i -z_j} \st_\nrmu^{(j)} \bigspc \nn \\
&\quad \quad \quad +\sum_j \left( \frac{\bz_j}{z_i} \right)^{\!\!\bar{y}(r,u)} \!\frac{f(z_i,\bz_j ;\bar{y}(r,u))}{z_i -\bz_j} \bar{\st}_\nrmu^{(j)} \BIG{)}
\otimes \st_{\mnrn ,-u}^{(i)} \nn \\
&\quad \quad \quad -\frac{1}{z_i} \left( \bar{y}(r,u) -\theta(\bar{y}(r,u) \geq 1) \right) \st_\nrmu^{(i)} \st_{\mnrn ,-u}^{(i)} \BIG{]} \label{Eq 4.24b} \\
\hat{\bar{W}}_{\!i} (\st,\bz,z,\s) \equiv & {\cL}_{\sgb(\s)}^{\nrm;\mnrn}(\s) \sum_{u=0}^1 \BIG{[} \BIG{(} \sum_{j\neq i} \left( \frac{\bz_j}{\bz_i} 
  \right)^{\!\!\bar{y}(r,u)} \!\frac{f (\bz_i,\bz_j ;\bar{y}(r,u))}{\bz_i -\bz_j} \bar{\st}_\nrmu^{(j)} \nn \\
&\quad \quad \quad +\sum_{j} \left( \frac{z_j}{\bz_i} \right)^{\!\!\bar{y}(r,u)} \!\frac{f (\bz_i,z_j ;\bar{y}(r,u))}{\bz_i -z_j} \st_\nrmu^{(j)} \BIG{)} 
   \otimes \bar{\st}_{\mnrn ,-u}^{(i)} \nn \\
&\quad \quad \quad -\frac{1}{\bz_i} \left( \bar{y}(r,u) -\theta(\bar{y}(r,u) \geq 1) \right) \bar{\st}_\nrmu^{(i)} \bar{\st}_{\mnrn ,-u}^{(i)} \BIG{]} \label{Eq 4.24c}
\end{align}
\begin{gather}
\tilde{\hat{A}}_\s (\st,\bz,z) \left( \sum_{i=1}^n (\st_{0,\m,0}^{(i)} \!\otimes \!\one +\one \!\otimes \!\bar{\st}_{0,\m,0}^{(i)}) \right) =0 ,\quad \forall \m   \label{Eq 4.24d} \\
\tilde{\hat{A}}_\s (\st,\bz,z) \left( \sum_{i=1}^n (\st_{\r(\s)/2,\m,1}^{(i)} \!\otimes \!\one +\one \!\otimes \!\bar{\st}_{\r(\s)/2,\m,1}^{(i)}) \right) =0 ,\quad \forall \m ,\quad 
   \text{for } \r(\s) \text{ even} \label{Eq 4.24e} \\
\st^{(i)} \equiv \st (T^{(i)}\!,\s) ,\quad \bar{\st}^{(i)} \equiv \bar{\st}(\bT^{(i)}\!,\s) ,\quad \pl_i \equiv \frac{\pl}{\pl z_i} ,\quad \bpl_i
  \equiv \frac{\pl}{\pl \bz_i} ,\quad i=1,\ldots ,n \\
\bar{y}(r,u) =\srac{\bar{n}(r)}{\r(\s)} +\srac{\bar{u}}{2} ,\quad f (z,w,\bar{y}) = 1 +\frac{z-w}{w} \theta (\bar{y} \geq 1) 
\end{gather}
\end{subequations}
In this form, we see that the global Ward identities ($4.24$d,e) select {\it singlets} under the residual symmetry algebra 
\eqref{Eq 3.28} of open-string sector $\hat{h}_\s$.

For reference, we collect here the forms of the total twisted representation matrices
\begin{subequations}
\label{Eq4.25}
\begin{gather}
\st_\nrmu (T,\s) \!=\!\st_\nrm (T,\s) \tau_u ,\quad \bar{\st}_\nrmu (\bT,\s) \!=\!\st_\nrm (\bT,\s) (-1)^u \tau_u ,\quad u\!=\!0,1 \\
\st_\nrm (T,\s) \!=\! \schisig_\nrm U(\s)_\nrm{}^a U(T,\s) T_a U\hc (T,\s) \label{Eq 4.25b} \\
\st_\nrm (\bT,\s) \!=\! \schisig_\nrm U(\s)_\nrm{}^a U(\bT,\s) \bT_a U\hc (\bT,\s) =-\st_\nrm (T,\s)^t \label{Eq 4.25c} \\
[\st_\nrmu (T,\s) ,\st_\nsnv (T,\s)] =i\scf_{\nrm;\nsn}{}^{n(r)+n(s),\de}(\s) \st_{n(r)+n(s),\de,u+v}(T,\s) \\
[\bar{\st}_\nrmu (\bT,\s) ,\bar{\st}_\nsnv (\bT,\s)] =i\scf_{\nrm;\nsn}{}^{n(r)+n(s),\de}(\s) \bar{\st}_{n(r)+n(s),\de,u+v}(\bT,\s) 
\end{gather}
\end{subequations}
which appear in the one-sided form \eqref{Eq4.24} of the twisted open-string KZ equations. Here $\vec{\tau}$ are the Pauli matrices, $T$ is any matrix rep 
of $g$ and we have also included the definitions of the ordinary twisted representation matrices $\st_\nrm (T), \st_\nrm (\bT)$. We remark in particular 
that both sets $\st_\nrmu (T)$, $\bar{\st}_\nrmu (\bT)$ satisfy the same total orbifold Lie algebra $\hg_O (\s)$ in \eqref{Eq 2.26c}. Moreover 
$\st_{0,\m,0}(T) =\st_{0,\m}(T) ,\,\bar{\st}_{0,\m,0}(\bT)= \st_{0,\m}(\bT)$ so that the global Ward identity \eqref{Eq 4.24d} has exactly the same form as 
the chiral global Ward identities \cite{Big} of ordinary space-time orbifolds $A_g (H)/H$.

In the one-sided notation, the constraints \eqref{Eq3.8} take the form
\begin{subequations}
\label{Eq4.26}
\begin{gather}
\tilde{\hg} (\st(\bT,\s),z,\bz,\s)^t = \tilde{\hg} (\st(T,\s),\bz,z,\s) (\tau_3 \otimes \tau_3) \bigspc \\
\Rightarrow \tilde{\hg} (\st(\bT,\s),z,\bz,\s)^{\Nrm u;\Nsn v} =(-1)^{u+v} \tilde{\hg} (\st(T,\s),\bz,z,\s)^{\Nsn v;\Nrm u}  
\end{gather}
\end{subequations} \vspace{-0.2in}
\begin{subequations}
\label{Eq4.27}
\begin{gather}
[\tau_1 ,\hg(\st,\bz,z,\s)]\!=\!0 \rightarrow \tilde{\hg}(\st,\bz,z,\s) (\one \otimes \tau_1 ) =\tilde{\hg}(\st,\bz,z,\s) (\tau_1 \otimes \one ) \quad \quad \\
\quad \quad \Rightarrow \tilde{\hg}(\st,\bz,z,\s) (\tau_1 \otimes \tau_1 )= \tilde{\hg}(\st,\bz,z,\s) (\one \otimes (\tau_1)^2 )=\tilde{\hg}(\st,\bz,z,\s) \label{Eq 4.27b} \\
    \Rightarrow \tilde{\hg}(\st,\bz,z,\s)^{\Nsn v; \Nrm u} =\tilde{\hg}_{(u+v)}(\st,\bz,z,\s)^{\Nsn;\Nrm} \label{Eq 4.27c}
\end{gather}
\end{subequations}
where \eqref{Eq4.26} is the behavior of the (one-sided) twisted affine primary field under world-sheet parity and \eqref{Eq 4.27c} is the solution of 
the $\tau_1$ constraint. As discussed for the two-sided formulation in App.~C, the one-sided $\pl_i$ and $\bpl_i$ twisted KZ equations are redundant 
under world-sheet parity \eqref{Eq4.26}.

We emphasize that one-sided open-string KZ systems of this general type \cite{Giusto}
\begin{subequations}
\label{Eq4.28}
\begin{gather}
\pl_i \tilde{F} =\tilde{F} 2L_g^{ab} \Big{(} \sum_{j\neq i} \frac{T_a^{(i)} \otimes T_b^{(j)}}{z_i -z_j} +\sum_j \frac{T_a^{(i)} \otimes 
   \bT_b^{(j)}}{z_i -\bz_j} \Big{)} 
\end{gather}
\begin{gather}
\bpl_i \tilde{F} =\tilde{F} 2L_g^{ab} \Big{(} \sum_{j\neq i} \frac{\bT_a^{(i)} \otimes \bT_b^{(j)}}{\bz_i -\bz_j} +\sum_j \frac{\bT_a^{(i)} 
   \otimes T_b^{(j)}}{\bz_i -z_j} \Big{)} \\
\tilde{F} \sum_{i=1}^n (T_a^{(i)} \!\otimes \!\one +\one \!\otimes \!\bT_a^{(i)}) =0
\end{gather}
\end{subequations}
are known in the theory of untwisted open WZW strings. Here and in the twisted KZ equations \eqref{Eq4.24}, one clearly sees the interaction of charges at 
$z_i$ with image charges at $\bz_j$.

On the other hand, there is something quite special about our set of open-string KZ equations: The relation \eqref{Eq 4.27b} tells us that the 
interactions between the charge at $z_i$ and its image charge at $\bz_i$ are {\it non-singular} in the twisted KZ system, in distinction to the singular 
interactions in the untwisted system \eqref{Eq4.28}. To see this, consider for example the $j=i$ term of the one-sided connection \eqref{Eq 4.24b} 
and follow the steps:
\begin{subequations}
\label{Eq4.29}
\begin{gather}
\tilde{\hat{A}}_\s \tau_1^{(i)} \otimes \tau_1^{(i)} =\tilde{\hat{A}}_\s \label{Eq 4.29a} \\
\tilde{\hat{A}}_\s \sum_{u=0}^1 \left( \frac{\bz_i}{z_i} \right)^{\!\!\bar{y}(r,u)} \!\frac{f (z_i,\bz_i ;\bar{y}(r,u))}{z_i -\bz_i} 
  \bar{\st}_\nrmu^{(i)} \otimes \st_{\mnrn ,-u}^{(i)} \bigspc \bigspc \bigspc \nn \\
\quad \quad =\tilde{\hat{A}}_\s \sum_{u=0}^1 \left( \frac{\bz_i}{z_i} \right)^{\!\!\bar{y}(r,u)} \!\frac{f (z_i,\bz_i ;\bar{y}(r,u))}{z_i -\bz_i} 
   (-1)^u \st_\nrm (\bT^{(i)}) \otimes \st_{\mnrn}(T^{(i)}) \\
\sum_{u=0}^1 (-1)^u \left( \frac{\bz_i}{z_i} \right)^{\!\!\bar{y}(r,u)} \!\frac{f (z_i,\bz_i ; \bar{y}(r,u))}{z_i -\bz_i} =\left( \frac{\bz_i}{z_i} 
   \right)^{\frac{\bar{n}(r)}{\r(\s)}} 
\left\{ \begin{array}{ll} \frac{1 -(\bz_i/z_i )^{\frac{1}{2}}}{z_i-\bz_i} & \text{when } \srac{\bar{n}(r)}{\r(\s)} < \srac{1}{2} \\ 
    \frac{1 -(z_i/\bz_i )^{\frac{1}{2}}}{z_i-\bz_i} & \text{when } \srac{\bar{n}(r)}{\r(\s)} \geq \srac{1}{2} \end{array} \right. \label{Eq 4.29c} \\
\bigspc \bigspc =\left\{ \begin{array}{ll} \frac{1}{2z_i} +\Ord (\bz_i -z_i ) &\text{when } \srac{\bar{n}(r)}{\r(\s)} < \srac{1}{2} \\
   -\frac{1}{2z_i} +\Ord (\bz_i -z_i) & \text{when } \srac{\bar{n}(r)}{\r(\s)} \geq \srac{1}{2} \end{array} \right. \,. 
\end{gather}
\end{subequations}
The sum in \eqref{Eq 4.29c} is explicitly non-singular as $\bz_i \rightarrow z_i$, and similarly for the $j=i$ term in the one-sided connection 
\eqref{Eq 4.24c}. This is in fact the same conclusion reached in Subsec.~$4.2$, since the $j=i$ terms of the twisted KZ equations arise directly from the 
non-singular $f$ terms in the twisted vertex operator equations.

We finally mention that the one-sided form of the twisted KZ system above is simpler for computational purposes (see Subsec.~$5.3$) than the two-sided 
form of the system in Subsec.~$4.3$.

\subsection{The Orbifold Form of the Open-String Twisted KZ System}

Orientation orbifolds are special cases of orbifolds, and Appendix \ref{OrbNotApp} rewrites a number of the major results of our orientation-orbifold 
theory in the standard notation of the orbifold program [3,5,13-16]. The basic device is the {\it super-index notation} $\tnrm$:
\begin{gather}
\tnrm \equiv \nrmu ,\quad \srac{\hat{n}(r)}{\tr} \equiv \srac{n(r)}{\r(\s)}+\srac{u}{2} ,\quad \srac{\overline{\hat{n}(r)}}{\tr} \equiv \srac{\hat{n}(r)}{\tr} 
  -\Big{\lfloor} \srac{\hat{n}(r)}{\tr} \Big{\rfloor} \,.  \label{Eq4.30}
\end{gather}
For the results below, we will need only the following super-index identities
\begin{subequations}
\label{Eq4.31}
\begin{gather}
\srac{\overline{\hat{n}(r)}}{\tr} = \bar{y}(r,u) -\theta(\bar{y}(r,u) \geq 1) ,\quad \bar{y}(r,u) =\srac{\bar{n}(r)}{\r(\s)} +\srac{\bar{u}}{2} \\
\left( \frac{w}{z} \right)^{\frac{\overline{\hat{n}(r)}}{\tr}} =\left( \frac{w}{z} \right)^{\bar{y}(r,u)} f(z,w;\bar{y}(r,u)) 
\end{gather}
\end{subequations}
which are derived in App.~\ref{OrbNotApp}.

With these identities, we find the {\it orbifold form} of the twisted open-string KZ equations:
\vspace{-0.10in}
\begin{subequations}
\label{Eq4.32}
\begin{align}
&\!\!\pl_i \tilde{\hat{A}}_\s (\st,\bz,z) \!=\!\tilde{\hat{A}}_\s (\st,\bz,z) \hat{W}_i(\st,\bz,z,\s) ,\,\,\, \bpl_i \tilde{\hat{A}}_\s (\st,\bz,z) 
   \!=\!\tilde{\hat{A}}_\s (\st,\bz,z) \hat{\bar{W}}_{\!i} (\st,\bz,z,\s) 
\end{align}\vspace{-0.10in}
\begin{align}
\hat{W}_i (\st,\bz,z,\s) =& 2{\cL}_{\sgb(\s)}^{\tnrm;-\hat{n}(r),\hat{\n}}(\s) \BIG{[} \BIG{(} \!\sum_{j\neq i} \left( \frac{z_j}{z_i} 
  \right)^{\!\frac{\overline{\hat{n}(r)}}{\tr}} \!\!\frac{1}{z_i -z_j} \st_\tnrm^{(j)} \bigspc \bigspc \nn \\
& +\!\sum_j \left( \frac{\bz_j}{z_i} \right)^{\!\frac{\overline{\hat{n}(r)}}{\tr}} \!\!\frac{1}{z_i -\bz_j} \bar{\st}_\tnrm^{(j)} \BIG{)} \!\otimes 
  \!\st_{-\hat{n}(r),\hat{\n}}^{(i)} -\frac{1}{z_i} \srac{\overline{\hat{n}(r)}}{\tr} \st_\tnrm^{(i)} \st_{-\hat{n}(r),\hat{\n}}^{(i)} \BIG{]} \\
\hat{\bar{W}}_{\!i} (\st,\bz,z,\s) =& 2{\cL}_{\sgb(\s)}^{\tnrm;-\hat{n}(r),\hat{\n}}(\s) \BIG{[} \BIG{(} \!\sum_{j\neq i} \left( \frac{\bz_j}{\bz_i} 
  \right)^{\!\frac{\overline{\hat{n}(r)}}{\tr}} \!\!\frac{1}{\bz_i-\bz_j} \bar{\st}_\tnrm^{(j)} \bigspc \bigspc \nn \\
&+\! \sum_j \left( \frac{z_j}{\bz_i} \right)^{\!\frac{\overline{\hat{n}(r)}}{\tr}} \!\!\frac{1}{\bz_i -z_j} \st_\tnrm^{(j)} \BIG{)} \!\otimes 
   \!\bar{\st}_{-\hat{n}(r),\hat{\n}}^{(i)} -\frac{1}{\bz_i} \srac{\overline{\hat{n}(r)}}{\tr} \bar{\st}_\tnrm^{(i)} \bar{\st}_{-\hat{n}(r),\hat{\n}}^{(i)} 
   \BIG{]}
\end{align}
\begin{gather}
\tilde{\hat{A}}_\s (\st,\bz,z) \sum_{i=1}^n (\st_{0,\hat{\m}}^{(i)} \otimes \one +\one \otimes \bar{\st}_{0,\hat{\m}}^{(i)} )=0 ,\quad \forall \hat{\m} \,. \label{Eq 4.32d}
\end{gather}
\end{subequations} 
The twisted tensors in \eqref{Eq4.32} are nothing but the total quantities $\st ,\bar{\st}$ and ${\cL}$, now written in super-index notation 
(see Eq.~\eqref{B.3}). It is also explained in App.~B that the global Ward identity \eqref{Eq 4.32d} includes both sets of global Ward identities 
in Eq.~\eqref{Eq4.24}. This form of the twisted open-string KZ equations is recognized as an apparently ordinary {\it chiral orbifold KZ system} [13-16]
\vspace{-0.10in}
\begin{subequations}
\label{Eq4.33}
\begin{gather}
\hat{A}_\s (\st ,\{ z\}) \equiv \tilde{\hat{A}}_\s (\st,\bz,z) ,\quad \pl_\kappa \hat{A}_\s (\st,\{z\}) =\hat{A}_\s (\st,\{z\}) \hat{W}_\kappa (\st,\{z\},\s) 
\end{gather}
\begin{align}
\hat{W}_\kappa (\st,\{z\},\s) &= 2{\cL}_{\sgb(\s)}^{\tnrm;-\hat{n}(r),\hat{\n}}(\s) \BIG{[} \!\sum_{\r \neq \kappa} \left( \frac{z_\r}{z_\kappa} 
  \right)^{\!\frac{\overline{\hat{n}(r)}}{\tr}} \!\!\frac{1}{z_{\kappa \r}} \st_\tnrm^{(\r)} \otimes \st_{-\hat{n}(r),\hat{\n}}^{(\kappa)} \bigspc \nn \\
& \bigspc \bigspc \bigspc \quad \quad -\frac{1}{z_\kappa} \srac{\overline{\hat{n}(r)}}{\tr} \st_\tnrm^{(\kappa)} \st_{-\hat{n}(r),\hat{\n}}^{(\kappa)} \BIG{]} 
\end{align}\vspace{-0.25in}
\begin{gather}
\hat{A}_\s (\st,\{z\}) \sum_{\kappa =1}^{2n} \st_{0,\hat{\m}}^{(\kappa )} =0 ,\quad \forall \hat{\m} \\
\pl_\kappa \equiv \frac{\pl}{\pl z_\kappa} ,\quad z_{\kappa \r} \equiv z_\kappa -z_\r ,\quad z_\kappa \equiv \left\{ \begin{array}{ll} z_\kappa , 
   &\kappa =1\ldots n, \\ \bz_{\kappa -n} ,&\kappa =n+1 \ldots 2n \end{array} \right. \\
\st_\tnrm^{(\kappa)} \!\equiv \!\left\{ \begin{array}{ll} \st_\nrmu (T^{(\kappa)}\!,\s) =\st_\nrm (T^{(\kappa)}\!,\s) \tau_u^{(\kappa)} &\kappa =1\ldots n \\
   \bar{\st}_\nrmu (\bT^{(\kappa -n)}\!,\s) \!=\!\st_\nrm (\bT^{(\kappa -n)}\!,\s) (-1)^u \tau_u^{(\kappa -n)} & \kappa =n+1\ldots 2n  \end{array} \right. 
   \label{Eq 4.33e}
\end{gather}
\end{subequations}
on the $2n$ variables $\{ z_\kappa \}$, with the particular choice of twisted representation matrices in Eq.~\eqref{Eq 4.33e}. It remains to discuss whether 
the system \eqref{Eq4.33} is in fact realized in ordinary space-time orbifold theory.

As a first step in this discussion, we note that total twisted representation matrices of the form 
\vspace{-0.12in}
\begin{gather}
\st_{orb} = \{ \st_\nrm (T,\s) \tau_u ,\, \st_\nrm (\bT,\s) \tau_u \} \label{Eq4.34}
\end{gather}
occur in the ordinary space-time WZW orbifold 
\vspace{-0.07in}
\begin{gather}
A_{g\oplus g} (H)/H ,\quad H \!=\!\Zint_2 \times \tilde{H} ,\quad \tilde{H}\!\subset \!Aut(g)  \label{Eq4.35}
\end{gather}
where the $Z_2$ permutes\footnote{The characteristic matrices $t_{\hat{j}j}(\s),\, \hat{j}=0,\ldots, f_j(\s)-1$ of the permutation orbifolds
[13-15] reduce precisely to $\tau_0$ and $\tau_1$ for $\Zint_2$ cyclic orbifolds.} the two copies of $g$. The twisted representation matrices 
$\st_\nrm (T,\s)$, $\st_\nrm (\bT,\s)$ are the ordinary orbifold matrices in Eqs.~\eqref{Eq 4.25b}, \eqref{Eq 4.25c}, with $T$ any rep of $g$. Moreover, the 
extra orientation-orbifold factor $(-1)^u$ in the matrices $\bar{\st}(\bT,\s)$ (see \eqref{Eq 4.33e}) can be removed with the identity \eqref{t3-Id},
and one finds that the $n$-point correlators of the redefined twisted affine primary fields
\begin{gather}
\hg (\st(T,\s),\{z\},\s) \equiv \tilde{\hg} (\st(T,\s),\bz,z,\s) (\one \otimes \tau_3 )  \label{Eq4.36}
\end{gather}
satisfy exactly the same chiral orbifold KZ system as the corresponding $2n$-point space-time orbifold correlators of $A_{g\oplus g}(H)/H$ with the twisted 
representation matrices in \eqref{Eq4.34}. 

Put another way, our twisted KZ equations (and their global Ward identities) are ``doubled" but {\it ordinary chiral orbifold KZ systems} 
(although the open-string system comes with other constraints such as the world-sheet parity and the constraint \eqref{Eq 4.29a}). The Giusto-Halpern 
open-string KZ system \eqref{Eq4.28} is similarly an ordinary chiral KZ system on $2n$ variables, so one may expect that more general twisted 
open-string KZ equations will also be of the orbifold type.
 
In spite of the apparent simplicity of the orbifold form \eqref{Eq4.33} of the twisted KZ equations, it is not useful computationally because the
super-index notation masks important structure.

\section{More About the Case $h_\s^2 =1$}

\subsection{Overview}

In what follows, we further discuss the relatively simple open-string orientation-orbifold sectors with $h_\s^2=1$. In this case, many of our 
results above simplify, including the following:
\begin{subequations}
\begin{gather}
\ws^2 =\one ,\quad \srac{\bar{n}(r)}{\r(\s)} \in \{0,\srac{1}{2} \} ,\quad \bar{x}(r,u) \in \{0,\srac{1}{2} ,1\} \\
W(h_\s ;T)^2 =\one ,\quad \srac{\bar{N}(r)}{R(\s)} \in \{ 0,\srac{1}{2} \} ,\quad E^\ast (T,\s) =E(T,\s) \\
E(T,\s) \st_\nrm (T,\s) E(T,\s) =(-1)^{n(r)} \st_\nrm (T,\s) \label{ETE=-1T}
\end{gather}\vspace{-0.07in}
\begin{gather}
f(z,w; \bar{x}=0,\srac{1}{2}) =1 ,\quad f(z,w; \bar{x}=1) =\frac{z}{w} \\
h(z,w;\bar{x}=0) =1 ,\quad h(z,w;\bar{x}=\srac{1}{2}) =\frac{z-w}{2w} ,\quad h(z,w;\bar{x}=1) =\frac{z}{w} \\
\langle \hj_\nrmu (z,\s) \hj_\nsnv (w,\s) \rangle_\s =\left( \frac{w}{z} \right)^{\!\bar{x}(r,u)} \!\left( \frac{1}{(z-w)^2} +\frac{\bar{x}(r,u)}{z-w}
   \right)  \times \bigspc \bigspc \quad \quad \nn \\
\bigspc \bigspc \times 2 \de_{u+v,0\,\text{mod }2} \de_{n(r)+n(s),0\,\text{mod }\r(\s)} \sG_{\nrm;\mnrn}(\s) 
\end{gather}
\end{subequations}\vspace{-0.1in}
\begin{subequations}
\label{GSCfWt-Spec}
\begin{gather}
\gscfwt = \frac{c_g}{16}  ,\quad c_g = \frac{2k \text{dim }g}{2k +Q_\gfraks} \\
\text{dim}[\bar{n}(r)=0] +\text{dim}[\bar{n}(r)=1] =\text{dim} g \,.
\end{gather}
\end{subequations}
The twist-field conformal weights in Eq.~\eqref{GSCfWt-Spec} are given only for the special case \eqref{Eq3.37} when $g$ is permutation-invariant.

The open-string sectors with $h_\s^2=1$ can also be divided into the basic open-string sector with $h_\s=1$ and $\bar{n}(r)=0$ only (see Subsec.~$5.2$),
and the generic cases with $h_\s \neq 1$. For the generic cases the integral affine subalgebra takes the form \eqref{Eq3.27}:
\begin{subequations}
\begin{gather}
[ \hj_{0,\m,0} (m) ,\hj_{0,\n,0}(n)] =i\scf_{0,\m ;0,\n}{}^{0,\de}(\s) \hj_{0,\de,0}(m\!+\!n) +2m\de_{m+n,0} \sG_{0,\m;0,\n}(\s) \\
[ \hj_{0,\m,0} (m) ,\hj_{1,\n,1}(n \!+\! 1)] =i\scf_{0,\m;1,\n}{}^{1,\de}(\s) \hj_{1,\de,1} (m\!+\!n\!+\!1) \\
[ \hj_{1,\m,1} (m\!+\!1) ,\hj_{1,\n,1}(n \!+\! 1)] = i\scf_{1,\m;1,\n}{}^{\!\!0,\de}(\s) \hj_{0,\de,0}( m\!+\!n\!+\!2) \bigspc \bigspc \nn \\
   \bigspc \bigspc \quad \quad \quad+2(m\!+\!1) \de_{m+n+2,0} \sG_{1,\m;1,\n}(\s) \,.
\end{gather}
\end{subequations}
In this case, the global Ward identities ($4.24$d,e) may also be written as 
\begin{subequations}
\begin{gather}
\hat{A}_\s Q_\nrm (\st,\s)=0 ,\,\,\,\forall \m ,\, \bar{n}(r)=0,1 \\
Q_{0\m}(\st,\s) \equiv \sum_i \left( \st_{0\m}(T^{(i)}\!,\s) \!\otimes \!\one +\one \!\otimes \!\st_{0\m}(\bT^{(i)}\!,\s) \right) \\
Q_{1\m}(\st,\s) \equiv \sum_i \left( \st_{1\m}(T^{(i)}\!,\s) \tau_1^{(i)} \!\otimes \!\one -\one \!\otimes \!\st_{1\m}(\bT^{(i)}\!,\s) \tau_1^{(i)} \right) \label{5.4c} \\
[Q_{\nrm} (\st,\s) ,Q_{\nsn}(\st,\s) ]=i\scf_{\nrm ;\nsn}{}^{n(r)+n(s),\de} (\s) Q_{n(r)+n(s) ,\de}(\st,\s) \label{445d} 
\end{gather}
\end{subequations}
so that the twisted correlators are singlets under the residual symmetry algebra \eqref{445d}.
Moreover, each term in the twisted KZ equations contains only the four terms corresponding to $\bar{n}(r)=0,1$ and $\bar{u}=0,1$.

\subsection{Example: The Basic Open-String Sector $\hat{h}_\s =\tau_1 \times \one$}

We give here further details in the case of the single twisted open-string sector\footnote{This open-string sector can also appear in other 
orientation orbifolds with many sectors.} of the {\it basic} WZW orientation orbifold on Lie $g$. In this case, the orientation-reversing automorphism 
$\tau_1 \!\times \thickone$ is nothing but the action of world-sheet parity itself: 
\begin{subequations}
\begin{gather}
h_\s =1: \quad \ws = W(h_\s;T) =W(h_\s;\bT)= \thickone ,\quad \r(\s)=1 \\
J \,\leftrightarrow \, \bJ ,\quad g(T,\bz,z) \,\leftrightarrow \, g(\bT,z,\bz)^t \,.
\end{gather}
\end{subequations}
Then the general eigendata above reduces to the following
\begin{subequations}
\begin{gather}
U(\s) =U(T,\s) =\sU (\s) =\one ,\quad E(\s) =E(T,\s) =\one ,\quad \schisig =1 \\
\nrmu \rightarrow au: \quad \bar{n}(r)=0 ,\quad \m= a=1,\ldots ,\text{dim }g ,\,\,\,\, \bar{u}=0,1 \\
\Nrm u \rightarrow \a u: \quad \bar{N}(r)=0 ,\quad \m=\a =1,\ldots ,\text{dim }T 
\end{gather}
\begin{gather}
\sG_{\nrmu;\nsnv}(\s) \rightarrow G_{au;bv} =2\de_{u+v,0\,\text{mod }2} G_{ab} ,\quad G_{ab} =\oplus_I k_I \eta^I_{a(I)b(I)} \\
\scf_{\nrm;\nsn}{}^\ntd (\s) \rightarrow f_{ab}{}^c =\oplus_I f_{a(I)b(I)}^{c(I)} ,\quad {\cL}_{\sgb(\s)}^{\nrm;\nsn}(\s) \rightarrow L_g^{ab} =\oplus_I \frac{\eta_I^{a(I)b(I)}}{2k_I +Q_I} \label{cL-Basic} \\
\st_\nrmu(T) \rightarrow \st_{au}(T) =T_a \tau_u ,\quad \bar{\st}_\nrmu (\bT) \rightarrow \bar{\st}_{au}(\bT) =\bT_a (-1)^u \tau_u ,\quad \bT_a =-T_a^t\\
[T_a ,T_b] =if_{ab}{}^c T_c ,\quad \bar{y}(r,u) =\srac{\bar{u}}{2} ,\quad f(z,w;\bar{y}(r,u)) =1
\end{gather}
\end{subequations}
where $G_{ab}$ and $f_{ab}{}^c$ are the generalized Killing metric and structure constants of $g$, and $T$ is any matrix rep of $g$. In what follows we 
give results exclusively in the one-sided or chiral form of Subsec.~$4.4$, and we will drop most $\s$ labels since we are working in a single fixed open-string sector.

The explicit form of the twisted operator algebra in this sector is
\begin{subequations}
\label{Basic-Alg}
\begin{gather}
\hat{L}_u (m\!+\!\srac{u}{2}) =\srac{1}{2} L_g^{ab} \sum_{u=0}^1 \sum_{p \in \Zint} :\!\hj_{av} (p\!+\!\srac{v}{2}) \hj_{b,u-v}(m\!-\!p\!+\! \srac{u-v}{2})
   \!:_M +\de_{m+\frac{u}{2},0} \gscfwt \label{LMod-Basic} \\
]\hj_{au} (m\!+\!\srac{u}{2}) ,\hj_{bv} (n\!+\!\srac{v}{2})] =if_{ab}{}^c \hj_{c,u+v} (m\!+\!n\!+\! \srac{u+v}{2}) +2(m\!+\!\srac{u}{2}) 
   \de_{m+n+\frac{u+v}{2},0} G_{ab} \\
[\hat{L}_u (m\!+\!\srac{u}{2}) ,\hat{L}_v (n\!+\!\srac{v}{2}) ] =(m\!-\!n \!+\!\srac{u-v}{2}) \hat{L}_{u+v} 
   (m\!+\!n \!+\! \srac{u+v}{2}) \bigspc \quad \quad \nn \\
\bigspc \quad \quad +\de_{m+n+\srac{u+v}{2},0} \frac{2c_g}{12} (m\!+\!\srac{u}{2}) ((m\!+\!\srac{u}{2} )^2 -1) 
\end{gather}
\begin{gather}
[\hat{L}_u (m\!+\!\srac{u}{2}) ,\hj_{av} (n\!+\!\srac{v}{2})] =-(n\!+\!\srac{v}{2}) \hj_{a,u+v} (m\!+\!n\!+\!\srac{u+v}{2}) \\
[\hat{L}_u (m\!+\!\srac{u}{2}) ,\hg(\st(T,\s),\bz,z,\s)] \!=\!\hg(\st(T,\s),\bz,z,\s) \tau_u \Big{(} (\lpl \!z \!+\! (m\!+\!\srac{u}{2} \!+\!1) D_g(T)) z^{m+\frac{u}{2}} \quad \quad \nn \\
\bigspc \bigspc + (-1)^u (\overleftarrow{\bpl} \!\bz +(m\!+\!\srac{u}{2} \!+\!1) D_g(T)) \bz^{m+\frac{u}{2}} \Big{)} \label{Basic-Lg} \\
[\hj_{au}(m\!+\!\srac{u}{2}) ,\hg(\st(T),\bz,z)] =\hg(\st(T),\bz,z) \left( T_a \tau_u z^{m+\frac{u}{2}} + \bT_a (-1)^u \tau_u \bz^{m+\frac{u}{2}} \right) \label{Basic-Jg}
\end{gather}
\begin{gather}
D_g(T) =L_g^{ab} T_a T_b = L_g^{ab} \bT_a \bT_b ,\quad \gscfwt =\frac{c_g}{16} ,\quad c_g =\sum_I \frac{2k_I \text{dim }\gfrak^I}{2k_I +Q_I} \label{Dg-Basic}
\end{gather}
\end{subequations}
where $D_g(T)$ is the conformal weight matrix of rep $T$ under the affine-Sugawara construction on $g$, and $\gscfwt$ is the scalar twist-field 
conformal weight of the sector.

Moreover, the twisted vertex operator equations of the basic sector have the simple form:
\begin{subequations}
\label{TVOE-Basic}
\begin{align}
\pl \tilde{\hg}(\st(T),\bz,z) =&L_g^{ab} \sum_{u=0}^1 :\!\hj_{au} (z) \tilde{\hg}(\st(T),\bz,z) \!:_M T_b \tau_u \nn \\
&+\tilde{\hg}(\st(T),\bz,z) \left( \frac{1- (\bz/z )^{\frac{1}{2}}}{z-\bz} L_g^{ab} T_a \otimes \bT_b - \frac{D_g (T)}{2z} \right) 
\end{align}
\begin{align}
\bpl \tilde{\hg}(\st(T),\bz,z) =&L_g^{ab} \sum_{u=0}^1 :\!\hj_{au} (\bz) \tilde{\hg}(\st(T),\bz,z) \!:_M \bT_b (-1)^u \tau_u \nn \\
&+\tilde{\hg}(\st(T),\bz,z) \left( \frac{1- (z/\bz )^{\frac{1}{2}}}{\bz-z} L_g^{ab} T_a \otimes \bT_b - \frac{D_g (T)}{2\bz} \right) \,.
\end{align}
\end{subequations}
As a check on these equations, we have used the commutators \eqref{Basic-Lg} and \eqref{Basic-Jg} to compute $(\pl \!+\!\bpl) \hg$ from the mode form 
of $\hat{L}_\s (-1)$ in \eqref{LMod-Basic}
\begin{align}
&(\pl +\bpl )\tilde{\hg}(\st,\bz,z) =[\hat{L}_\s (-1) ,\tilde{\hg}(\st,\bz,z)] \nn \\
& \quad \quad \quad = L_g^{ab} \sum_{u=0}^1 :\!\left( \hj_{au} (z) \tilde{\hg}(\st,\bz,z) T_b +\hj_{au} (\bz) \tilde{\hg}(\st,\bz,z) \bT_b (-1)^u 
   \right) \!:_M \! \tau_u \nn \\
& \bigspc \quad +\tilde{\hg}(\st,\bz,z) \left\{ \frac{1}{z-\bz} \left( \left( \frac{z}{\bz} \right)^{\!\frac{1}{2}} \!-\!\left( \frac{\bz}{z} 
   \right)^{\!\frac{1}{2}} \right) L_g^{ab} T_a \otimes \bT_b -\frac{D_g(T)}{2} \left( \frac{1}{z} +\frac{1}{\bz} \right) \right\}
\end{align}
finding after some algebra that the result agrees with the sum of the twisted vertex operator equations in Eq.~\eqref{TVOE-Basic}.

Finally, the explicit form of the twisted KZ equations and the global Ward identity is:
\begin{subequations}
\begin{align}
&\!\!\pl_i \tilde{\hat{A}}_\s (\st,\bz,z) = \tilde{\hat{A}}_\s (\st,\bz,z) \hat{W}_i (\st,\bz,z) ,\,\,\,
  \bpl_i \tilde{\hat{A}}_\s (\st,\bz,z) = \tilde{\hat{A}}_\s (\st,\bz,z) \hat{\bar{W}}_{\!i} (\st,\bz,z) 
\end{align}
\begin{align}
\hat{W}_i (\st,\bz,z) =& L_g^{ab} \sum_{u=0}^1 \sum_{j\neq i} \BIG{[} \left( \frac{z_j}{z_i} \right)^{\!\frac{u}{2}} \!\frac{1}{z_i -z_j}
   T_a^{(j)} \tau_u^{(j)} \otimes T_b^{(i)} \tau_u^{(i)} \nn \\
& \bigspc \quad \quad +\left( \frac{\bz_j}{z_i} \right)^{\!\frac{u}{2}} \!\frac{(-1)^u}{z_i -\bz_j} T_a^{(i)} 
  \tau_u^{(i)} \otimes \bT_b^{(j)} \tau_u^{(j)} \BIG{]} \nn \\
& \bigspc \quad \quad + \frac{1 -(\bz_i /z_i )^{\frac{1}{2}}}{z_i -\bz_i} L_g^{ab} T_a^{(i)} \otimes \bT_b^{(i)} -\frac{1}{2z_i} D_g (T^{(i)}) 
\end{align}
\begin{align}
\hat{\bar{W}}_i (\st,\bz,z) =& L_g^{ab} \sum_{u=0}^1 \sum_{j\neq i} \BIG{[} \left( \frac{\bz_j}{\bz_i} \right)^{\!\frac{u}{2}} 
   \!\frac{1}{\bz_i -\bz_j} \bT_a^{(j)} \tau_u^{(j)} \otimes \bT_b^{(i)} \tau_u^{(i)} \nn \\
& \bigspc \quad \quad +\left( \frac{z_j}{\bz_i} \right)^{\!\frac{u}{2}} \!\frac{(-1)^u}{\bz_i -z_j} T_a^{(j)} \tau_u^{(j)} \otimes \bT_b^{(i)} 
  \tau_u^{(i)} \BIG{]} \nn \\
& \bigspc \quad \quad + \frac{1 -(z_i /\bz_i )^{\frac{1}{2}}}{\bz_i -z_i} L_g^{ab} T_a^{(i)} \otimes \bT_b^{(i)} -\frac{1}{2\bz_i} D_g (T^{(i)}) 
\end{align}
\begin{gather}
\tilde{\hat{A}}_\s (\st(T),\bz,z) \sum_{i=1}^n (T_a^{(i)} \!\otimes \!\one +\one \!\otimes \!\bT_a^{(i)}) =0 ,\quad a=1\ldots \text{dim }g \,.\label{Basic-GWI}
\end{gather}
\end{subequations}
In this case, the global Ward identity in \eqref{Basic-GWI} has the standard form found in untwisted chiral KZ theory. One must also enforce 
the $\tau_1$-constraint \eqref{Eq4.14} and the world-sheet parity \eqref{Eq4.16}, \eqref{Eq4.17}. We refer the reader to our discussion 
(see Subsec.~$4.3$ and App.~C) of redundancy in these systems.

\subsection{Example: The Open-String One-point Correlators for all $h_\s^2 =1$}

The twisted one-point correlators of open-string sector $\hat{h}_\s$
\begin{gather}
\tilde{\hat{A}}_\s (\st(T,\s),\bz,z) =\langle \tilde{\hg} (\st(T,\s),\bz,z,\s) \rangle_\s
\end{gather}
are in fact pinned three-point correlators -- counting the scalar twist fields. In this subsection, we solve the twisted KZ systems of these one-point
correlators for all $h_\s^2 =1$. 

For the basic open-string sector with $h_\s =1$, the one-point correlator satisfies the following twisted KZ system:
\begin{subequations}
\begin{gather}
\tilde{\hat{A}}_\s \equiv \tilde{\hat{A}}_\s (\st(T),\bz,z) = \langle \tilde{\hg}(\st(T),\bz,z) \rangle_\s \\
\pl \tilde{\hat{A}}_\s =\tilde{\hat{A}}_\s \left( \frac{1- (\bz/z )^{1/2}}{z-\bz} L_g^{ab} T_a \otimes \bT_b -\frac{D_g (T)}{2z} \right) \\
\bpl \tilde{\hat{A}}_\s =\tilde{\hat{A}}_\s \left( \frac{1-(z/\bz )^{1/2}}{\bz -z} L_g^{ab} T_a \otimes \bT_b -\frac{D_g (T)}{2\bz} \right) 
\end{gather}
\begin{gather}
\tilde{\hat{A}}_\s (T_a \!\otimes \!\one +\one \!\otimes \!\bT_a ) =0 \label{1pt-GWI} ,\quad a=1\ldots \text{dim }g \\
\tilde{\hat{A}}_\s \tau_1 \otimes \tau_1 =\tilde{\hat{A}}_\s \label{1pt-t1} \\
\tilde{\hat{A}}_\s (\st(T),\bz,z)^t = \tilde{\hat{A}}_\s (\st(\bT),z,\bz) \tau_3 \otimes \tau_3 \,. \label{1pt-WSPar}
\end{gather}
\end{subequations}
Here $L_g^{ab}$ is again the inverse inertia tensor \eqref{cL-Basic} of the affine-Sugawara construction [6,7,31-33,12] on $g$ and $D_g (T)$ is the 
conformal weight matrix \eqref{Dg-Basic} of rep $T$ of $g$. 

The solution of this system is 
\begin{subequations}
\begin{gather}
\tilde{\hat{A}}_\s (\st(T),\bz,z)^{\a u;\be v} = C_{(u+v)}(T) \de^{\gamma \de} \Big{(} |z|^{-D_g(T)}
   ((\sqrt{\bz} + \sqrt{z})^2 )^{L_g^{ab} T_a \otimes \bT_b} \Big{)}_{\gamma \de}^{\,\,\,\a \be} \nn \\
= C_{(u+v)}(T) \de^{\gamma \de} \left( (|z| (\sqrt{\bz} +\sqrt{z})^2 )^{-D_g(T)} \right)_{\gamma \de}^{\,\,\,\a \be} \label{1pt-Soln} 
\end{gather}
\begin{gather}
\tilde{\hat{A}}_\s (\st(\bT),z,\bz)^{\a u;\be v} = C_{(u+v)}(\bT) \de^{\gamma \de} \Big{(} |z|^{-D_g(\bT)}
   ((\sqrt{z} +\sqrt{\bz})^2 )^{L_g^{ab} \bT_a \otimes T_b} \Big{)}_{\gamma \de}^{\,\,\,\a \be}  \label{bT-1pt-Soln} \\
= C_{(u+v)}(\bT) \de^{\gamma \de} \left( (|z| (\sqrt{\bz} +\sqrt{z})^2 )^{-D_g(T)} \right)_{\gamma \de}^{\,\,\,\a \be} \\
C_{(w)} (\bT) =(-1)^w C_{(w)}(T) ,\quad \bar{u},\bar{v},\bar{w} \in \{ 0,1\} \label{WSPar-Const} 
\end{gather}
\end{subequations}
where the Kronecker factor $\de^{\gamma \de}$ enforces the global Ward identity in \eqref{1pt-GWI}. The form $C_{(u+v)}$ of the constant multiplier enforces 
the $\tau_1$ constraint \eqref{1pt-t1}, and the relation in \eqref{WSPar-Const} enforces the world-sheet parity \eqref{1pt-WSPar}. We note in particular 
that (in contrast to the normal situation in \eqref{Eq 4.19c}) the matrix $\bT_a$ in Eq.~\eqref{bT-1pt-Soln} acts on the left index $\gamma$ of the Kronecker 
factor. This solution has trivial monodromy and is symmetric under $z\leftrightarrow \bz$. 

For the generic open-string sector with $h_\s^2 \!=\!1 ,\,h_\s \!\neq \!1$, the twisted KZ system for the one-point correlator is:
\begin{subequations}
\label{KZConn-hs2}
\begin{gather}
\pl \tilde{\hat{A}}_\s \!=\!\tilde{\hat{A}}_\s \BIG{(} \frac{1- (\bz/z )^{1/2}}{z-\bz} \Big{(} {\cL}_{\sgb(\s)}^{0\m ;0\n}(\s) \st_{0\m}(T,\s) 
  \!\otimes \!\st_{0\n}(\bT,\s) \bigspc \bigspc \nn \\
  \bigspc -\!{\cL}_{\sgb(\s)}^{1\m ;1\n}(\s) \st_{1\m}(T,\s) \!\otimes \!\st_{1\n}(\bT,\s) \Big{)} -\frac{\D_{\sgb(\s)} (\st(T,\s))}{2z} \BIG{)} \label{514a} 
\end{gather}
\begin{gather}
\bpl \tilde{\hat{A}}_\s \!=\!\tilde{\hat{A}}_\s \BIG{(} \frac{1- (z/\bz )^{1/2}}{\bz -z} \Big{(} {\cL}_{\sgb(\s)}^{0\m ;0\n}(\s) \st_{0\m}(T,\s) 
  \!\otimes \!\st_{0\n}(\bT,\s) \bigspc \bigspc \nn \\
  \bigspc -\!{\cL}_{\sgb(\s)}^{1\m ;1\n}(\s) \st_{1\m}(T,\s) \!\otimes \!\st_{1\n}(\bT,\s) \Big{)} -\frac{\D_{\sgb(\s)} (\st(\bT,\s))}{2\bz} \BIG{)} \label{514b} 
\end{gather}
\begin{gather}
\D_{\sgb(\s)}(\st(T,\s)) \!=\!{\cL}_{\sgb(\s)}^{\nrm;\mnrn}(\s) \st_\nrm(T,\s) \st_\mnrn (T,\s) \bigspc \nn \\
\quad \quad \quad \quad =\!U(T,\s) D_g(T) U\hc (T,\s) ,\quad \quad D_g(T) \!=\! L_g^{ab} T_a T_b \label{514c} \\
\tilde{\hat{A}}_\s (\st(T,\s),\bz,z,\s) \tilde{Q}_\nrm (\st,\s)=0 ,\,\,\,\forall \m ,\,\,\,\,\bar{n}(r)=0,1 \label{1pt-GWI2} \\
\tilde{Q}_\nrm (\st(T,\s),\s) \equiv \st_\nrm(T,\s) \otimes \one +(-1)^{n(r)} \one \otimes \st_\nrm (\bT,\s) \label{tQ-Defn} \\
\tilde{\hat{A}}_\s (\st(T,\s),\bz,z,\s) \tau_1 \otimes \tau_1 =\tilde{\hat{A}}_\s (\st(T,\s),\bz,z,\s) \\
\tilde{\hat{A}}_\s (\st(T,\s),\bz,z,\s)^t = \tilde{\hat{A}}_\s (\st(\bT,\s),z,\bz,\s) \tau_3 \otimes \tau_3 \,.
\end{gather}
\end{subequations}
Here $\D_{\sgb(\s)}$ is the twisted conformal weight matrix \cite{Big}, and we have noticed in \eqref{tQ-Defn} that, in the case of the one-point correlators, the 
$\tau_1$'s c
an be stripped from the $Q_{1u}$ global Ward identity \eqref{5.4c}. 

In spite of the apparent complexity of Eqs.~\eqref{514a}, \eqref{514b}, the connections can be simplified by using the global Ward identities 
\eqref{1pt-GWI2} to eliminate $\st(\bT)$ in favor of $\st(T)$. This leads to the simple solutions
\begin{subequations}
\label{Gen-1pt-Fn}
\begin{gather}
\tilde{\hat{A}}_\s (\st(T,\s),\bz,z,\s) =\langle \hg(\st(T,\s),\bz,z,\s) \rangle_\s \bigspc \nn \\
  ={{\cal C}}(\st(T,\s)) \left( |z| (\sqrt{z} +\sqrt{\bz})^2 \right)^{-\D_{\sgb(\s)} (\st(T,\s))} \\
{{\cal C}} (\st(T))^{\Nrm u;\Nsn v} ={{\cal C}}_{(u+v)} (\st(T))^{\Nrm ;\Nsn} \\
{{\cal C}}_u (\st(\bT))^{\Nrm ;\Nsn} =(-1)^u {{\cal C}}_u (\st(T))^{\Nsn ;\Nrm} \\
{{\cal C}} (\st(T)) \tilde{Q}_\nrm (\st(T,\s),\s) =0 ,\,\,\, \forall \m ,\,\,\bar{n}(r)=0,1  \label{GWI-1pt} 
\end{gather}
\end{subequations}
each of which has trivial monodromy and $z\leftrightarrow \bz$ symmetry. The reduced form of the global Ward identity in Eq.~\eqref{GWI-1pt} tells us that 
the constants ${{\cal C}}$ are singlets under the algebra of $\tilde{Q}$ (which is the same as that of $Q$ in \eqref{445d}). However, because many different 
open-string sectors are involved here, we have not further analyzed this condition. 

We remind the reader that in the special case \eqref{Eq3.14} of a permutation-invariant system, the twisted conformal weight matrix reduces to 
\begin{gather}
\D_{\sgb(\s)}(\st(T,\s)) =D_g (T) =\Delta_{\gfraks} (T) \thickone
\end{gather}
where $\Delta_{\gfraks}(T)$ is the conformal weight of rep $T$ of $\gfrak$.

\section{Classical and Free-Boson Avatars}

\subsection{The Classical Limit and Clashing Monodromies}

The classical (high-level) limit of the twisted affine primary fields are the so-called {\it group orbifold elements}, which satisfy the classical system: 
\begin{subequations}
\begin{gather}
\pl \hg(\st,\bz,z,\s) \!=\!\hg(\st,\bz,z,\s) \hj(\st,z,\s) ,\quad \bpl \hg(\st,\bz,z,\s) \!=\! -\hjb (\st,\bz,\s) \hg(\st,\bz,z,\s) \label{Cl-EOMs} \\
\hj(\st,z,\s) \equiv \sG^{\nrmu;\nsn v}(\s) \hj_\nrmu (z,\s) \st_{\nsn v} \\
\hjb(\st,\bz,\s) \equiv \sG^{\nrmu;\nsn v}(\s) \hj_\nrmu (\bz,\s) \tilde{\st}_{\nsn v} \\ 
\hj_\nrmu (z,\s) = \sum_{m \in \Zint} \hj_\nrmu (m\!+\!\nrrs \!+\! \srac{u}{2}) z^{-(m+\nrrsf +\frac{u}{2}) -1} 
\end{gather}
\begin{gather}
\sG^{\nrmu ;\nsn v}(\s) =\srac{1}{2} \de_{u+v,0\,\text{mod }2} \sG^{\nrm;\nsn}(\s) \label{Full-G^} \\
\st_\nrmu = \st_\nrm (T,\s) \tau_u ,\quad \tilde{\st}_\nrmu =\st_\nrm (T,\s) (-1)^u \tau_u \\
\hg (\st (\bT,\s),z,\bz,\s)^t = \tau_3 \hg(\st(T,\s),\bz,z,\s) \tau_3 \label{61g} \\ 
\hg_1 (\st,\bz,z,\s) \hg_2 (\st,\bz,z,\s) =\hg_3 (\st,\bz,z,\s) \,.\label{438e} 
\end{gather}
\end{subequations}
The classical equations of motion in \eqref{Cl-EOMs} are the high-level limit (${\cL} \rightarrow \sG^{\bullet} /2$) of the twisted vertex operator 
equations \eqref{Eq3.10}, where the $\sG^{\bullet}$ in ($6.1$b,c) is the inverse of the total twisted metric in Eq.~\eqref{Eq3.7}. The quantities 
$\hj ,\hjb$ are the classical twisted matrix currents, and Eq.~\eqref{61g} is the behavior of the group orbifold elements under the level-independent 
world-sheet parity \eqref{Eq 3.8b}. We also assume that the group orbifold elements form a group, and the relation in \eqref{438e} expresses the 
requirement that the product of any two group orbifold elements is in the group. 

In the classical limit, we also find the twisted tangent-space form \cite{Big,Fab,Geom} of the group orbifold element:
\begin{subequations}
\label{6.2}
\begin{gather}
g(T,\bz,z) =e^{ix^a(\bz,z) T_a} \label{6.2a} \\
x^a (\bz,z)' = -x^b (z,\bz) \ws_b{}^a ,\quad a=1\ldots \text{dim }g \label{6.2b} \\
x^{a0}(\bz,z) \equiv x^a (\bz,z) ,\quad x^{a1} (\bz,z) \equiv -x^a (z,\bz) \label{6.2c} \\
\sx^\nrmu (\bz,z,\s) = x^{a\Id} (\bz,z) \schisig_\nrm^{-1} U\hc (\s)_a^\nrm (\srac{1}{\sqrt{2}} U\hc {}_\Id{}^u) \,\dual \, \hx_\s^\nrmu (\bz,z) \label{6.2d} \\
\hg(\st(T,\s),\bz,z,\s) = e^{i \hx_\s^\nrmu (\bz,z) \st_\nrmu (T,\s)} \,. \label{TwTS-Form}
\end{gather}
\end{subequations}
Starting from the tangent-space form of the untwisted group element $g(T)$ in \eqref{6.2a}, one may easily check with Eqs.~\eqref{6.2b} and \eqref{Eq 2.5e} 
that the action \eqref{Eq 2.5b} of the orientation-reversing automorphism on $g(T)$ can now be written in the classical form \eqref{Eq2.7}. Moreover, 
the twisted tangent-space form of $\hg$ in \eqref{TwTS-Form} realizes the group multiplication \eqref{438e} because the total twisted representation matrices 
satisfy the total orbifold Lie algebra $\hg_O (\s)$ in Eq.~\eqref{Eq 2.26c}.

The twisted tangent-space form of $\hg$ also allows us to write the action \eqref{Eq 3.8b} of world-sheet parity on the group orbifold element in the
alternate form
\begin{subequations}
\begin{gather}
\hg(\st(\bT,\s),z,\bz,\s)^t =e^{i\hx_\s^\nrmu (z,\bz) \st_\nrmu (\bT,\s)^t}  \bigspc \bigspc \quad \nn \\
  \bigspc \bigspc =e^{-i\hx_\s^\nrmu (z,\bz) \st_\nrmu (T,\s)} =\hg^{-1} (\st(T,\s),z,\bz,\s) \label{6.3a} \\
\Rightarrow \hg^{-1} (\st(T,\s),z,\bz,\s) =\tau_3 \hg (\st(T,\s),\bz,z,\s) \tau_3 \quad \quad \label{6.3b}\\
\Leftrightarrow \hx_\s^\nrmu (z,\bz) =(-1)^{u+1} \hx_\s^\nrmu (\bz,z) \label{6.3c}
\end{gather}
\end{subequations}
where we have used the transpose identity in \eqref{Eq2.21} to obtain Eq.~\eqref{6.3a}. We emphasize that the forms in \eqref{6.3b}, \eqref{6.3c} 
show {\it world-sheet identifications} for each $\st(T)$, so that if desired, one may restrict attention to the upper half-plane. 

We turn next to the question of monodromy. The monodromies of the classical index currents
\begin{subequations}
\label{hJ-Monos2}
\begin{gather}
\hj_\nrmu (ze^\tp ,\s) \!=\! e^{-\tp (\nrrsf +\frac{u}{2})} \hj_\nrmu (z,\s) \\
\hj_\nrmu (\bz e^{-\tp},\s) \!=\! e^{\tp (\nrrsf +\frac{u}{2})} \hj_\nrmu (\bz,\s) 
\end{gather}
\end{subequations}
are the same as the operator index currents in \eqref{Eq 3.6a}, so the monodromies of the matrix currents are
\begin{subequations}
\label{MatJ-Mono}
\begin{gather}
\hj(\st(T,\s), z e^\tp,\s) =\tau_3 E(T,\s) \hj(\st(T,\s),z,\s) E^\ast (T,\s) \tau_3 \\
\hjb (\st(T,\s),\bz e^{-\tp},\s) =\tau_3 E^\ast (T,\s) \hjb(\st(T,\s),\bz,\s) E(T,\s) \tau_3 
\end{gather}
\end{subequations}
where $E(T,\s)$ is the eigenvalue matrix of the extended $H$-eigenvalue problem in Eq.~\eqref{Eq2.16}. Here we have used the $\st$-selection rule 
\eqref{Eq 2.20e} and the identity \eqref{t3-Id} to translate Eq.~\eqref{hJ-Monos2} into Eq.~\eqref{MatJ-Mono}. 

We also find that the twisted matrix currents are related by world-sheet parity as follows
\vspace{-0.1in}
\begin{subequations}
\label{hJ-WSPar}
\begin{gather}
\hj (\st,r,\xi,\s) \equiv \hj (\st , z=re^{i\xi} ,\s) ,\quad \hjb (\st,r,\xi,\s) \equiv \hjb (\st,\bz =re^{-i\xi} ,\s) \\
\hjb (\st,r,\xi,\s) = \tau_3 \hj (\st,r,-\xi,\s) \tau_3 
\end{gather}
\end{subequations}
where $r=|z|$. Then the twisted current boundary conditions on the real axis
\begin{subequations}
\begin{gather}
\hjb (\st,r,0,\s) =\tau_3 \hj (\st,r,0,\s) \tau_3 \\
\hjb (\st (T,\s),r,\pi,\s) =E^\ast (T,\s) \hj (\st(T,\s) ,r,\pi,\s) E(T,\s)
\end{gather}
\end{subequations}
follow from Eq.~\eqref{hJ-WSPar} and the monodromies \eqref{MatJ-Mono}. 

The general solution to the classical PDEs in \eqref{Cl-EOMs} is 
\begin{subequations}
\begin{gather}
\hg(\st,\bz,z,\s) =\bar{\Omega} (\st,\bz,\s) \hg(\st,0,0,\s) \Omega (\st,z,\s) \\
\Omega (\st,z,\s) \equiv \Ord \exp (\int_0^z \!\!dz' \hj(\st,z' ,\s)) ,\quad \bar{\Omega} (\st,\bz,\s) \equiv \Ord^{\ast} \exp (-\!\int_0^{\bz} \!\!d\bz ' 
   \hjb(\st,\bz ',\s)) \\
\Omega (\st(T,\s),ze^{\tp} ,\s) =\tau_3 E(T,\s) \Omega (\st(T,\s),z,\s) E^\ast (T,\s) \tau_3 \\
\bar{\Omega} (\st(T,\s),\bz e^{-\tp},\s) =\tau_3 E^\ast (T,\s) \bar{\Omega} (\st(T,\s),\bz,\s) E(T,\s) \tau_3 
\end{gather}
\end{subequations}
where $\Ord$ and $\Ord^\ast$ are ordering in $z$ and anti-ordering in $\bz$ respectively.

We next inquire when it is possible for the group orbifold element to satisfy world-sheet parity \eqref{Eq 3.8b} with a {\it definite $2\pi$-monodromy}.
The necessary and sufficient conditions for this are the boundary conditions at the origin
\begin{subequations}
\begin{gather}
\hg(\st(\bT,\s) ,0,0,\s)^t =\tau_3 \hg(\st(T,\s),0,0,\s) \tau_3 \\
 \hg^{-1} (\st(T,\s),0,0,\s) E(T,\s) \hg(\st(T,\s),0,0,\s) =E^\ast (T,\s) 
\end{gather}
\end{subequations}
which gives the $2\pi$-monodromy:
\begin{gather}
\hg(\st(T,\s) ,\bz e^{-\tp},ze^\tp ,\s)= \tau_3 E^\ast (T,\s) \hg(\st(T,\s),\bz,z,\s) E^{\ast}(T,\s) \tau_3 \,. \label{hg-Mono}
\end{gather}
But this monodromy violates the group multiplication \eqref{438e} -- unless in fact 
\begin{gather}
E^\ast (T,\s)^2 =\thickone
\end{gather}
which implies $h_\s^2 =1$. The same conclusion is seen immediately on expansion of $\hg$ on both sides of Eq.~\eqref{hg-Mono}, using the twisted tangent-space 
form \eqref{TwTS-Form} at leading order in $\hx$.

We emphasize the central role in this argument of the {\it clashing monodromies} of the twisted currents in Eqs.~\eqref{hJ-Monos2}, \eqref{MatJ-Mono}, which
tell us that the group orbifold elements of the open-string sectors {\it cannot} have a definite $2\pi$-monodromy when $h_\s^2 \neq 1$. Indeed, one easily 
checks near the origin
\begin{subequations}
\label{FBM}
\begin{gather}
\hg(\st ,\bz,z,\s) \simeq \one + i\hx_\s^\nrmu (\bz,z) \st_\nrmu \\
\pl \hx_\s^\nrmu (\bz,z) \simeq -i\sG^{\nrmu ;\nsn v}(\s) \hj_{\nsn v} (z,\s) \\
\bpl \hx_\s^\nrmu (\bz,z) \simeq i\sG^{\nrmu ;\nsn v}(\s) \hj_{\nsn v} (\bz,\s) \\
\pl \bpl \hx_\s^\nrmu (\bz,z) \simeq 0 
\end{gather}
\end{subequations}
that $\hx$ and hence $\hg$ have {\it mixed} monodromy for $h_\s^2 \!\neq \!1$, with terms reflecting the monodromies of both $\hj$ and $\hjb$. Because the 
left- and right-mover currents of open strings are constructed from the same set of current modes, such clashing current monodromies and mixed monodromies
of $\hg$ and $\hx$ are also expected in more general twisted open strings. 

The arguments above substantiate the intuitive notion that definite $2\pi$-monodromy is not a natural concept for open strings. The surprise here, on the 
other hand, is that the open-string sectors with $h_\s^2 =1,\, E(T,\s)^2 =1$ can be described by the definite monodromy
\begin{gather}
\hg(\st(T,\s) ,\bz e^{-\tp},ze^\tp ,\s) =\tau_3 E(T,\s) \hg(\st(T,\s),\bz,z,\s) E^\ast (T,\s) \tau_3
\end{gather}
which would in fact be obtained by {\it local isomorphisms} from the automorphic response \eqref{Eq 2.24b} of the eigenfield $\sg$. On the basis of this 
observation, we will return elsewhere \cite{Orient2} to a more complete classical description of WZW and sigma-model orientation-orbifold sectors with 
$h_\s^2 =1$.

We close this subsection by discussing the monodromy relation \eqref{hg-Mono} vis-\'{a}-vis the one-point correlators of the previous subsection. 
In the case $E(T,\s)=1$, the monodromy \eqref{hg-Mono} gives the additional constraint on the basic one-point correlators of Subsec.~$5.3$
\begin{gather}
\tilde{\hat{A}}_\s (\st(T),\bz e^{-\tp},ze^\tp )=\tilde{\hat{A}}_\s (\st(T),\bz,z) (\tau_3 \otimes \tau_3)
\end{gather}
which is satisfied only when we further choose the constant in the solution \eqref{1pt-Soln} as:
\begin{gather}
C_{(1)}(T) =0 \,.
\end{gather}
For the generic one-point correlators in Eq.~\eqref{Gen-1pt-Fn}, the monodromy \eqref{hg-Mono} with $E^2(T,\s)=1$ gives the constraint
\begin{gather}
\tilde{\hat{A}}_\s (\st(T,\s),\bz e^{-\tp},ze^\tp )=\tilde{\hat{A}}_\s (\st(T,\s),\bz,z) (\tau_3 E^\ast (T,\s) \otimes \tau_3 E^\ast(T,\s))
\end{gather}
which holds iff we choose the constant matrix ${{\cal C}}(\st(T))$ to satisfy:
\begin{gather}
{{\cal C}}(\st(T))^{\Nrm u;\Nsn v} =0 \text{ unless } u\!+\!v\!+\!N(r)\!+\!N(s) =0\text{ mod }2 \,.
\end{gather}
To see that this condition is sufficient, one needs the identity
\begin{gather}
E^\ast (T,\s) \D_{\sgb(\s)} (\st(T,\s)) E^\ast (T,\s) =\D_{\sgb(\s)} (\st(T,\s))
\end{gather}
which follows from the selection rule \eqref{ETE=-1T} because the twisted conformal weight matrix $\D_{\sgb(\s)}(\st)$ in \eqref{514c} is quadratic 
in $\st$. The consistency of the classical monodromy \eqref{hg-Mono} with the one-point quantum correlators is in fact more than we would have 
expected, since only the constant factors survive in the high-level limit.

\subsection{Example: Free Twisted Open Strings on Abelian $g$}

As simple examples of our construction, we turn next to the open-string sectors of the orientation orbifolds on abelian $g$.

When the original Lie algebra $g$ is taken to be abelian, the approximate equations \eqref{FBM} found near the origin for $\hx^\nrmu$ become the 
{\it exact} description of a twisted {\it free} boson system $\pl \bpl \hx =0$:
\begin{subequations}
\begin{gather}
f_{ab}{}^c =0 ,\quad L_g^{ab} =\frac{G^{ab}}{2} ,\quad g=e^{ix^a T_a} ,\quad [T_a ,T_b] =0 \\
x^a (\bz,z)' = -x^b (z,\bz) w\hc (h_\s)_b{}^a ,\quad a,b=1\ldots N ,\quad \bar{c}=c=N \\
\Longrightarrow \scf =0 ,\quad {\cL}^{\bullet} =\frac{\sG^{\bullet}}{2} ,\quad \hg =e^{i\hx^\nrmu \st_\nrmu} ,\quad [\st_\nrmu ,\st_{\nsn v}]=0 \\
\hat{c} =2N ,\quad \sum_r \text{dim} [n(r)] =N \,.
\end{gather}
\end{subequations}
Here the relations among the various coordinates are the same as those given in Eqs.~\eqref{6.2c}, \eqref{6.2d}, and the twisted abelian ``momenta" 
$\st$ are still related to the untwisted abelian ``momenta" $T$ by the same formula \eqref{Eq 2.26a}. 

For pedagogical purposes, we continue our discussion of these twisted free-boson systems in Minkowski-space $(\xi,t)$, where the equations of motion 
take the form:
\begin{subequations}
\label{6.20}
\begin{gather}
z \rightarrow e^{i(t+\xi)} ,\quad \bz \rightarrow e^{i(t-\xi)} \\
\hx^\s_\nrmu \equiv \sG_{\nrmu ;\nsn v}(\s) \hx_\s^{\nsn v} = 2 \sG_{\nrm;\mnrn}(\s) \hx_\s^{\mnrn ,-u} \\
\pl_t \hx^\s_\nrmu (\xi,t) =\hj_\nrmu (\xi,t,\s) +(-1)^{u+1} \hj_\nrmu (-\xi,t,\s) \label{plt-hx} \\
\pl_\xi \hx^\s_\nrmu (\xi,t) =\hj_\nrmu (\xi,t,\s) +(-1)^u \hj_\nrmu (-\xi,t,\s) \label{plx-hx} \\
\hj_\nrmu (\xi,t,\s) \equiv \sum_{m\in \Zint} \hj_\nrmu (m\!+\!\nrrs \!+\!\srac{u}{2}) e^{-i (m+\nrrsf +\frac{u}{2}) (t+\xi)} 
\end{gather}
\begin{gather}
\hj_\nrmu (\xi+2\pi ,t,\s) = e^{-\tp (\nrrsf +\frac{u}{2})} \hj_\nrmu (\xi,t,\s) \label{Cyl-Mono} \\
(\pl_t^2 -\pl_\xi^2 )\hx^\s_\nrmu (\xi,t) =0 ,\quad \hx^\s_\nrmu (-\xi,t) =(-1)^{u+1} \hx^\s_\nrmu (\xi,t) \,. \label{hx-WSPar}
\end{gather}
\end{subequations}
The world-sheet parity of the string coordinate $\hx$ in \eqref{hx-WSPar} shows a particularly transparent form of the world-sheet identification 
$\xi \leftrightarrow -\xi$. Since the free-boson dynamics is linear, we may consider the system \eqref{6.20} either at the operator level, including the 
orbifold Virasoro generators and twisted current algebra
\begin{subequations}
\begin{align}
&\!\!\!\hat{L}_u (m\!+\!\srac{u}{2})= \bigspc \bigspc \bigspc \bigspc \bigspc \bigspc \\
&\quad \,\srac{1}{4} \sG^{\nrm ;\mnrn}(\s) \sum_{v=0}^1 \sum_{p\in \Zint} :\!\hj_{\nrm v} (p\!+\!\nrrs \!+\!\srac{v}{2}) 
   \hj_{\mnrn ,u-v} (m\!-\!p \!-\!\nrrs \!+\!\srac{u-v}{2}) \!: \quad \,\, \nn \\
&[\hj_\nrmu (m\!+\!\nrrs \!+\!\srac{u}{2}) ,\hj_{\nsn v} (n\!+\!\srac{n(s)}{\r(\s)} \!+\!\srac{v}{2})] = \bigspc \bigspc \nn \\
& \bigspc  =(m\!+\!\nrrs \!+\!\srac{u}{2}) \de_{m+n+\frac{n(r)+n(s)}{\r(\s)} +\frac{u+v}{2},0} (2\de_{u+v,0\,\text{mod }2}) \sG_{\nrm;\mnrn} (\s) 
\end{align}
\end{subequations}
or at the classical level by ignoring normal ordering and replacing commutators with brackets.

The partial differential equations \eqref{plt-hx}, \eqref{plx-hx} and the $\hj$ monodromy \eqref{Cyl-Mono} imply the following boundary conditions or
{\it orientation-orbifold branes} for the free bosons on the strip $0\! \leq \!\xi \!\leq \pi$ 
\begin{subequations}
\label{hx-BCs}
\begin{gather}
(\pl_t \hx_\s^{\nrm 0}) (0,t) =0 ,\quad (\pl_\xi \hx_\s^{\nrm 1}) (0,t) =0 \\
\cos (\nrrs \pi) (\pl_t \hx_\s^\nrmu )(\pi,t) -i\sin (\nrrs \pi) (\pl_\xi \hx_\s^\nrmu )(\pi,t) =0
\end{gather}
\end{subequations}
so that, for example, the $\bar{n}(r)\!=\!0$ free bosons $\hx_\s^{0\m u}$ are Dirichlet-Dirichlet and Neumann-Dirichlet\footnote{Half-integer moded scalar 
fields \cite{2faces2} and the corresponding free twisted open strings with N-D or D-N boundary conditions \cite{CF,WS} are well-known as the first examples
of twisted sectors of orbifolds.} for $\bar{u}=0$ and $\bar{u}=1$ respectively. For $\srac{\bar{n}(r)}{\r(\s)} \!=\!\srac{1}{2}$, the bosons 
$\hx_\s^{\r(\s)/2 ,\m ,0}$ and $\hx_\s^{\r(\s)/2 ,\m ,1}$ are Dirichlet-Neumann and Neumann-Neumann respectively. In the generic case (when $\bar{n}(r)$ is 
not equal to $0$ or $\srac{\r(\s)}{2}$) the boundary conditions at $\xi =\pi$ are {\it mixed}.

Moreover, we can explicitly solve the PDEs in \eqref{6.20} (or the wave equation in \eqref{hx-WSPar} with the boundary conditions \eqref{hx-BCs}) to 
obtain the free bosons in terms of the twisted current modes
\begin{subequations}
\label{hx-Mode}
\begin{gather}
\hx^\s_{0\m 0}(\xi,t) = 2\Big{\{} \hj_{0\m 0}(0) \xi +\sum_{m \neq 0} \frac{\hj_{0\m 0}(m)}{m} e^{-imt} \sin (m\xi ) \Big{\}} \label{Spec-hx-1} 
\end{gather}
\begin{gather}
\hx^\s_{\r(\s)/2 ,\m ,1}(\xi,t) =\hat{q}_{\r(\s)/2,\m ,1} +2\hj_{\r(\s)/2,\m 1}(0)t + \bigspc \bigspc \bigspc \quad \quad \quad \nn \\
   \quad \quad \bigspc +2i\sum_{m\neq -1} \frac{\hj_{\r(\s)/2,\m 1}(m+1)}{m+1} e^{-i(m+1)t} \cos ((m\!+\!1)\xi ) \label{Spec-hx-2} 
\end{gather}
\begin{gather}
\bar{n}(r) \neq 0: \,\,\hx^\s_{\nrm 0}(\xi,t) =2\sum_{m\in \Zint} \frac{\hj_{\nrm 0}(m\!+\!\nrrs)}{m\!+\!\nrrs} e^{-i(m+\nrrsf)t} \sin ((m\!+\!\nrrs) \xi) \\
\bar{n}(r) \!\neq \!\srac{\r(\s)}{2} \!: \,\,\hx^\s_{\nrm ,1}(\xi,t) \!= \!\hat{q}_{\nrm ,1} +\bigspc \bigspc \bigspc \bigspc \bigspc \quad \quad \quad \nn \\
  \bigspc \bigspc +\!2i\sum_{m\in \Zint} \frac{\hj_{\nrm 1}(m\!+\!\nrrs \!+\!\srac{1}{2})} {m\!+\!\nrrs \!+\!\srac{1}{2}} 
  e^{-i(m+\nrrsf +\frac{1}{2})t} \cos ((m\!+\!\nrrs \!+\!\srac{1}{2})\xi )
\end{gather}
\end{subequations}
where the form in Eq.~\eqref{Spec-hx-2} applies only when the order $\r(\s)$ of $h_\s$ is even. To obtain these results we have also imposed the 
world-sheet parity in \eqref{hx-WSPar}, whose only effect is to set $(\xi,t)$-independent terms $\hat{q}$ to zero when $\bar{u}\!=\!0$. The expansions in 
\eqref{hx-Mode} are immediately recognized as twisted open strings\footnote{As argued above, the twisted open strings have mixed monodromy when 
$\srac{\bar{n}(r)}{\r(\s)}$ is not equal to $0$ or $\srac{1}{2}$.}.

At this point, it is helpful to note another algebraic requirement on the coordinates of the twisted open strings
\begin{gather}
[\hj_\nrmu (m\!+\! \nrrs \!+\!\srac{u}{2}) ,\hx_\s^{\nsn v}(\xi,t)] =\bigspc \bigspc \bigspc \bigspc \nn \\
   \quad \quad =-i\de_\nrmu{}^{\!\!\nsn v} \Big{(} e^{i (m+\nrrsf +\frac{u}{2})(t+\xi)} + (-1)^{u+1} e^{i (m+\nrrsf +\frac{u}{2})(t-\xi)} \Big{)} \label{Jx-Comm}
\end{gather}
which follows from the abelian analogue of Eq.~\eqref{Eq 3.22b}. We find that the mode expansions \eqref{hx-Mode} are consistent with Eq.~\eqref{Jx-Comm} only
when we choose the $[\hj ,\hat{q}]$ commutators as:
\begin{subequations}
\begin{gather}
\bar{n}(s) \neq \srac{\r(\s)}{2} \!:\,\, [\hj_\nrmu (m\!+\! \nrrs \!+\!\srac{u}{2}) ,\hat{q}_{\nsn 1}] =0 \\
[\hj_\nrmu (m\!+\! \nrrs \!+\!\srac{u}{2}) ,\hat{q}_{\r(\s)/2 ,\n, 1}] =-2i\sG_{\nrmu ;\r(\s)/2 ,\n,1}(\s) \de_{m+\nrrsf +\frac{u}{2},0} \bigspc \nn \\
\bigspc  =-4i \de_{u+1,0\,\text{mod }2} \de_{n(r)+\frac{\r(\s)}{2} ,0\,\text{mod }\r(\s)} \sG_{-\frac{\r(\s)}{2},\m ;\frac{\r(\s)}{2},\n}(\s) 
   \de_{m+\nrrsf +\frac{u}{2},0} \,.
\end{gather}
\end{subequations} 
Jacobi identities then tell us that the twisted currents commute with the commutators of any two $\hat{q}$'s.

To study the twisted open strings further, we will also introduce the twisted momenta: 
\begin{gather}
\hat{p}_\nrmu (\xi,t) \equiv \frac{1}{4\pi} \pl_t \hx_\nrmu^\s (\xi,t) =\frac{1}{4\pi} (\hj_\nrmu (\xi,t) +(-1)^{u+1} \hj_\nrmu (-\xi,t)) \,.
\end{gather}
With this definition and the mode expansions \eqref{hx-Mode}, we find after some computation the following quasi-canonical equal-time algebra on the 
strip $0\leq \xi,\eta \leq \pi$:
\begin{subequations}
\begin{align}
& [\hat{p}^\s_\nrmu (\xi,t) ,\hat{p}^\s_{\nsn v} (\eta,t)] \nn \\
&\bigspc \quad \quad \quad =\frac{(-1)^{u+1}}{4\pi} \sG_{\nrmu ;\nsn v}(\s) \pl_\xi \Big{(} \sin ((\xi +\eta) \bar{y}(r,u))\, 
   \de(\xi+\eta) \Big{)} \label{pp-Comm} \\
&[\hx_\s^\nrmu (\xi,t) ,\hat{p}^\s_{\nsn v}(\eta,t)] \bigspc \bigspc \bigspc \bigspc \nn \\
& \bigspc \quad = i\de_{\nsn v}{}^{\!\!\nrmu} \Big{(} \de(\xi -\eta) +(-1)^{u+1} \cos ((\xi+\eta) \bar{y}(r,u)) \,\de(\xi+\eta) \Big{)} \label{xp-Comm} \\
& [\hx_\s^\nrmu (\xi,t) ,\hx_\s^{\nsn v}(\eta,t)] = -2\pi \sG^{\nrmu ;\nsn v}(\s) \left\{ \begin{array}{cc} 0 & \text{if } \xi=\eta=0, \\ \sin (2\pi \nrrs) &
   \text{if } \xi=\eta =\pi , \\ 0 & \text{otherwise,} \end{array} \right. \label{xx-Comm} \\
&\sG^{\nrmu ;\nsn v}(\s) =\srac{1}{2} \de_{u+v,0\, \text{mod }2} \sG^{\nrm;\nsn}(\s) ,\quad \bar{y}(r,u) =\srac{\bar{n}(r)}{\r(\s)} +\srac{\bar{u}}{2} \,.
\end{align}
\end{subequations}
The $\de (\xi+\eta)$ terms in \eqref{pp-Comm}, \eqref{xp-Comm} are familiar (see e.~g. Ref.~\cite{Giusto}) as interactions of charges in the strip with {\it 
image charges} outside the strip. Moreover, Eq.~\eqref{xx-Comm} defines a set of new {\it non-commutative geometries}, which generalize known [37-41,24] 
non-commutative open-string geometries. (The non-commutative geometries in \eqref{xx-Comm} are realized at vanishing twisted $B$ field, and moreover the 
geometry is commutative for $\bar{n}(r) =0,\srac{\r(\s)}{2}$, which includes N-N, D-D, D-N and N-D boundary conditions.)

To obtain these results, we used the summation identities in App.~A and the following $[\hat{q},\hat{q}]$ commutators (for all $n(s)$):
\begin{subequations}
\label{626}
\begin{gather}
[\hat{q}_{\frac{\r(\s)}{2} ,\m,1} ,\hat{q}_{\nsn 1}] =0 \\
\bar{n}(r) \! \neq \!\srac{\r(\s)}{2} \!: \,\,\, [\hat{q}_{\nrm 1} ,\hat{q}_{\nsn 1}] =4\pi \sG_{\nrm 1;\nsn 1}(\s) \tan (\pi \nrrs) \,. \label{627b}
\end{gather}
\end{subequations}
Following Ref.~\cite{Giusto}, this choice is necessary to insure that the $[\hx ,\hx ]$ commutators are zero (i.~e. canonical) in the bulk\footnote{If any of 
the commutators in \eqref{626} are relaxed (e.~g.~choosing the commutator \eqref{627b} to vanish), the corresponding $[\hx ,\hx]$ commutator is shifted by an overall constant, and is therefore no longer canonical in the bulk.}. We call attention to the only non-zero commutator in Eq.~\eqref{626}
\begin{gather}
\bar{n}(r) \neq \srac{\r(\s)}{2} \!: \,\,\, [\hat{q}_{\nrm 1} ,\hat{q}_{\mnrn 1}] =8\pi \sG_{\nrm;\mnrn}(\s) \tan (\pi \nrrs)
\end{gather}
which, so far as we know, is a novel feature of this construction.

The simplest examples of our twisted open-string sectors are obtained by starting from a single (closed-string) free boson $x(\xi,t)$. We mention 
first the basic orientation-reversing automorphism 
\begin{gather}
c=N=1 ,\quad x(\xi,t)' =-x(-\xi,t) ,\quad \ws =1 ,\quad \r(\s)=1 ,\quad ,\bar{n}(r)=0
\end{gather}
which orbifolds to an open-string sector with the twisted bosons $\hx_\s^{00u}(\xi,t) ,\,\bar{u}=0,1$ and $\hat{c}=2$. These twisted bosons have
boundary conditions D-D for $\bar{u}=0$ and N-D for $\bar{u}=1$. Another example is the automorphism  
\begin{gather}
c=N=1  ,\quad x(\xi,t)' =x(-\xi,t) ,\quad \ws =-1 ,\quad \r(\s) =2 ,\quad \bar{n}(r)=1
\end{gather}
which orbifolds to an open-string sector with the twisted bosons $\hx_\s^{10u}(\xi,t) ,\,\bar{u}=0,1$ and $\hat{c}=2$. In this case the boundary conditions
are D-N for $\bar{u}=0$ and N-N for $\bar{u}=1$. This second situation is not an automorphism in the non-abelian case.

\subsection{Closed- and Open-String Descriptions of Open WZW Strings}

We close this paper with a brief taxonomy of open WZW strings, untwisted and twisted, with emphasis on the distinction between open-string orientifold
sectors and our new open-string orientation-orbifold sectors.

The open-string sectors of conventional WZW orientifolds [25-27,42-55] arise by tadpole cancellation in unoriented closed-string theories, and do not 
involve fractionally-moded fields. In the {\it closed-string picture} of open WZW strings, the WZW orientifolds are included in the study of boundary states 
\cite{Ishi,FS2} constructed from the general left- and right-mover untwisted current modes:
\begin{gather}
(J_a (m) +w_a{}^b \bJ_b (-m) )|B\rangle =0 ,\quad w \in Aut(g) \,.
\end{gather} 
Correspondingly, in the {\it open-string picture} \cite{Cardy1,Cardy2} of open WZW strings \cite{Giusto}, the blocks of the WZW orientifolds are included 
in the general untwisted open-string KZ system
\begin{subequations}
\begin{gather}
[J_a(m) ,J_b(n)] =if_{ab}{}^c J_c(m+n) +G_{ab} m\de_{m+n,0} \label{ALA} \\
[J_a (m) ,\tg (T,\bz,z)] =\tg (T,\bz,z) (T_a z^{-m} +\bT_a^{'} \bz^{-m}) 
\end{gather}
\begin{gather}
\pl_i \tilde{F} =\tilde{F} 2L_g^{ab} \Big{(} \sum_{j\neq i} \frac{T_a^{(i)} \otimes T_b^{(j)}}{z_i -z_j} +\sum_j \frac{T_a^{(i)} \otimes 
   \bT_b^{(j)'}}{z_i -\bz_j} \Big{)} \\
\bpl_i \tilde{F} =\tilde{F} 2L_g^{ab} \Big{(} \sum_{j\neq i} \frac{\bT_a^{(i)'} \otimes \bT_b^{(j)'}}{\bz_i -\bz_j} +\sum_j \frac{\bT_a^{(i)'} 
   \otimes T_b^{(j)}}{\bz_i -z_j} \Big{)} 
\end{gather}
\begin{gather}
\tilde{F} \sum_{i=1}^n (T_a^{(i)} \!\otimes \!\one + \one \!\otimes \!\bT_a^{(i)'} )=0 \\
\bT_a^{'} \equiv w_a{}^b \bT_b ,\quad w \in Aut(g) \\
[T_a ,T_b] =if_{ab}{}^c T_c ,\quad [\bT_a^{'} ,\bT_b^{'} ]=if_{ab}{}^c \bT_c^{'}
\end{gather}
\end{subequations} 
where $w\neq 1$ describes systems which are T-dual to the case $w=1$ discussed explicitly by Giusto and Halpern \cite{Giusto}. 

On the other hand, the new twisted open-string sectors of our WZW orientation orbifolds will be included in the study of boundary states constructed 
from general left- and right-mover {\it twisted} current modes \cite{Big}
\begin{gather}
\left( \hj_\nrm (\mnrrs) + \tilde{w} (n(r),\s)_\m{}^\n \hjb_{n(r),\n} (\mnrrs) \right) |B \rangle =0 
\end{gather}
where $\tilde{w} \in Aut (\gfrakh (\s))$ is any mode-number-preserving automorphism of the twisted right-mover current algebra. Correspondingly, these 
new sectors can be described in the open-string picture by generalizing the Giusto-Halpern construction to include {\it twisted} open WZW strings.

\vspace{-.02in}
\bigskip

\noindent
{\bf Acknowledgements}

For helpful discussions, we thank J.~de Boer, O.~Ganor, P.~Ho\v{r}ava, K.~Hori, J.~Polchinski, A.~Sagnotti and C.~Schweigert.  

This work was supported in part by the Director, Office of Energy Research, Office of High Energy and Nuclear
Physics, Division of High Energy Physics of the U.S. Department of Energy under Contract DE-AC03-76SF00098 
and in part by the National Science Foundation under grant PHY00-98840.

\pagebreak
\appendix

\section{Useful Identities}

We collect here some identities
\begin{equation}
 \tau_1 \r_\Id \tau_1 =\one -\r_\Id = (\tau_1 )_\Id {}^\Jd \r_\Jd \label{A1} 
\end{equation}
\begin{subequations}
\begin{gather}
\sqrt{2} \sum_{\Id =0}^1 U_u{}^\Id U\r_\Id U\hc = \tau_u ,\quad \sqrt{2} \sum_{\Id =0}^1 U_u{}^\Id U(\one -\r_\Id ) U\hc =(-1)^u \tau_u
  ,\quad u=0,1 \label{A2} \\
\sqrt{2} \sum_{\Id =0}^1 U_u{}^\Id U_v{}^\Id (U\hc)_\Id{}^w =\de_{u+v+w,0\, \text{mod }2} \\
\tau_3 \tau_u \tau_3 =(-1)^u \tau_u  \label{t3-Id} 
\end{gather}
\end{subequations}
\begin{subequations}
\label{Inf-Sum}
\begin{align}
&\frac{1}{z} \left( \frac{w}{z} \right)^{\nrrs +\srac{u}{2}} \sum_{m \in \Zint} \theta (m\!+\!\nrrs \!+\!\srac{u}{2} \geq 0) \left( 
   \frac{w}{z} \right)^m \bigspc \bigspc \nn \\
& \bigspc \bigspc =\frac{1}{z} \left( \frac{w}{z} \right)^{\bar{y}(r,u)} \sum_{m^{'} \in \Zint} \theta (m^{'} \!+\!\bar{y}(r,u) \geq 0) 
   \left( \frac{w}{z} \right)^{m^{'}}  \\
& \bigspc \bigspc = \left( \frac{w}{z} \right)^{\bar{y}(r,u)} \frac{1}{z-w} \left( 1 +\frac{z-w}{w} \theta (\bar{y}(r,u) \geq 1) \right)  ,\quad 
   |z| > |w| \\
& \quad \quad \bar{y}(r,u) \equiv \srac{\bar{n}(r)}{\r(\s)} +\srac{\bar{u}}{2} ,\quad 0\leq \bar{y}(r,u) \leq \srac{3}{2} -\srac{1}{\r(\s)} 
   <\srac{3}{2} ,\quad m^{'} \equiv m + \lfloor \nrrs \rfloor + \lfloor \srac{u}{2} \rfloor
\end{align}
\end{subequations}
which we found useful in the development of the text.

For the free-boson computations of Subsec.~$6.2$, we also needed the identities:
\begin{subequations}
\begin{gather}
\sin (\nrrs (\xi -\eta)) \de(\xi -\eta) =0 ,\quad 0\leq \xi,\eta \leq \pi \\
\cos (\nrrs (\xi -\eta)) \de(\xi -\eta) =\de(\xi-\eta) ,\quad 0\leq \xi,\eta \leq \pi \\
\sin (\nrrs (\xi +\eta)) \de(\xi +\eta) =0 \,\text{ except at } \xi=\eta=\pi \\
\sum_{m\neq -1} \frac{1}{m+1} \cos ((m\!+\!1)\xi )\,\cos ((m\!+\!1) \eta )= \sum_{m\neq 0} \frac{1}{m} \sin (m\xi) \,\sin (m\eta) =0
\end{gather}
\begin{align}
&\bar{n}(r) \neq 0\!: \,\,\,\sum_{m\in \Zint} \frac{1}{m\!+\!\nrrsf} \sin ((m\!+\!\nrrs )\xi) \,\sin ((m\!+\!\nrrs )\eta )= \bigspc \nn \\
&\bigspc  = \pi \int_{\xi-\eta}^{\xi +\eta} \!d\eta ' \sin (\nrrs \eta ') \de (\eta ') = \left\{ \begin{array}{cc} 0&\text{if } \xi =\eta =0 \\ 
   \srac{\pi}{2} \sin (\frac{2\pi n(r)}{\r(\s)} ) & \text{if } \xi=\eta= \pi \\ 0 & \text{otherwise }  \end{array} \right.
\end{align}
\begin{align}
&\bar{n}(r) \!\neq \!\srac{\r(\s)}{2} \!: \,\,\,\sum_{m\in \Zint} \frac{1}{m\!+\!\nrrsf \!+\!\frac{1}{2}} \cos ((m\!+\!\nrrs \!+\!\srac{1}{2})\xi )
   \,\cos ((m\!+\!\nrrs \!+\!\srac{1}{2})\eta ) \bigspc \bigspc \nn \\
&\bigspc \bigspc =-\pi\tan (\pi \nrrs) -\pi \left( \!\int_0^{\xi +\eta} \!+\!\int_0^{\xi -\eta} \right) \!d\eta ' 
   \sin ((\nrrs \!+\!\srac{1}{2})\eta ') \,\de (\eta ') \\
&\bigspc \bigspc = -\pi\tan (\pi \nrrs) + \left\{ \begin{array}{cc} 0&\text{if } \xi =\eta =0 \\ 
   \srac{\pi}{2} \sin (2\pi \nrrs) & \text{if } \xi=\eta= \pi \\ 0 & \text{otherwise } \end{array} \right. \,.
\end{align}
\end{subequations}
Following standard procedure in distribution theory, all sums here were defined by temporarily inserting a smearing function $\exp (-\eps m^2 ),\,\eps 
\rightarrow 0$, and in particular the smearing function was used to rearrange the conditionally convergent $(\cos \cdot \cos)$-type sums. Moreover, the sum 
\begin{gather}
\bar{n}(r) \neq \srac{\r(\s)}{2} \!:\,\,\,(\nrrs \!+\!\srac{1}{2}) \sum_{m \in \Zint} \frac{1}{m^2 -(\nrrsf +\frac{1}{2})^2} = -\pi \cot (\pi (\nrrs 
   +\srac{1}{2})) =\pi \tan (\pi \nrrs)
\end{gather}
was evaluated by a standard contour trick.

\section{Orientation Orbifolds in the Standard Orbifold Notation \label{OrbNotApp}}

As mentioned in the text, the major results of this paper can be expressed in the standard notation of the orbifold program \cite{Dual,More,Big}.
In this notation, we write the total action of each orientation-reversing automorphism $\hat{h}_\s =\tau_1 \!\times \!h_\s$ as
\begin{subequations}
\begin{gather}
J_{a\Id} (z)' = \tws_{a\Id}{}^{b\Jd} J_{b\Jd} (z) ,\quad \tilde{g}(T,z)' = \tWsT \tilde{g}(T,z) \hat{W}\hc (\hat{h}_\s;T) \\
\tws \equiv \tau_1 \otimes \ws ,\quad \tWsT \equiv \tau_1 \otimes W(h_\s ;T) 
\end{gather}
\end{subequations}
where $J$ and $\tilde{g}$ are respectively the two-component currents and the matrix affine primary fields of the text. The ordinary transformation
matrices $\ws$ and $W(h_\s ;T)$ are the same as those given in Eq.~\eqref{Eq2.13}. 

Similarly, the total $H$-eigenvalue problem and the total extended $H$-eigenvalue problem appear in the notation of the orbifold program as
\begin{subequations}
\begin{gather}
\tws_{a\Id}{}^{b\Jd} \tU\hc (\s)_{b\Jd}{}^{\tnrm} =\tU\hc (\s)_{a\Id}{}^{\tnrm} \tE_{\tnrm}(\s) 
  ,\quad \tE_{\tnrm}(\s) = e^{-\tp \srac{\hat{n}(r)}{\tr} } \\
\!\!\tWsT_{\a \Id}{}^{\!\be \Jd} \tU\hc (T,\s)_{\be \Jd}{}^{\!\tNrm} \!=\!\tU\hc (T,\s)_{\a \Id}{}^{\!\tNrm} 
   \tE_{\tNrm} (T,\s) \\
\tE_{\tNrm} (T,\s) \!=\!e^{\!-\tp \srac{\hat{N}(r)}{\tR}} \quad 
\end{gather}
\begin{gather}
\tU\hc (\s)_{a\Id}{}^{\tnrm} = U\hc (\s)_a{}^\nrm (U\hc )_\Id{}^u ,\quad \tU\hc (T,\s)_{\a \Id}{}^{\tNrm} 
   = U\hc (T,\s)_\a{}^\Nrm (U\hc )_\Id{}^u \\
\tnrm \equiv \nrmu ,\quad \tNrm \equiv \Nrm u ,\quad \frac{\hat{n}(r)}{\tr} \equiv \frac{n(r)}{\r(\s)} +\frac{u}{2} ,\quad 
   \frac{\hat{N}(r)}{\tR} \equiv \frac{N(r)}{R(\s)} +\frac{u}{2}
\end{gather}
\end{subequations}
where $\tnrm ,\tNrm$ are {\it super-indices} which label the energies and degeneracies of the total eigenvalue problems. The ordinary eigenvector matrices 
$U, U(\s)$ and $U(T,\s)$ are those which appear in the text. We mention in particular that $\tr =2\r(\s)$ for $\r(\s)$ odd, $\tr=\r(\s)$ for $\r(\s)$ 
even, and similarly for $\tR$.

In this notation, the total duality transformations in Eqs.~\eqref{Eq 2.26a}, \eqref{Eq3.7} and \eqref{Eq 3.18a} 
are nothing but the standard formulae \cite{Dual,More,Big} for the basic duality transformations of the orbifold program:
\begin{subequations}
\label{B.3}
\begin{gather}
\sG_{\tnrm;\hat{n}(s)\hat{\n}} (\s) \!\equiv \!\hat{\schi}(\s)_\tnrm \hat{\schi}(\s)_{\hat{n}(\s)\hat{\n}}
   \tU (\s)_\tnrm{}^{a\Id} \tU (\s)_{\hat{n}(\s)\hat{\n}}{}^{b\Jd} G_{a\Id;b\Jd} \nn \\
   =\sG_{\nrmu;\nsnv} (\s) =2 \de_{u+v,0\,\text{mod }2} \sG_{\nrm;\nsn}(\s) \\
\!\!\scf_{\tnrm ;\hat{n}(s)\hat{\n}}{}^{\!\!\hat{n}(t)\hat{\de}}(\s) \!\equiv \!\hat{\schi}(\s)_\tnrm \hat{\schi}(\s)_{\hat{n}(s)\hat{\n}} 
   \hat{\schi}(\s)^{-1}_{\hat{n}(t)\hat{\de}} \hU(\s)_\tnrm{}^{a\Id} \hU(\s)_{\hat{n}(s)\hat{\n}}{}^{b\Jd} \bigspc \nn \\
\bigspc \bigspc \times f_{a\Id ;b\Jd}{}^{c\Kd} \hU\hc (\s)_{c\Kd}{}^{\hat{n}(t)\hat{\de}} \nn \\
\quad \quad \quad =\scf_{\nrmu;\nsnv}{}^{\ntd w}(\s) =\de_{u+v-w,0\,\text{mod }2} \scf_{\nrm;\nsn}{}^\ntd (\s)
\end{gather}
\begin{gather}
{\cL}_{\sgb(\s)}^{\tnrm ;\hat{n}(s)\hat{\n}}(\s) \!\equiv \!\hat{\schi}(\s)_\tnrm^{-1} \hat{\schi}(\s)_{\hat{n}(s)\hat{\n}}^{-1} 
   L_g^{a\Id ;b\Jd} \hU\hc (\s)_{a\Id}{}^\tnrm \hU\hc (\s)_{b\Jd}{}^{\hat{n}(s)\hat{\n}} \bigspc \nn \\
   \bigspc ={\cL}_{\sgb(\s)}^{\nrmu ;\nsnv} (\s) =\srac{1}{2} \de_{u+v,0\,\text{mod }2} {\cL}_{\sgb(\s)}^{\nrm;\nsn}(\s) \\
\st_\tnrm (T,\s) \!\equiv \!\hat{\schi}(\s)_\tnrm \hU(\s)_\tnrm{}^{a\Id} \hU(T,\s) T_{a\Id} \hU\hc (T,\s) \nn \\
 = \st_\nrmu (T,\s) = \st_\nrm (T,\s) \tau_u 
\end{gather}
\begin{gather}
\hat{\schi}(\s)_\tnrm \equiv \sqrt{2} \schisig_\nrm ,\quad T_{a\Id} =T_a \r_\Id ,\quad G_{a\Id;b\Jd} =\de_{\Id \Jd} G_{ab} \\
f_{a\Id ;b\Jd}{}^{c\Kd} = \de_{\Id \Jd} \de_\Jd{}^\Kd f_{ab}{}^c ,\quad L_g^{a\Id ;b\Jd} =\de^{\Id \Jd} L_g^{ab} \,.
\end{gather}
\end{subequations}
All the duality transformations above are super-index periodic $\hat{n}(r) \!\rightarrow \!\hat{n}(r) \!\pm \!\tr$, $\hat{N}(r) \!\rightarrow \!\hat{N}(r) 
\!\pm \!\tR$ as a consequence of the separate periodicities \eqref{Eq3.9} of $n(r), N(r)$ and $u$.

As another example, the exact twisted current-current operator product \eqref{Eq3.31} in open-string sector $\hat{h}_\s$ of the orientation orbifold 
takes the super-index form 
\begin{align}
&\hj_\tnrm (z,\s) \hj_{\hat{n}(s)\hat{\n}}(w,\s) =\left( \frac{w}{z} \right)^{\frac{\overline{\hat{n}(r)}}{\tr}} \Big{\{} \left[
   \frac{1}{(z-w)^2} +\frac{\overline{\hat{n}(r)} /\tr}{w(z-w)} \right] \sG_{\tnrm ;\hat{n}(s)\hat{\n}}(\s) \label{hJ-hJ-OrbNot} \\
&\quad \quad +\frac{i \scf_{\tnrm ;\hat{n}(s)\hat{\n}}{}^{\hat{n}(r)+\hat{n}(s),\hat{\de}}(\s) \hj_{\hat{n}(r)+\hat{n}(s),\hat{\de}}
   (w,\s)}{z-w} \Big{\}} + :\!\hj_\tnrm (z,\s) \hj_{\hat{n}(s)\hat{\n}}(w,\s) \!:_M \nn
\end{align}
which is recognized as nothing but the exact form of the twisted current-current operator product given for space-time orbifolds in Refs.~\cite{More,Big}, now with $\nrm \rightarrow \tnrm$. Similarly, 
the twist-field conformal weight \eqref{Eq 3.36c} can be written as 
\begin{gather}
\gscfwt = \sum_{r,\hat{\m},\hat{\n}} \!{\cL}_{\sgb(\s)}^{\tnrm ;-\hat{n}(r),\hat{\n}}(\s) \sG_{\tnrm ;-\hat{n}(r),\hat{\n}}
  (\s) \srac{\overline{\hat{n}(r)}}{2\hat{\r}(\s)} \left( 1-\srac{\overline{\hat{n}(r)}}{\hat{\r}(\s)} \right) \label{gscfwt-OrbNot}
\end{gather}
which is the standard form \cite{Big,Big'} in the orbifold program. 

To check the identity \eqref{hJ-hJ-OrbNot}, one first needs the following relation
\begin{subequations}
\label{nhatbar}
\begin{align}
\srac{\overline{\hat{n}(r)}}{\hat{\r}(\s)} &\equiv \srac{\hat{n}(r)}{\hat{\r}(\s)} - \Big{\lfloor} \srac{\hat{n}(r)}{\hat{\r}(\s)} \Big{\rfloor} 
   =\srac{n(r)}{\r(\s)} +\srac{u}{2} -\Big{\lfloor} \srac{n(r)}{\r(\s)} +\srac{u}{2} \Big{\rfloor} \\
& =\srac{\bar{n}(r)}{\r(\s)} +\srac{\bar{u}}{2} -\Big{\lfloor} \srac{\bar{n}(r)}{\r(\s)} +\srac{\bar{u}}{2} \Big{\rfloor} \label{B6b} \\
& =\bar{y}(r,u) -\theta (\bar{y}(r,u) \geq 1) ,\quad \bar{y}(r,u) = \srac{\bar{n}(r)}{\r(\s)} +\srac{\bar{u}}{2}  \label{B6c}
\end{align}
\end{subequations}
where $\lfloor x \rfloor$ is the floor of $x$. Here \eqref{B6b} holds because $\srac{n(r)}{\r(\s)} +\srac{u}{2}$ differs from 
$\srac{\bar{n}(r)}{\r(\s)} +\srac{\bar{u}}{2}$ by an integer and $\lfloor x+n \rfloor =\lfloor x\rfloor +n$. Then the non-trivial identities
\begin{subequations}
\label{f-Ids}
\begin{gather}
\left( \frac{w}{z} \right)^{\frac{\overline{\hat{n}(r)}}{\tr}} \left[ 1+ \srac{\overline{\hat{n}(r)}}{\tr} \frac{(z-w)}{w} \right] =
   \left( \frac{w}{z} \right)^{\bar{y}(r,u)} h(z,w;\bar{y}(r,u)) \\
\left( \frac{w}{z} \right)^{\frac{\overline{\hat{n}(r)}}{\tr}} =\left( \frac{w}{z} \right)^{\bar{y}(r,u)} f(z,w;\bar{y}(r,u)) \label{f-Ids1} 
\end{gather}
\end{subequations}
are verified from \eqref{nhatbar} and the definitions of $h ,f$ in Eq.~\eqref{Eq3.31}. The form of the operator product in \eqref{hJ-hJ-OrbNot} follows 
immediately from \eqref{f-Ids}, \eqref{Eq3.7} and \eqref{Eq 3.31a}. 

The conformal weight  identity \eqref{gscfwt-OrbNot} is somewhat more difficult to establish, and the reader may find the following steps helpful:
\begin{subequations}
\begin{align}
& \sum_{r,\hat{\m},\hat{\n}} \!{\cL}_{\sgb(\s)}^{\tnrm ;-\hat{n}(r),\hat{\n}}(\s) \sG_{\tnrm ;-\hat{n}(r),\hat{\n}}
  (\s) \srac{\overline{\hat{n}(r)}}{2\hat{\r}(\s)} \left( 1-\srac{\overline{\hat{n}(r)}}{\hat{\r}(\s)} \right) \nn \\
& \quad \,\, =\!\! \sum_{r,\m,\n,u,v} \!\left( \srac{1}{2} \de^{uv} {\cL}_{\sgb(\s)}^{\nrm ;\mnrn}(\s) \right) 
  \left( 2\de_{uv} \sG_{\nrm ;\mnrn}(\s) \right) \srac{1}{2} \!\left[ -\bar{y}(r,u)^2 +\bar{y}(r,u) \quad \right. \nn \\
& \bigspc \bigspc \bigspc \left. +2\bar{y}(r,u) \theta (\bar{y}(r,u) \geq 1) -2\theta (\bar{y}(r,u) \geq 1) \right] \label{B8a} \\
& \quad \,\, =\!\sum_{r,\m,\n,u} \!{\cL}_{\sgb(\s)}^{\nrm ;\mnrn}(\s) \sG_{\nrm ;\mnrn}(\s) (1-\bar{y}(r,u)) \left( \srac{1}{2} \bar{y}(r,u) -\theta 
   (\bar{y}(r,u) \!\geq \!1) \right) \,. \label{B8b}
\end{align}
\end{subequations}
Here one uses the identity in \eqref{B6c} and the forms in \eqref{B.3} to obtain Eq.~\eqref{B8a}. The result in 
Eq.~\eqref{B8b} agrees with the form \eqref{Eq 3.36b} of the conformal weight given in the text.

Finally, the two sets of global Ward identities in Eqs.~\eqref{Eq 4.24d}, \eqref{Eq 4.24e} can be combined in the super-index notation to read
\begin{subequations}
\begin{gather}
\tilde{\hat{A}}_\s (\st,\bz,z) \sum_{i=1}^n (\st_{\hat{0},\hat{\m}}^{(i)} \!\otimes \!\one +\one \!\otimes \!\bar{\st}_{\hat{0},\hat{\m}}^{(i)} ) =0 
  ,\quad \forall \m \\
\{ \st_{\hat{0} ,\hat{\m}} \} = \{ \st_{0,\m,0} \} \text{  for } \r(\s) \text{ odd,} \quad \{ \st_{\hat{0} ,\hat{\m}} \} =\{ \st_{0,\m,0} ,\, 
   \st_{\r(\s)/2 ,\m ,1} \} \text{  for } \r(\s) \text{ even}
\end{gather}
\end{subequations}
because $\srac{\overline{\hat{n}(r)}}{\tr} =0$ for the case $(\srac{\bar{n}(r)}{\r(\s)}=\srac{1}{2} ,\,\bar{u}=1)$ as well as for the case
$(\bar{n}(r)=0 ,\,\bar{u}=0)$.

The super-index form of the open-string twisted KZ equations is given in Sec.~$4.5$.

\section{Redundancy in the Open-String Twisted KZ Systems}

In the text we asserted that the $\pl$ and $\bpl$ differential equations of WZW orientation orbifold theory are redundant under the world-sheet parity 
transformation \eqref{Eq 3.8b} of the twisted affine primary field $\hg$, and we sketch here the proof of this statement.

We begin with the simpler case of the twisted vertex operator equations. The $\bpl \hg$ equation \eqref{Eq 3.10b} is obtained from the $\pl \hg$ 
equation \eqref{Eq 3.10a} by following the steps
\begin{subequations}
\begin{align}
&\bpl \hg(\st(T),\bz,z) =\tau_3 \pl_w \hg(\st(\bT),\bw,w)^t \tau_3 \Big{|}_{w \rightarrow \bz} \label{E1a} \\
&\quad \quad = \tau_3 {\cL}_{\sgb(\s)}^{\nrm;\nsn}(\s) \sum_{u=0}^1 \left( :\!\hj_\nrmu (w) \hg(\st(\bT),\bw,w) \st_\nrmu (\bT) \!: \right)^t \!\tau_3 
   \Big{|}_{w \rightarrow \bz} \label{E1b} \\
&\quad \quad ={\cL}_{\sgb(\s)}^{\nrm;\nsn}(\s) \sum_{u=0}^1 :\!\hj_\nrmu (w) (\tau_3 \st_\nrmu (\bT)^t \tau_3 )(\tau_3 \hg(\st(\bT),\bw,w)^t \tau_3 )\!: 
   \Big{|}_{w \rightarrow \bz} \\
&\quad \quad ={\cL}_{\sgb(\s)}^{\nrm;\nsn}(\s) \sum_{u=0}^1 :\!\hj_\nrmu (\bz) (\tau_3 \st_\nrmu (\bT)^t \tau_3) \hg(\st(\bT),\bz,z) \!: \label{E1d} \\
&\quad \quad =-{\cL}_{\sgb(\s)}^{\nrm;\nsn}(\s) \sum_{u=0}^1 :\!\hj_\nrmu (\bz) \tilde{\st}_\nrmu (T) \hg(\st(\bT),\bz,z) \!: \label{E1e}
\end{align}
\end{subequations}
where the world-sheet parity of $\hg$ was used in \eqref{E1a}, and \eqref{Eq 3.10a} is used in \eqref{E1b}. To obtain \eqref{E1e} from \eqref{E1d} we 
used Eqs.~\eqref{Eq2.21} and \eqref{t3-Id}.

Similarly (see Eq.~\eqref{Eq4.18}), one may use the world-sheet parity \eqref{Eq 3.8b} to derive the $\bpl_i$ twisted KZ equations from the $\pl_i$ twisted KZ equations, 
and vice-versa. This demonstration is more involved, and we begin with the proof of Eq.~\eqref{Eq4.17} in the text
\begin{subequations}
\begin{align}
& \hat{A}_\s(\st(T),\bz,z) \!=\!\langle \hg(\st(T^{(1)}),\bz_1,z_1) \cdots \hg(\st(T^{(i)}),\bz_i,z_i) \cdots \hg(\st(T^{(n)}),\bz_n,z_n) \rangle_\s \\
&\quad =\!\langle \hg(\st(T^{(1)}),\bz_1,z_1) \cdots \tau_3^{(i)} \hg(\st(\bT^{(i)}),z_i,\bz_i)^t \tau_3^{(i)} \cdots \hg(\st(T^{(n)}),\bz_n,z_n) 
   \rangle_\s \nn \\
&\quad = \tau_3^{(i)} \hat{A}_\s (\st (T'),\bw,w)_{\ast}{}^{t_{(i)}} \tau_3^{(i)} \label{A-WSPar} \\
& \bigspc \bigspc \hat{A}_\s (\st (T'),\bw,w)_{\ast} \equiv \hat{A}_\s (\st (T'),\bw,w) \BIG{|}_{\stackrel{T^{(i)}{}' =\bT^{(i)} ,\,\,w_i =\bz_i 
  \quad \quad}{\tyny{ T^{(j)}{}' =T^{(j)} ,\,\,w_j =z_j ,\,\, j\neq i}}} \label{Astar-Defn}
\end{align}  
\end{subequations}
where superscript $t_{(i)}$ is matrix transpose in the $i$th subspace. This allows us to express $\bpl_i \hat{A}_\s$ in the following form:
\begin{gather}
\bpl_i \hat{A}_\s (\st(T),\bz,z) = \tau_3^{(i)} \pl_i \hat{A}_\s (\st(T'),\bw,w)_{\ast}{}^{t_{(i)}} \tau_3^{(i)} \,. \label{E2d}
\end{gather}
We will also need the identity
\begin{align}
&\pl_i \hat{A}_\s (\st(T'),\bw,w)_{\ast} ={\cL}_{\sgb(\s)}^{\nrm;\mnrn}(\s) \times \nn \\
&\quad \times \sum_{u=0}^1 \BIG{[} \sum_{j\neq i} \left( \frac{z_j}{\bz_i} \right)^{\!\!\bar{y}(r,u)} \! \frac{f(\bz_i ,z_j 
   ;\bar{y}(r,u))}{\bz_i -z_j} \hat{A}_\s (\st(T'))_{\ast} \st_\nrmu (T^{(j)}) \st_{\mnrn ,-u}(\bT^{(i)}) \nn \\
& \bigspc +\sum_{j\neq i} \left( \frac{\bz_j}{\bz_i} \right)^{\!\!\bar{y}(r,u)} \! \frac{f(\bz_i ,\bz_j ;\bar{y}(r,u))}{\bz_i -\bz_j} 
   \tilde{\st}_\nrmu (T^{(j)}) \hat{A}_\s (\st(T'))_{\ast} \st_{\mnrn ,-u}(\bT^{(i)}) \nn \\
&\bigspc -\left( \frac{z_i}{\bz_i} \right)^{\!\!\bar{y}(r,u)} \! \frac{f(\bz_i ,z_i ;\bar{y}(r,u))}{\bz_i -z_i} \tilde{\st}_\nrmu (\bT^{(i)}) 
   \hat{A}_\s (\st(T'))_{\ast} \st_{\mnrn ,-u}(\bT^{(i)}) \nn \\
&\bigspc -\frac{1}{\bz_i} (\bar{y}(r,u) -\theta (\bar{y}(r,u) \geq 1)) \hat{A}_\s (\st(T'))_{\ast} \st_\nrmu (\bT^{(i)}) \st_{\mnrn ,-u}(\bT^{(i)})
   \BIG{]} \label{Step1}
\end{align}
which follows from the $\pl_i \hat{A}_\s$ equation in \eqref{Eq 4.10a} and the definition of $\hat{A}_\ast$ in \eqref{Astar-Defn}. Then using 
Eq.~\eqref{Step1} in \eqref{E2d}, we find after some algebra that
\begin{subequations}
\begin{align}
&\bpl_i \hat{A}_\s (\st(T),\bz,z) =\tau_3^{(i)} \pl_i \hat{A}(\st(T'),\bw,w)_{\ast}{}^{t_{(i)}} \tau_3^{(i)} \nn \\
&\,\, =\!{\cL}_{\sgb(\s)}^{\nrm;\mnrn}(\s) \times \nn \\
& \quad \times \!\!\sum_{u=0}^1 \!\BIG{[} \!-\!\!\sum_{j\neq i} \!\left( \frac{z_j}{\bz_i} \right)^{\!\!\bar{y}(r,u)} \!\!\frac{f(\bz_i ,z_j 
   ;\bar{y}(r,u))}{\bz_i -z_j} \tilde{\st}_{\mnrn ,-u} (T^{(i)}) \hat{A}_\s (\st(T),\bz,z) \st_\nrmu (T^{(j)}) \quad \nn \\
&\bigspc +\!\!\sum_{j\neq i} \!\left( \frac{z_i}{\bz_i} \right)^{\!\!\bar{y}(r,u)} \!\!\frac{f(\bz_i ,z_i ;\bar{y}(r,u))}{\bz_i -z_i} \tilde{\st}_{\mnrn ,-u} 
   (T^{(i)}) \tilde{\st}_\nrmu (T^{(j)}) \hat{A}_\s (\st(T),\bz,z) \nn \\
&\bigspc -\left( \frac{z_i}{\bz_i} \right)^{\!\!\bar{y}(r,u)} \frac{f(\bz_i ,z_i ;\bar{y}(r,u))}{\bz_i -z_i} \tilde{\st}_{\mnrn ,-u}(T^{(i)}) \hat{A}_\s 
   (\st(T),\bz,z) \st_\nrmu (T^{(i)}) \nn \\
& \bigspc -\frac{1}{\bz_i} (\bar{y}(r,u) -\theta (\bar{y}(r,u) \geq 1)) \tilde{\st}_{\mnrn ,-u}(T^{(i)}) \tilde{\st}_\nrmu(T^{(i)}) \hat{A}_\s 
   (\st(T),\bz,z) \BIG{]} \label{Step2}
\end{align}
\begin{gather}
\st_\nrmu (\bT^{(i)})^{t_{(i)}} =-\st_\nrmu (T^{(i)}) ,\quad \tilde{\st}_\nrmu (\bT^{(i)})^{t_{(i)}} =-\tilde{\st}_\nrmu (T^{(i)}) \\
\tau_3^{(i)} \st_\nrmu (T^{(i)}) \tau_3^{(i)} =\tilde{\st}_\nrmu (T^{(i)}) \label{stT-Ids}
\end{gather}
\end{subequations}
where we have used \eqref{A-WSPar} and Eqs.~\eqref{Eq2.21}, \eqref{t3-Id} in the form (C$.5$b,c). But combining the first and third terms in \eqref{Step2}, 
we see that this result is exactly the $\bpl_i \hat{A}_\s$ twisted KZ equation in \eqref{Eq 4.10b}.

\vskip .5cm 
\addcontentsline{toc}{section}{References} 
 
\renewcommand{\baselinestretch}{.4}\rm 
{\footnotesize 
 
\providecommand{\href}[2]{#2}\begingroup\raggedright\endgroup

\pagebreak

\end{document}